\documentclass{JHEP3}

\newcount\hour\newcount\minute
        \hour=\time \divide\hour by60 \minute=\time
        {\multiply\hour by60 \global\advance\minute by-\hour}
        \edef\militarytime{\number\hour:\ifnum\minute<10
0\fi\number\minute}

\makeatletter

\@addtoreset{equation}{section}

\def\asymp#1%
{\mathrel{\raisebox{-.4em}{$\widetilde{\scriptstyle #1}$}}}
\makeatother

\newcommand\Ref[1]     {Ref.\,\cite{#1}}
\newcommand\Refs[1]    {Refs.\,\cite{#1}}
\newcommand\eqn[1]     {Eq.\,(\ref{#1})}
\newcommand\eqns[2]    {Eqs.\,(\ref{#1}) and~(\ref{#2})}
\newcommand\eqnss[2]   {Eqs.\,(\ref{#1})--(\ref{#2})}

\newcommand\sect[1]    {Sect.\,{\ref{#1}}}

\newcommand\nn         {\nonumber}
\newcommand\kT[1]      {k_{#1\perp}}

\newcommand\as         {\ensuremath{\alpha_{\mathrm{s}}}}

\newcommand\gs         {\ensuremath{g_{\mathrm{s}}}}

\renewcommand\O        {{\mathrm O}}
\newcommand\from[2]    {}

\newcommand\bom[1]     {{\mbox{\boldmath $#1$}}}
\newcommand{\bA}[1]    {\bom{\mathrm A}_{#1}}
\newcommand{\bC}[1]    {\bom{\mathrm C}_{#1}}

\newcommand{\bS}[1]    {\bom{\mathrm S}_{#1}}
\newcommand{\bSqq}[1]  {\bom{\mathrm S}^{(q\qb)}_{#1}}
\newcommand{\bSgg}[1]  {\bom{\mathrm S}^{(gg)}_{#1}}
\newcommand{\bSab}[1]  {\bom{\mathrm S}_{#1}^{(\rm ab)}}
\newcommand{\bSnab}[1] {\bom{\mathrm S}_{#1}^{(\rm nab)}}
\newcommand{\bSA}[1]   {\bom{\mathrm S}_{#1}^{(\rm A)}}
\newcommand{\bSN}[1]   {\bom{\mathrm S}_{#1}^{(\rm N)}}
\newcommand{\bCS}[2]   {\bom{\mathrm C}_{#1}\bom{\mathrm S}_{#2}}

\newcommand{\bSCS}[1]  {\bom{\mathrm C}\kern-2pt\bom{\mathrm S}_{#1}}
\newcommand{\spa}[1]   {\la#1\ra}

\newcommand{\bT}       {\bom{T}}

\newcommand{\cS}       {{\cal S}}
\newcommand{\cT}       {{\cal T}}

\newcommand{\half}     {\frac12}
\newcommand{\sixth}    {\frac16}
\newcommand{\eighth}   {\frac18}
\newcommand\cAtree[1]  {{\cal A}_{#1}^{\rm tree}}
\newcommand\Atree[1]   {A_{#1}^{\rm tree}}
\newcommand\Sp[3]      {{\rm Split}^{\rm tree}_{#1}(#2,#3)}
\newcommand\bra[3]     {\la {\cal M}_{#1}^{#2}#3|}
\newcommand\ket[3]     {|{\cal M}_{#1}^{#2}#3\ra}
\newcommand\SME[3]     {|{\cal M}_{#1}^{(#2)}{(#3)}|^2}
\newcommand\M[2]       {\ensuremath{|{\cal{M}}_{#1}^{#2}|^2}}
\newcommand\qb         {{\bar q}}

\newcommand\cF         {{\cal F}}

\newcommand\la         {\langle}
\newcommand\ra         {\rangle}
\newcommand\epas       {8\pi\as\mu^{2\eps}}

\def\beq{\begin{equation}}
\def\eeq{\end{equation}}
\def\beeq{\begin{eqnarray}}
\def\eeeq{\end{eqnarray}}
\def\ldot{\!\cdot\!}
\def\aand{\!\!\!\!&&}

\def\hP{\hat{P}}

\def\Pab{P^{{\rm (ab)}}}
\def\Pnab{P^{{\rm (nab)}}}
\def\nn{\nonumber}

\def\AP{Altarelli--Parisi }
\def\ID{1 \kern -.45 em 1}

\newcommand{\ri}{{\mathrm{i}}}
\newcommand{\rd}{{\mathrm{d}}}
\newcommand\tsig[1]   {\sigma^{\mathrm{#1}}}
\newcommand\dsig[1]   {\rd\sigma^{{\rm #1}}}
\newcommand\dsiga[2]  {\rd\sigma^{{\rm #1,A}_{\scriptscriptstyle #2}}}

\newcommand{\eps}{\varepsilon}                            

\newcommand{\cM}{{\cal M}}

\newcommand{\alps}{\alpha_{\mathrm{s}}}
\newcommand{\CF}{C_{\mathrm{F}}}
\newcommand{\CA}{C_{\mathrm{A}}}
\newcommand{\TR}{T_{\mathrm{R}}}                                                
\newcommand{\Nc}{N_{\mathrm{c}}}

\newcommand\mathswitchr[1]{\relax\ifmmode{\mathrm{#1}}\else$\mathrm{#1}$\fi}

\def\draftdate{\relax}
\def\mda{\relax}
\def\mua{\relax}
\def\mla{\relax}
\def\draft{
\def\thtystars{******************************}
\def\sixtystars{\thtystars\thtystars}
\typeout{}
\typeout{\sixtystars**}
\typeout{* Draft mode!
         For final version remove \protect\draft\space in source file *}
\typeout{\sixtystars**}
\typeout{}
\def\draftdate{\today}
\def\mua{\marginpar[\boldmath\hfil$\uparrow$]%
                   {\boldmath$\uparrow$\hfil}%
                    \typeout{marginpar: $\uparrow$}\ignorespaces}
\def\mda{\marginpar[\boldmath\hfil$\downarrow$]%
                   {\boldmath$\downarrow$\hfil}%
                    \typeout{marginpar: $\downarrow$}\ignorespaces}
\def\mla{\marginpar[\boldmath\hfil$\rightarrow$]%
                   {\boldmath$\leftarrow $\hfil}%
                    \typeout{marginpar: $\leftrightarrow$}\ignorespaces}
\overfullrule 5pt
\oddsidemargin -15mm
\marginparwidth 29mm
}

\def\stars{\strut\leaders\hbox{*}\hfill\strut}
\def\starline{\hfil\strut\hfil\hbox to \textwidth {\stars}\hfil}

\title{Matching of Singly- and Doubly-Unresolved Limits \\[.5em]
of Tree-level QCD Squared Matrix Elements}

\author{G\'abor Somogyi and Zolt\'an Tr\'ocs\'anyi\\
University of Debrecen and \\Institute of Nuclear Research of
the Hungarian Academy of Sciences\\ H-4001 Debrecen, PO Box 51, Hungary\\
E-mail: \email{z.trocsanyi@atomki.hu}}

\author{Vittorio Del Duca\\
Istituto Nazionale di Fisica Nucleare, Sez. di Torino\\
via P. Giuria, 1 - 10125 Torino, Italy\\
E-mail: \email{delduca@to.infn.it}}

\abstract{We describe how to disentangle the singly- and doubly-unresolved 
(soft and/or collinear) limits of tree-level QCD squared matrix elements.
Using the factorization formulae presented in this paper, we outline a
viable general subtraction scheme for computing next-to-next-to-leading
order corrections for electron-positron annihilation into jets.}

\preprint{ {hep-ph/0502226}\\ {DFTT 05/05} }

\begin{document}

\clearpage

\renewcommand{\thefootnote}{\fnsymbol{footnote}}

\section{Introduction}
\label{sec:intro}

QCD, the theory of strong interactions, is an important component of
the Standard Model of elementary particle interactions. It is
asymptotically free, which allows us to compute cross sections of 
elementary particle interactions at high energies as a perturbative
expansion in the running strong coupling $\as(\mu_R)$. However,
the running coupling $\as(\mu_R)$ remains rather large at the energies
relevant at recent and future colliders. In addition, to leading order
in the perturbative expansion, the coupling varies sizeably with the
choice of the (unphysical) renormalisation scale $\mu_R$.  In
hadron-initiated processes, the situation above is worsened by the
dependence of the cross section on the (also unphysical) factorisation 
scale $\mu_F$, which separates the long-distance from the short-distance
part of the strong interaction. Thus, a leading-order evaluation of the
cross section yields rather unreliable predictions for most processes in
the theory. To improve this situation, in the past 25 years the
radiative corrections at the next-to-leading order (NLO) accuracy have
been computed. These efforts have culminated, when process-independent
methods were presented for computing QCD cross sections to NLO
accuracy, namely the slicing~\cite{Giele:1991vf,Giele:1993dj},
subtraction~\cite{Frixione:1995ms,Nagy:1996bz,Frixione:1997np}
and dipole subtraction~\cite{Catani:1996vz} methods.  In some cases,
though, the NLO corrections were found to be disturbingly large,
and/or the dependence on $\mu_R$ (and eventually $\mu_F$) was found to
be still sizeable, thus casting doubts on the applicability of the
perturbative predictions. When the NLO corrections are found to
be of the same order as the leading-order prediction, the only way to 
assess the reliability of QCD perturbation theory is the computation
of the next-to-next-to-leading order (NNLO) corrections.

In recent years severe efforts have been made to compute the NNLO
corrections to the parton distribution functions
\cite{Moch:2004pa}
and important basic processes, such as vector boson production 
\cite{Hamberg:1990np,Harlander:2002wh,Anastasiou:2003yy,Anastasiou:2003ds}
and Higgs production 
\cite{Harlander:2002wh,Anastasiou:2002yz,Anastasiou:2004xq,Anastasiou:2005qj}
in hadron collisions and jet production in electron-positron annihilation
\cite{Anastasiou:2004qd,Gehrmann-DeRidder:2004tv,Gehrmann-DeRidder:2004xe}.
These computations evaluate also the phase space integrals in $d$
dimensions, thus, do not follow the process-independent methods
used to compute the NLO corrections. 
Presently it is not clear whether those techniques 
\cite{Heinrich:2002rc,Anastasiou:2003gr,Binoth:2004jv,Heinrich:2004jv}
can be directly applied for processes with more
complex final states.  

The more traditional approach relies on defining approximate
cross sections which match the singular behaviour of the QCD cross
sections in all the relevant unresolved limits. Various attempts 
were made in this direction in 
\Refs{Kosower:1997zr,Campbell:1998nn,Weinzierl:2003fx,Weinzierl:2003ra,Kosower:2003bh,Gehrmann-DeRidder:2003bm,Frixione:2004is,Gehrmann-DeRidder:2005hi,Ridder:2005aw}.  
In general, the definition
of the approximate cross sections must rely on the single
and double unresolved limits of the QCD squared matrix elements. 
Although the infrared limits of QCD matrix elements have been extensively 
studied~\cite{Campbell:1997hg,Catani:1998nv,DelDuca:1999ha,Kosower:2002su}, 
the formulae presented in the literature do not lend themselves directly
for devising the approximate cross sections for two reasons.  The first
problem is that the various single and double soft and/or collinear
limits overlap in a very complicated way and the infrared factorization
formulae have to be written in such forms that these overlaps can be
disentangled so that double subtraction is avoided.  The second problem
is that even if the factorization formulae are written such that double
subtraction does not happen, the expressions cannot straightforwardly
be used as subtraction formulae, because the momenta of the partons in
the factorized matrix elements are unambiguously defined only
in the strict soft and collinear limits. In order to define the
approximate cross sections one also has to factorize the phase space of
the unresolved partons such that the singular factors can be integrated
and the remaining expressions can be combined with the virtual correction
leading to cross sections which are finite and integrable in four
dimensions.

In this paper we present a solution to the first problem. Due to
momentum conservation constraints the expressions presented here are
not unique. Nevertheless, these expressions may be useful in making the
second step leading to a general subtraction scheme for computing QCD
cross sections at the NNLO accuracy. We outline such a scheme for
processes without coloured partons in the initial state, but do not
explicitly define the approximate cross sections, which we leave for
later work.  

The paper is organised as follows: after setting the notation in 
\sect{sec:notation}, we review in \sect{sec:singleunresolved}
the singly-unresolved limits and the subtraction terms relevant to
NLO cross sections. In \sect{sec:doubleunresolved} we review the
doubly-unresolved limits and introduce the corresponding
subtraction term. In \sect{sec:matchingsingledoubleunresolved}
we introduce the subtraction term that regularizes the squared matrix
element in all the unresolved regions of the phase space relevant to
NNLO computations. That subtraction term is the keystone of this paper.
In Sects.~\ref{sec:iteratedsingleunresolved} and
\ref{sec:singlyunresolvedA2} we derive the iterated singly-unresolved
limits and the singly-unresolved limits of the doubly-unresolved
factorization formulae, respectively. Those singly-unresolved limits are
used in the construction of the several contributions to the subtraction 
term presented in \sect{sec:matchingsingledoubleunresolved}. In 
\sect{sec:matching} we sketch the proof of the validity of the
subtraction term of \sect{sec:matchingsingledoubleunresolved} as a correct
regulator of the divergences that occur in all the unresolved regions 
of the phase space relevant to NNLO computations. In \sect{sec:method}
we outline a possible general
subtraction method for computing NNLO corrections to jet cross sections.
In \sect{sec:concl} we draw our conclusions.

\section{Notation}
\label{sec:notation}

We consider matrix elements of processes with $m+2$ coloured particles
(partons) in the final-state.  Any number of additional non-coloured
particles is allowed, too, but they will be suppressed in the notation.
 Resolved partons will be labelled by $i,j,k,\dots$, unresolved ones
are $r$ and $s$.  

The colour indices of the partons are denoted by $c_i$, which range
over $1,\dots,N_c^2-1$ for gluons (or any other partons, such as
gluinos, in the adjoint representation of the gauge group) and over
$1,\dots,N_c$ for quarks and antiquarks (or any other partons, such as
squarks, in the fundamental representation).  Spin indices are
generically denoted by $s_i$. As in \Ref{Catani:1996vz}, 
we formally introduce an orthogonal basis of unit vectors 
$|c_1,\dots,c_m\ra\otimes|s_1,\dots,s_m\ra$ in the space of colour and
spin, in such a way that an amplitude of a process involving $m$ external
partons, $\cM_m^{\{c_I,s_I\}}(\{p_I\})$ with definite colour, spin and
momenta $\{p_I\}$ can be written as 
\beq
\cM_m^{\{c_I,s_I\}}(\{p_I\}) \equiv
\Big(\la c_1,\dots,c_m| \otimes \la
s_1,\dots,s_m|\Big)|\cM_m(\{p_I\})\ra\:.
\label{ketnotation}
\eeq
Thus $\ket{m}{}{}$ is an abstract vector in colour and spin space, and
its normalization is fixed such that the squared amplitude summed over
colours and spins is
\beq
\label{eq:M2}
|\cM_m|^2 = \bra{m}{}{}\ket{m}{}{}\:.
\eeq
The matrix element has the following formal loop expansion:
\beq
\ket{}{}{} = \ket{}{(0)}{} + \ket{}{(1)}{} + \ket{}{(2)}{} + \dots\,,
\label{FormalLoopExpansion}
\eeq
where $\ket{}{(0)}{}$ denotes the tree-level contribution,
$\ket{}{(1)}{}$ is the one-loop contribution, $\ket{}{(2)}{}$ is the 
two-loop expression and the dots stand for higher-loop contributions.
The amplitude $\ket{}{}{}$ is assumed to be renormalized. 
In this paper we study the infrared behaviour of the tree-level
contribution, therefore, renormalization concerns us only to the extent
of regularization. We use tree amplitudes obtained in conventional
dimensional regularization.  

Colour interactions at the QCD vertices are represented by associating
colour charges $\bT_i$ with the emission of a gluon from each
parton $i$. The colour charge $\bT_i= \{T_i^n \} $ is a vector
with respect to the colour indices $n$ of the emitted gluon and an
$SU(N_c)$ matrix with respect to the colour indices of the parton $i$.
More precisely, for a final-state parton $i$ the action onto the colour
space is defined by
\beq
\label{eq:colmat}
\la c_1,\dots,c_i,\dots,c_m| T_i^n | b_1,\dots,b_i,\dots,b_m \ra 
= \delta_{c_1b_1} \dots T_{c_ib_i}^n \dots \delta_{c_m b_m} \:,
\eeq
where $T_{c b}^n$ is the colour-charge matrix in the representation of
the final-state particle $i$, i.e.\ $T_{c b}^n=\ri f_{cnb}$ if $i$ is a
gluon or a gluino, $T_{\alpha\beta}^n=t_{\alpha\beta}^n$ if $i$ is a (s)quark
and $T_{\alpha\beta}^n=-t_{\beta\alpha}^n$ if $i$ is an anti(s)quark. 
Using this notation, we define the two-parton colour-correlated squared
tree-amplitudes, $|\cM^{(0)}_{(i,k)}|^2$, as
\beeq
|\cM^{(0)}_{(i,k)}(\{p_I\})|^2 \aand \equiv
\bra{}{(0)}{(\{p_I\})}
\,\bT_i \ldot \bT_k \,
\ket{}{(0)}{(\{p_I\})}
\nn \\ \aand =
\left[ \cM^{(0);\ldots a'_i  \ldots a'_k  \ldots}(\{p_I\}) \right]^*
\, T_{a'_ia_i}^n \, T_{a'_ka_k}^n
\, {\cM}^{(0);\ldots a_i  \ldots a_k  \ldots}(\{p_I\})
\:,
\label{eq:colam2}
\eeeq
and similarly the doubly two-parton colour-correlated squared
tree-amplitudes $|\cM^{(0)}_{(i,k),(j,l)}|^2$,
\beeq
|\cM^{(0)}_{(i,k)(j,l)}|^2 \aand \equiv
\bra{}{(0)}{}
\,\{\bT_i \cdot \bT_k,\bT_j \cdot \bT_l\} \,
\ket{}{(0)}{}\,,
\label{eq:colam22}
\eeeq
where the anticommutator 
$\{\bT_i \cdot \bT_k,\bT_j \cdot \bT_l\}$
is non-trivial only if $i=j$ or $k=l$, see \eqn{eq:colalg}.

In our notation, each vector $\ket{}{}{}$ is a colour-singlet
state, so colour conservation is simply
\beq
\biggl(\sum_j \bT_j \biggr) \,\ket{}{}{} = 0\,,
\label{eq:colorconserv}
\eeq
where the sum over $j$ extends over all the external partons of the
state vector $\ket{}{}{}$, and the equation is valid order by order in
the loop expansion of \eqn{FormalLoopExpansion}.

The colour-charge algebra for 
$\sum_n T_i^n T_k^n \equiv \bT_i \ldot \bT_k$ is
\beq
\bT_i \ldot \bT_k =\bT_k \ldot \bT_i \quad  {\rm if}
\quad i \neq k; \qquad \bT_i^2= C_i\:.
\label{eq:colalg}
\eeq
Here $C_i$ is the quadratic Casimir operator in the representation of
particle $i$ and we have $\CF= \TR(\Nc^2-1)/\Nc= (\Nc^2-1)/(2\Nc)$ in
the fundamental and $\CA=2\,\TR \Nc=\Nc$ in the adjoint representation,
i.e.~we are using the customary normalization $\TR=1/2$.

\section{Factorization in the singly-unresolved collinear and soft limits}
\label{sec:singleunresolved}

As explained in the introduction in order to devise the approximate
matrix elements, we have to study the factorization properties of the
relevant squared matrix elements when one, or two partons become soft or
collinear to another parton. The factorization of the tree-level matrix
elements when one parton becomes soft or collinear to another one are
well known \cite{BCM,AP}. In order to introduce a new notation, in this
section we recall the relevant formulae for a tree-level matrix element
$\ket{m+2}{(0)}{}$ with $m+2$ massless QCD partons in the final state
and two massless and colourless particles in the initial state.

\subsection{The collinear limit}
\label{sec:singlecoll}

We define the collinear limit of two final-state momenta $p_i$ and $p_r$
with the help of an auxiliary light-like vector $n_{ir}^\mu$
($n_{ir}^2=0$) in the usual way,
\beq
p_i^\mu = z_i p_{ir}^\mu - \kT{r}^\mu 
- \frac{\kT{r}^2}{z_i}\frac{n_{ir}^\mu}{2p_{ir}n_{ir}}\:, \quad
p_r^\mu = z_r p_{ir}^\mu + \kT{r}^\mu 
- \frac{\kT{r}^2}{z_r}\frac{n_{ir}^\mu}{2p_{ir}n_{ir}}\:, \quad
\label{eq:sudakov}
\eeq
where $p_{ir}^\mu$ is a light-like momentum that points towards the
collinear  direction and $\kT{r}$ is the momentum component that is
orthogonal to both $p_{ir}$ and $n_{ir}$ ($p_{ir}\cdot \kT{r}=
n_{ir}\cdot \kT{r}=0$). Momentum conservation requires that
$z_i + z_r = 1$. The two-particle invariant masses
($s_{ir} \equiv 2 p_i p_r$) of the final-state partons are 
\beq
\label{eq:scdouble}
s_{ir} = -\frac{\kT{r}^2}{z_i z_r}\:.
\eeq
The collinear limit is defined by the uniform rescaling 
\beq
\kT{r}\to \lambda \kT{r}\,,
\eeq
and taking the limit $\lambda \to 0$, when the squared matrix element of
an $(m+2)$-parton process has the following asymptotic form,
\beq
|\cM_{m+2}^{(0)}{(p_i,p_r,\dots)}|^2 \simeq
8\pi\as\mu^{2\eps}\,\frac{1}{s_{ir}}\,
\bra{m+1}{(0)}{(p_{ir},\dots)} \hP_{f_if_r}^{(0)}(z_i,z_r,\kT{r};\eps)
\ket{m+1}{(0)}{(p_{ir},\dots)}\:.
\label{Rcollfact}
\eeq
In \eqn{Rcollfact} the meaning of the $\simeq$ sign is that we have
neglected subleading terms (in this case those that are less singular
than $1/\lambda^2$).  The $(m+1)$-parton matrix elements on the
right-hand side of \eqn{Rcollfact} are obtained from the $(m+2)$-parton
matrix elements by removing partons $i$ and $r$ and replacing them with
a single parton denoted by $ir$.  The parton $ir$ carries the quantum
numbers of the pair $i+r$ in the collinear limit: its momentum is
$p_{ir}^\mu$ and its other quantum numbers (flavour, colour) are
obtained according to the following rule: anything~+~gluon gives
anything and quark~+~antiquark gives gluon. The kernels
$\hP_{f_if_r}^{(0)}$ are the $d$-dimensional Altarelli-Parisi splitting
functions, which depend on the momentum fractions of the decay products
and on the relative transverse momentum of the pair.  For the sake of
simplicity, we label the momentum fraction belonging to a certain
parton flavour with the corresponding label of the squared matrix
element, $z_{f_i} = z_i$. In the case of splitting into a pair,
only one momentum fraction is independent, 
yet, we find it more convenient to keep the functional
dependence on both $z_i$ and $z_r$. 
Depending on the $f_i$ flavours of the splitting
products the explicit functional forms are 
\beeq
\la\mu|\hP_{g_ig_r}^{(0)}(z_i,z_r,\kT{};\eps)|\nu\ra \aand=
2\CA\left[-g^{\mu\nu}\left(\frac{z_i}{z_r}+\frac{z_r}{z_i}\right)-2(1-\eps)
z_i z_r \frac{\kT{}^{\mu}\kT{}^{\nu}}{\kT{}^2}\right]\,,
\label{P0gg}
\\
\la\mu|\hP_{\qb_i q_r}^{(0)}(z_i,z_r,\kT{};\eps)|\nu\ra \aand=
\TR\left[
-g^{\mu\nu}+4z_i z_r \frac{\kT{}^{\mu}\kT{}^{\nu}}{\kT{}^2}
\right]\,,
\label{P0qq}
\\
\la r|\hP_{q_ig_r}^{(0)}(z_i,z_r;\eps)|s\ra \aand=
\delta_{rs}\CF\left[\frac{1+z_i^2}{z_r}-\eps z_r\right]
\equiv \delta_{rs} P^{(0)}_{q_ig_r}(z_i,z_r;\eps)\,,
\label{P0qg}
\eeeq
where in the last equation we introduced our notation for the
spin-averaged splitting function,
\beq
P_{f_if_r}(z_i,z_r;\eps) \equiv \la\hP_{f_if_r}(z_i,z_r,\kT{};\eps)\ra\:.
\label{eq:Pav}
\eeq
The gluon-gluon and quark-antiquark splittings are symmetric in the
momentum fractions of the two decay products, while the quark-gluon
splitting is not. Nevertheless, we do not distinguish the flavour
kernels $\hP_{qg}$ and $\hP_{gq}$.  The ordering of the flavour indices
and arguments of the \AP kernels has no meaning in our notation, i.e.,
\beq
\hP_{f_if_r}(z_i,z_r;\eps) = \hP_{f_rf_i}(z_r,z_i;\eps) \,.
\label{eq:convention}
\eeq
Thus, it is sufficient to record the kernel belonging to one ordering.
We keep this convention throughout.

In order to simplify further discussion, we introduce a symbolic
operator $\bC{ir}$ that performs the action of taking the collinear
limit of the squared matrix element, keeping the leading singular term.
Thus we can write \eqn{Rcollfact} as
\beq
\bC{ir}|\cM_{m+2}^{(0)}{(p_i,p_r,\dots)}|^2 =
8\pi\as\mu^{2\eps}\,\frac{1}{s_{ir}}\,
\bra{m+1}{(0)}{(p_{ir},\dots)} \hP_{f_if_r}^{(0)}
\ket{m+1}{(0)}{(p_{ir},\dots)}\:.
\label{Rcollfactnew}
\eeq

\subsection{The soft limit}
\label{sec:singlesoft}

The soft limit is defined by parametrizing the soft momentum as
$p_r=\lambda q_r$ and letting $\lambda\to 0$ at fixed $q_r$. Neglecting
terms that are less singular than $1/\lambda^2$, the soft limit of the squared
matrix element can be written as
\beq
|\cM_{m+2}^{(0)}{(p_r,\dots)}|^2 \simeq
-8\pi\as\mu^{2\eps} \sum_{i,k} \frac12 \cS_{ik}(r)
|\cM_{m+1;(i,k)}^{(0)}(\dots)|^2 \,,
\label{Rsoftfact}
\eeq
where 
\beq
\cS_{ik}(r) = \frac{2 s_{ik}}{s_{ir} s_{rk}}
\label{eq:Sikr}
\eeq
is the eikonal factor.
In \eqn{Rsoftfact} the $(m+1)$-parton matrix element on the right-hand
side is obtained from the $(m+2)$-parton matrix element on the
left-hand side by simply removing the soft parton.  

Similarly to the $\bC{ir}$ operator, we
introduce the symbolic operator $\bS{r}$ that performs the action of
taking the soft limit of the squared matrix element, keeping the
leading singular terms. Thus we can write \eqn{Rsoftfact} as
\beq
\bS{r}|\cM_{m+2}^{(0)}{(p_r,\dots)}|^2 =
-8\pi\as\mu^{2\eps} \sum_{i,k} \frac12 \cS_{ik}(r)
|\cM_{m+1;(i,k)}^{(0)}(\dots)|^2 \,,
\label{Rsoftfactnew}
\eeq
if $r$ is a gluon and $\bS{r}|\cM_{m+2}^{(0)}{(p_r,\dots)}|^2 = 0$ if $r$
is a quark. 

\subsection{Matching the singly-unresolved limits}
\label{sec:matchingsingleunresolved}

If we want to regularize the squared matrix elements in all
singly-unresolved regions of the phase space then we have to subtract all
possible collinear and soft limits, i.e. subtract the sum
\beq
\sum_r\left(\sum_{i\ne r} \half \bC{ir} + \bS{r}\right)
|\cM_{m+2}^{(0)}{(p_i,p_r,\dots)}|^2\,,
\label{A1candidate}
\eeq
where the $\half$ symmetry factor takes into account that in the
summation each collinear configuration is taken into account twice.
Subtracting \eqn{A1candidate} we perform a double subtraction in some
regions of the phase space where the soft and collinear limits
overlap. The collinear limit of \eqn{Rsoftfactnew} when the soft gluon
$r$ becomes simultaneously collinear to parton $i$ is
\beq
\bCS{ir}{r}|\cM_{m+2}^{(0)}{(p_i,p_r,\dots)}|^2 =
-8\pi\as\mu^{2\eps}
\frac{2}{s_{ir}} \sum_{k\ne i}\frac{z_i}{z_r}
|\cM_{m+1;(i,k)}^{(0)}(p_i,\dots)|^2\,.
\label{Rcsoftfact}
\eeq
The factor $z_i/z_r$ is independent of $k$, therefore, using colour
conservation (\eqn{eq:colorconserv}) we can perform the summation and obtain,
\beq
\bCS{ir}{r}|\cM_{m+2}^{(0)}{(p_i,p_r,\dots)}|^2 =
8\pi\as\mu^{2\eps}
\frac{2}{s_{ir}}\frac{z_i}{z_r}
\bT_i^2 |\cM_{m+1}^{(0)}(p_i,\dots)|^2\,.
\label{Rcsoftfact2}
\eeq
Similarly, the soft limit of \eqn{Rcollfactnew} when $r$ is a gluon and
$z_r \to 0$ is
\beq
\bS{r}\bC{ir}|\cM_{m+2}^{(0)}{(p_i,p_r,\dots)}|^2 =
8\pi\as\mu^{2\eps}\,\frac{1}{s_{ir}}\,
\bT_i^2 \frac{2}{z_r}
|\cM_{m+1}^{(0)}{(p_{i},\dots)}|^2\:.
\label{Rscollfact}
\eeq
\eqns{Rcsoftfact2}{Rscollfact} differ by the factor $z_i = 1 - z_r$ in
the numerator of \eqn{Rcsoftfact2}, in which $z_r$ becomes subleading
if $r$ is soft. Therefore, \eqn{Rcsoftfact2} can be used to account for
the double subtraction: it cancels the soft subtraction in the collinear
limit by construction,
\beq
\bC{ir}(\bS{r} - \bCS{ir}{r})|\cM_{m+2}^{(0)}{}|^2 = 0\,,
\label{eq:CirSr-CSir}
\eeq
and the $\bC{ir} - \bCS{ir}{r}$ difference is subleading in the soft limit,
\beq
\bS{r}(\bC{ir} - \bCS{ir}{r})|\cM_{m+2}^{(0)}{}|^2 = 0\,.
\label{eq:SrCir-CSir}
\eeq
Accordingly, in order to remove the double subtraction from
\eqn{A1candidate}, we have to add terms like that in \eqn{Rcsoftfact2}.
That amounts to always take the collinear limit of the soft factorization
formula rather than the reverse (like terms in \eqn{Rscollfact}).  Thus
the candidate for a subtraction term for regularizing the squared
matrix element in all singly-unresolved limits is
\beeq
\!\!\!\!\!\!\!\!\!\!\!\!\!\!
\bA{1}|\cM_{m+2}^{(0)}|^2 &=&
\sum_{r} \left[\sum_{i\ne r} \half \bC{ir}
+ \left(\bS{r} - \sum_{i\ne r} \bCS{ir}{r}\right) \right]
|\cM_{m+2}^{(0)}{(p_i,p_r,\dots)}|^2
\nn \\
&=&8\pi\as\mu^{2\eps}
\sum_r \sum_{i\ne r}\Bigg\{
\half\,\frac{1}{s_{ir}}
\,\bra{m+1}{(0)}{(p_{ir},\dots)} \hP_{f_i f_r}^{(0)}
\ket{m+1}{(0)}{(p_{ir},\dots)}-
\nn \\ &&\qquad\qquad\qquad\quad\quad
-\delta_{f_rg}\sum_{k\ne i,r}
\left[\frac{s_{ik}}{s_{ir} s_{rk}} - \frac{2}{s_{ir}}\frac{z_i}{z_r}\right]
|\cM_{m+1;(i,k)}^{(0)}(\dots)|^2
\Bigg\}
\,.
\label{eq:A1}
\eeeq
Note that the cancellation of the collinear terms in the soft limit
actually requires the symmetry factor multiplying the collinear term, but
not the collinear-soft one. This feature will be valid throughout the
paper.

The formula in \eqn{eq:A1} cannot be used as a true subtraction term. 
The reason is that the factorization formulae are valid in the strict
collinear and/or soft limits. In order to define subtraction formulae
over a finite part of the phase space, we either have to specify the
momentum that becomes unresolved (general solutions are presented in
\Refs{Frixione:1995ms,Nagy:1996bz}), or exact momentum conservation
has to be implemented (as in \Ref{Catani:1996vz}). In the latter case
in addition to the poles that are shown explicitly, these subtraction
terms contain singularities in the $(m+1)$-parton matrix elements,
\beeq
\bC{js}\bA{1}\M{m+2}{(0)} \aand= \bC{js} \M{m+2}{(0)}
+ \Big(\bC{js}\bA{1} - \bC{js}\Big)\M{m+2}{(0)}
\,,
\label{eq:CjsA1}
\\
\bS{s}\bA{1}\M{m+2}{(0)} \aand= \bS{s} \M{m+2}{(0)}
+ \Big(\bS{s}\bA{1} - \bS{s}\Big)\M{m+2}{(0)}
\,,
\label{eq:SsA1}
\eeeq
where
\beeq
&&
\Big(\bC{js}\bA{1} - \bC{js}\Big)\M{m+2}{(0)} =
\nn \\ && \qquad
=\sum_{r\ne j,s}\bC{js} \Bigg(
\bS{r}
- \bC{sr} \bS{r} - \bC{jr} \bS{r}
+ \sum_{i\ne r,j,s}\Big( \half\,\bC{ir} - \bC{ir}\bS{r}\Big)
\Bigg) \M{m+2}{(0)}\,,
\qquad~
\label{eq:CjsA1-Cjs}
\\ &&
\Big(\bS{s}\bA{1} - \bS{s}\Big)\M{m+2}{(0)} =
\nn \\ && \qquad
=\sum_{r\ne s}\bS{s} \left(
\bS{r} - \bC{sr}\bS{r}
+ \sum_{i\ne r,s}\Big( \half\,\bC{ir} - \bC{ir}\bS{r}\Big)
\right) \M{m+2}{(0)}\,.
\label{eq:SsA1-Ss}
\eeeq
The terms in \eqns{eq:CjsA1-Cjs}{eq:SsA1-Ss} are the iterated
singly-unresolved limits that we shall define precisely in
\sect{sec:iteratedsingleunresolved}.  In a NLO calculation these
terms give singularities in the doubly-unresolved parts of the phase
space, where the measurement function, that defines the physical
quantity (see \sect{sec:method}), becomes zero, and thus screens the
divergencies.  The same is not true in a NNLO calculation, which we
shall tackle in \sect{sec:matching}.

\section{Doubly-unresolved limits}
\label{sec:doubleunresolved}

In this section we discuss all the various limits when partons $r$ and
$s$ are simultaneously unresolved, which are needed for constructing an
approximate cross section for regularizing the squared matrix element
in the doubly-unresolved regions of the phase space.  The infrared
factorization in the various doubly-unresolved regions was discussed in
\Ref{Catani:1999ss}, but without aiming at combining the different
factorization formulae such that double counting among the various
expressions relevant for single and double soft and/or collinear limits
is avoided.  In order to set our notation and also exhibit these common
terms, we rewrite the published formulae.

At NNLO, the dependence of the squared matrix element $|{\cM}_{m+2}|^2$
on the momenta of the final-state partons after integration over the
phase space leads to leading singularities in five different situations,
\vspace*{-12pt}
\begin{enumerate}
\itemsep=-2pt
\item
the emission of a collinear parton-triplet;
\item
the emission of two pairs of collinear partons;
\item
the emission of a soft gluon and a pair of collinear partons;
\item
the emission of a soft quark-antiquark pair;
\item 
the emission of two soft gluons.
\end{enumerate}
\vspace*{-8pt}
The first case is the triply-collinear one, when three final state
partons become collinear. The next one corresponds to the
doubly-collinear region, where two final-state partons become collinear
to two other final-state partons, but not to one another.  The third
case occurs in the soft-collinear region, where the momentum of a gluon
becomes soft in any fixed direction and, at the same time, two partons
become collinear. The last two situations occur in the soft region,
where either the momenta $p_r$ and $p_s$ of a final-state
quark-antiquark pair tend to zero, with the directions of $p_r$ and
$p_s$ as well as $p_r/p_s$ fixed, or the momenta of two final-state
gluons tend to zero in any fixed direction.  In the following we recall
the most singular behaviour of the squared matrix element in those
regions of phase space.

\subsection{Emission of a collinear-parton triplet}
\label{sec:tripleclim} 

We consider three final-state partons $i$, $r$ and $s$ that can be
produced in the splitting process $irs\to i+r+s$.  The
partons $i$, $r$ and $s$ have momenta $p_i$, $p_r$ and $p_s$.  We
introduce the following Sudakov parametrization of the parton momenta,
\beq
p_j^\mu = z_j p_{irs}^\mu + \kT{j}^\mu 
- \frac{\kT{j}^2}{z_j}\frac{n_{irs}^\mu}{2p_{irs}n_{irs}}\:, \quad
j = i, r, s\:,
\label{triplesudakov}
\eeq
where $p_{irs}^\mu$ is a light-like momentum that points towards the
collinear  direction, $n_{irs}^\mu$ is an auxiliary light-like vector
($n_{irs}^2=0$) and $\kT{j}$ are the momentum components that are
orthogonal to both $p_{irs}$ and $n_{irs}$ ($p_{irs}\cdot \kT{j}=
n_{irs}\cdot \kT{j}=0$).  The variables $z_j$ and $\kT{j}$ satisfy the
constraints $z_i + z_r + z_s = 1$ and $\kT{i} + \kT{r} + \kT{s} = 0$.

The two-particle invariant masses of the final-state partons are
\beq
\label{eq:sctriple}
(p_j + p_l)^2 =
-z_j z_l\left(\frac{\kT{j}}{z_j}-\frac{\kT{l}}{z_l}\right)^2
\:,\quad j,l = i,r,s\:.
\eeq

The triply-collinear region is identified by performing the uniform
rescaling 
\beq
\kT{j} \to \lambda \kT{j}
\:,\quad j = i,r,s\:,
\eeq
and studying the limit $\lambda \to 0$. Again, we introduce the symbolic
operator $\bC{irs}$ that keeps the leading singular ($\O(1/\lambda^4)$)
terms of the squared matrix element in this limit. According to 
\Ref{Catani:1999ss}, the squared matrix element $|{\cM}_{m+2}|^2$
behaves as,
\beeq
&&
\bC{irs}|\cM_{m+2}^{(0)}(p_i,p_r,p_s,\dots)|^2 =
\nn\\
\label{RR3collfact}
&&\qquad
(8\pi\as\mu^{2\eps})^2
\bra{m}{(0)}{(p_{irs},\dots)}
\,\frac{1}{s_{irs}^2}\hP_{f_if_rf_s}^{(0)}(\{z_j,s_{jl},\kT{j}\};\eps)\,
\ket{m}{(0)}{(p_{irs},\dots)}\:,
\eeeq
where $s_{irs} = (p_i+p_r+p_s)^2$ and $\ket{m}{(0)}{(p_{irs},\dots)}$
corresponds to the $m$-parton matrix element that is obtained from the
$(m+2)$-parton matrix element by replacing the three partons $i, r, s$
by the single parton $irs$\footnote{According to the convention of 
\eqn{eq:convention}, we have Altarelli-Parisi kernels for all the 
permutations of the flavour indices, which are obtained by permuting 
suitably the respective momenta in their arguments; {\it e.g.} $P_{gqg}$ 
is obtained from $P_{qgg}$ by swapping the first two momenta.}.

In general the kernel $\hP_{f_if_rf_s}(\{z_j,s_{jl},\kT{j}\};\eps)$ in
\eqn{RR3collfact} contains spin correlations of the parent
parton. However, in the case of splitting processes that involve a
fermion as a parent parton, the spin correlations are absent, therefore,
we can write the corresponding spin-dependent splitting function in terms
of its average over the polarizations of the parent fermion,
\beq
\la s| \hP_{q f_r f_s}(\{z_j,s_{jl},\kT{j}\};\eps)| s'\ra =
\delta_{ss'} P_{q f_r f_s}(\{z_j,s_{jl}\};\eps)\:,
\eeq
where we used our generic notation for the spin-averaged splitting
function (cf.~\eqn{eq:Pav}).  The spin-averaged splitting function is
presented in \Refs{Campbell:1997hg,Catani:1999ss} explicitly. In order to
better exhibit the soft structure of these kernels, we give different
forms of the same functions. For the $q \to q \qb' q'$ process we
write
\beeq
&&
\frac{1}{s_{irs}^2}
P_{q_i\qb_r'q_s'}(z_i,z_r,z_s,s_{ir},s_{is},s_{rs};\eps)
=
\nn \\ && \qquad\quad
=\CF \TR
\Bigg\{\frac{1}{s_{irs}s_{rs}}
\Bigg[\frac{z_i}{z_r + z_s} 
- \frac{s_{ir} z_s + s_{is} z_r}{s_{rs} (z_r + z_s)}
+ \frac{s_{ir} s_{is}}{s_{rs} s_{irs}}
+ \frac{s_{irs}}{s_{rs}} \frac{z_r z_s}{(z_r + z_s)^2}-
\nn \\ && \qquad\qquad\qquad\qquad
- \frac{z_r z_s}{z_r + z_s}
+ \frac{1 - \eps}{2}\left(z_r + z_s - \frac{s_{rs}}{s_{irs}}\right)  \Bigg]
+ (r \leftrightarrow s) \Bigg\}\:.
\label{eq:avPqqq}
\eeeq

We decompose the splitting functions that involve one, or two gluons
into abelian and non-abelian contributions. For the $q \to qgg$ splitting
we only need the spin-averaged functions,
\beq
P_{q g g} = \CF\,\Pab_{q g g} + \CA\,\Pnab_{q g g}\:,
\label{Pggq-decomp}
\eeq
where for the abelian part we write
\beeq
&&
\frac{1}{s_{irs}^2}
\Pab_{q_i g_r g_s}(z_i,z_r,z_s,s_{ir},s_{is},s_{rs};\eps) =
\nn \\ && \qquad
=\CF\Bigg\{\left(
  \frac{1-z_s}{s_{irs} s_{ir}}
+ \frac{1-z_r}{s_{irs} s_{is}}
+ \frac{z_i}{s_{ir} s_{is}}
\right)
\frac1{z_r}\left(\frac{1+z_i^2}{1 - z_i}
- \eps z_s - \eps (1 + \eps) \frac{z_r}2 \right)+
\nn \\ && \qquad\qquad\quad
+ \frac{1 - \eps}{s_{irs}^2}
  \left[\eps - \frac{s_{irs}}{s_{ir}} (1 + \eps) (3 - z_r - 2z_s)
- \frac{s_{is}}{s_{ir}}(1 - \eps)\right]
+ (r \leftrightarrow s) \Bigg\}\:,
\label{eq:avPggqab}
\eeeq
and for the non-abelian part we use
\beeq
&&
\frac{1}{s_{irs}^2}
\Pnab_{q_i g_r g_s}(z_i,z_r,z_s,s_{ir},s_{is},s_{rs};\eps) =
\nn \\ && \qquad
=\CF \Bigg\{\frac{1}{s_{irs}s_{rs}}\Bigg[(1-\eps)\left(
 \frac{s_{ir} z_s + s_{is} z_r}{s_{rs} (z_r + z_s)}
 - \frac{s_{ir} s_{is}}{s_{irs} s_{rs}}
 - \frac{s_{irs}}{s_{rs}} \frac{z_r z_s}{(z_r + z_s)^2}\right)-
\nn \\ && \qquad\qquad\qquad\quad
 - z_i \left(\frac{4}{z_r + z_s} - \frac1{z_r}\right)\Bigg]
- \frac{1}{s_{irs}s_{ir}} \frac{(1 - z_r)^2 + (1 - z_s)^2}{2 z_r (z_r+z_s)}-
\nn \\ && \qquad\quad
- \frac{1}{s_{ir} s_{is}} \frac{z_i}{2 z_r} \frac{1 + z_i^2}{z_r + z_s}
+ \frac1{2 s_{ir} s_{rs}} \left[\frac{1 + z_i^2}{z_s}
  +  \frac{1 + (1 - z_s)^2}{z_r + z_s}
 \right]+
\nn \\ && \qquad\quad
+ \frac{(1 - \eps)^2}{2 s_{irs}^2}
+ \frac1{s_{irs} s_{rs}} \left[(1 - \eps)
  \left(\frac2{z_r} - \frac1{z_r + z_s}\right)\frac{(z_r - z_s)^2}4 - 1\right]+
\nn \\ &&\qquad\quad
+ \frac1{2 s_{irs} s_{ir}} \left[\frac{1 + (1 - z_s)^2}{z_r}
  - \frac{4 - 2 z_s + z_s^2 - z_r}{z_r + z_s}\right]+
\nn \\ &&\qquad\quad
+ \frac\eps{2} \Bigg[
  \frac1{s_{irs} s_{ir}}
  \left((1-z_s) \left(\frac{z_r}{z_s} + \frac{z_s}{z_r} - \eps\right)
 - \frac{z_r^2 (1 - z_s)}{z_s (z_r + z_s)}\right)
+ \frac{z_i}{s_{ir} s_{is}}\times
\nn \\ &&\qquad\qquad\quad
\times 
  \left(\frac{z_r}{z_s} + \frac{1 + \eps}2\right)
 - \frac1{s_{ir} s_{rs}}
   \left(\frac{(z_r + z_s)^2}{z_s} + \frac{z_s^2}{z_r + z_s}\right)
\Bigg]
+ (r \leftrightarrow s) \Bigg\}\:.
\label{eq:avPggqnab}
\eeeq

In the case of the $g \to g q \qb$ process we decompose the
spin-dependent splitting function as
\beq
\la \mu| \hP_{g q \qb}|\nu\ra =
\CF\,\la \mu| \hP^{{\rm (ab)}}_{g q \qb}|\nu\ra  
+ \CA\,\la \mu| \hP^{{\rm (nab)}}_{g q \qb}|\nu\ra \:.
\label{Pgqq-decomp}
\eeq
The explicit form of the abelian term is
\beeq
&&
\frac{1}{s_{irs}^2}
\la \mu| \hP^{{\rm (ab)}}_{g_i q_r \qb_s}
(s_{ir},s_{is},s_{rs},\kT{i},\kT{r},\kT{s})
|\nu\ra =
\nn \\ && \qquad
=\TR \Bigg\{- g^{\mu\nu}
\left[\frac{1} {s_{ir} s_{is}} - \frac{1}{s_{irs}^2}
 - \frac{2}{s_{irs} s_{ir}}
   \left(1 - \frac{1 - \eps}{2}\,\frac{s_{ir} + s_{is}}{s_{irs}}\right)
\right]+
\nn \\ && \qquad\qquad
+\, \frac{2}{s_{irs} s_{ir} s_{is}}
\Big[ \kT{r}^\mu \kT{s}^\nu + \kT{r}^\nu \kT{s}^\mu
   - (1 - \eps) \kT{i}^\mu \kT{i}^\nu \Big]
+ (r \leftrightarrow s)\Bigg\}\,,
\label{eq:Pgqqab}
\eeeq
and that of the non-abelian term is
\beeq
&&
\frac{1}{s_{irs}^2}
\la \mu| \hP^{{\rm (nab)}}_{g_i q_r \qb_s}
(z_i,z_r,z_s,s_{ir},s_{is},s_{rs},\kT{i},\kT{r},\kT{s})
|\nu\ra =
\nn \\ && \qquad
=\TR \Bigg\{
- \frac{g^{\mu\nu}}{s_{irs}s_{rs}}
\Bigg[\frac{z_i}{z_r + z_s} 
- \frac{s_{ir} z_s + s_{is} z_r}{s_{rs} (z_r + z_s)}
+ \frac{s_{ir} s_{is}}{s_{rs} s_{irs}}
+ \frac{s_{irs}}{s_{rs}} \frac{z_r z_s}{(z_r + z_s)^2}\Bigg]+
\nn \\ && \qquad\qquad
+\, \frac{1}{s_{irs} s_{rs}}\frac{1}{z_i}
  \left(-g^{\mu\nu} + \frac{4 z_r z_s}{z_r + z_s}
        \frac{\kT{rs}^\mu \kT{rs}^\nu}{\kT{rs}^2}\right)
- \frac{g^{\mu\nu}}{2 z_i}
  \left(\frac{z_r}{s_{ir}s_{rs}} - \frac{1}{s_{irs}s_{rs}}\right)-
\nn \\ && \qquad\qquad
-\, g^{\mu\nu}\left[
 \frac{1-z_r}{2 s_{irs} s_{is}}\left(\frac{1}{z_i}+\frac{1}{z_r+z_s}\right)
- \frac{1}{2 s_{ir} s_{is}}
- \frac{1-\eps}{2 s_{irs}^2}
- \frac{z_r}{2 s_{ir} s_{rs}} \frac{1}{z_r + z_s}
\right]+
\nn \\ && \qquad\qquad
+\, \frac{1}{s_{irs} s_{ir} s_{rs}}
\Bigg[
2 (1 - \eps) \kT{r}^\mu \kT{r}^\nu
- \frac{4 z_r^2}{z_i (z_r + z_s)} \kT{s}^\mu \kT{s}^\nu +
\nn \\ && \qquad\qquad\qquad\qquad\quad
+ \left(\frac{2 z_r(z_s - z_i)}{z_i (z_r + z_s)} + 1 - \eps\right)
  \left(\kT{r}^\mu \kT{s}^\nu + \kT{s}^\mu \kT{r}^\nu\right)
\Bigg]-
\nn \\ && \qquad\qquad
-\, \frac{1}{s_{irs} s_{ir} s_{is}}
\Big[ \kT{r}^\mu \kT{s}^\nu + \kT{r}^\nu \kT{s}^\mu
   - (1 - \eps) \kT{i}^\mu \kT{i}^\nu \Big]
+ (r \leftrightarrow s)\Bigg\}\,,
\label{eq:Pgqqnab}
\eeeq
where we have introduced the abbreviation
\beq
\kT{rs}^\mu \equiv \frac{\kT{r}^\mu}{z_r} - \frac{\kT{s}^\mu}{z_s}\,.
\label{abbrev}
\eeq

The pure-gluon splitting kernel as given in Eq.~(66) of
\Ref{Catani:1999ss} is completely symmetric in its indices. Therefore,
without loss of definiteness, we can refer to that formula and obtain our
definition from that by making the change of indices $1 \to i$,
$2 \to r$ and $3 \to s$ and then the formal substitution
$\tilde{k}_{j\perp} \to \kT{j}$.

The triply-collinear factorization formulae of this section are valid in
the strict collinear limits. If these expressions are to be used as
subtraction terms, the momentum fractions $z_j$ ($j = i$, $r$, $s$) have
to be defined over the extended phase space region where the subtraction
is defined. This definition has to be such that $z_j$ vanishes only if
the corresponding parton $j$ becomes soft. Otherwise, in the
kernels $\hP_{qgg}$, $\hP^{({\rm nab})}_{gq\qb}$ and $\hP_{ggg}$ we introduce
spurious poles in those terms that contain $1/z_j$ factors. The
energy fractions $z_j = E_j/(E_i + E_r + E_s)$ fulfill such a requirement
over the whole phase space. Other definitions, for instance, in terms of
invariants, like $z_j = p_j\cdot n/[(p_i + p_r + p_s) \cdot n]$, with $n$
being a light-like momentum, e.g. the momentum of a hard parton $k$,
lead to spurious poles unless the region where the momentum $p_j$ is
collinear to the momentum $n$ is excluded, e.g. if the subtraction term is
defined there to be zero.

\subsection{Emission of two collinear pairs of partons}
\label{sec:doubleclim} 

The double collinear limit occurs when two pairs of collinear partons are
emitted. The limit is precisely defined by the usual Sudakov
parametrization \eqn{eq:sudakov} for the two pairs separately.  We
consider two pairs of final-state partons $i$, $r$ and $j$, $s$ that
can be produced in the splitting processes $ir\to i+r$ and $js\to j+s$.
The partons $i$ and $r$ have momenta $p_i$, $p_r$, their parametrization
is given in \eqn{eq:sudakov}.  The parametrization of the second
splitting is completely analogous in terms of momenta $p_{js}^\mu$,
$\kT{s}^\mu$, $n_{js}^\mu$ and parameter $z_j$.

The doubly-collinear region is identified by performing the uniform
rescaling 
\beq
\kT{r} \to \lambda \kT{r}\:,\quad
\kT{s} \to \lambda \kT{s}\:,
\label{eq:uniform2clim}
\eeq
and studying the limit $\lambda \to 0$. Neglecting terms that are less
singular than $1/\lambda^4$, the squared matrix element
$|{\cM}_{m+2}|^2$ behaves as \cite{Catani:1999ss}
\beeq
&&
\bC{ir;js}
|\cM_{m+2}^{(0)}(p_i,p_r,p_j,p_s,\dots)|^2 =
(8\pi\as\mu^{2\eps})^2\, \frac{1}{s_{ir} s_{js}}\times
\label{RR2collfact}
\\ &&\quad
\times
\bra{m}{(0)}{(p_{ir},p_{js},\dots)}
\,\hP_{f_if_r}^{(0)}(z_i,z_r,\kT{r};\eps)
\,\hP_{f_jf_s}^{(0)}(z_j,z_s,\kT{s};\eps)
\,\ket{m}{(0)}{(p_{ir},p_{js},\dots)}\:,
\hspace*{2em}
\nn
\eeeq
where $\ket{m}{(0)}{(p_{ir},p_{js},\dots)}$ corresponds to the
$m$-parton matrix element that is obtained from the $(m+2)$-parton
matrix element by replacing the two partons $i$ and $r$ by the single
parton $ir$ and the other two partons $j$, $s$ by the single parton
$js$. The $d$-dimensional Altarelli-Parisi splitting functions were given
in \eqnss{P0gg}{P0qg}. We see from \eqn{RR2collfact} that the
doubly-collinear limit factorizes completely as
\beq
\bC{ir;js} |\cM_{m+2}^{(0)}(p_i,p_r,p_j,p_s,\dots)|^2 =
\bC{ir} \bC{js} |\cM_{m+2}^{(0)}(p_i,p_r,p_j,p_s,\dots)|^2\,.
\eeq

\subsection{Soft-collinear limit}
\label{sec:s-clim} 

Next we consider the tree-level matrix element of $m+2$ partons in
the limit where the momentum $p_s$ of a final-state gluon becomes soft
and the momenta $p_i$ and $p_r$ of two {\em other} final-state partons
become collinear. The soft limit is defined by the rescaling $p_s =
\lambda_s q_s$ with letting $\lambda_s \to 0$. The collinear limit is
defined by the usual Sudakov parametrization of \eqn{eq:sudakov}.
The limits are taken uniformly, $\lambda_r = \lambda_s \equiv \lambda$. 
In this limit, the squared matrix element $|{\cM}_{m+2}|^2$ behaves
as \cite{Catani:1999ss}
\beeq
&&
\bSCS{ir;s}|\cM_{m+2}^{(0)}(p_i,p_r,p_s,\dots)|^2 =
\nn\\ && \qquad
=-(8\pi\as\mu^{2\eps})^2
\frac{1}{s_{ir}}\Bigg[
\sum_{j}
\sum_{l\ne j}
\frac12 \cS_{jl}(s)\times
\label{RRsoftcollfact}
\\ && \qquad\qquad\qquad\qquad\qquad\quad
\times
\bra{m}{(0)}{(p_{ir},\dots)}
\,\bT_j \ldot \bT_l
\,\hP_{f_if_r}^{(0)}(z_i,z_r,\kT{r};\eps)
\,\ket{m}{(0)}{(p_{ir},\dots)}
\Bigg]\, ,
\nn
\eeeq
where $\ket{m}{(0)}{(p_{ir},\dots)}$ corresponds to the $m$-parton
matrix element that is obtained from the $(m+2)$-parton matrix element
by omitting the gluon $s$ and replacing the two partons $i$, $r$ by the
single parton $ir$. In the terms of \eqn{RRsoftcollfact}, when either 
$j$ or $l$ is equal to $(ir)$, the colour charge operator is 
$\bT_{ir} = \bT_i + \bT_r$ and
\beq
\cS_{j(ir)}(s) =
\frac{2 s_{j(ir)}}{s_{js} s_{s(ir)}} \equiv
\frac{2(s_{ji} + s_{jr})}{s_{js}(s_{si} + s_{sr})}\:.
\label{Sj(ir)}
\eeq 
In the strict collinear limit 
\beq
\cS_{j(ir)}(s) = \cS_{ji}(s) = \cS_{jr}(s)\,,
\label{eq:Sj(ir)=Sji}
\eeq
which we shall use later.
Of course, $\bSCS{ir;s}|\cM_{m+2}^{(0)}(p_i,p_r,p_s,\dots)|^2 = 0$ if
parton $s$ is a quark.

\subsection{Emission of a soft $\qb q$-pair}
\label{sec:softqq}

In the region of soft $\qb q$-emission the momenta of the quark $p_r$
and that of the antiquark $p_s$ tends to zero such that $p_r=\lambda q_r$
and $p_s=\lambda q_s$ with $\lambda\to 0$ for fixed $q_r$ and $q_s$. In
this limit, the squared matrix element  diverges as $1/\lambda^4$ and
its most divergent part can be computed in terms of the eikonal current
of a soft gluon of momentum $p_r + p_s$ \cite{Catani:1999ss},
\beeq
&&
\bSqq{rs}
|\cM_{m+2}{(0)}(p_r,p_s,\dots)|^2 =
\label{RRsoftqqfact} \\ && \quad
=(8\pi\as\mu^{2\eps})^2
\frac{1}{s_{rs}^2}
\sum_{i \ne r,s} \sum_{k\ne i,r,s}
\left(\frac{s_{ir} s_{ks} + s_{kr} s_{is} - s_{ik} s_{rs}}
           {s_{i(rs)} s_{k(rs)}}
- 2 \frac{s_{ir} s_{is}}{s_{i(rs)}^2}\right)
\TR |\cM_{m;(i,k)}^{(0)}{(\dots)}|^2
\:,
\nn
\eeeq
where the matrix element on the right hand side is obtained from the
original $(m+2)$-parton matrix element by omitting the quark-antiquark
pair, having momenta $p_r$ and $p_s$.

\subsection{Emission of two soft gluons}
\label{sec:softgg} 

We consider the tree-level matrix element of $m+2$ final state partons
when a gluon of momentum $p_r$ and another gluon of momentum $p_s$
become simultaneously soft. The limit is precisely defined by rescaling
the soft-gluon momenta by an overall factor $\lambda$, $p_r=\lambda
q_r$ and $p_s=\lambda q_s$, and then performing the limit $\lambda\to 0$
for fixed $q_r$ and $q_s$. In this limit, the matrix element diverges
as $1/\lambda^2$ and its most divergent part can be computed in terms
of the two-gluon soft current $J^{\mu_1\mu_2}(q_r,q_s)$, given explicitly
in \Ref{Catani:1999ss}.

The singular behaviour of $\M{m+2}{(0)}$ at $\O(1/\lambda^4)$
can be written as a sum of an abelian and a non-abelian term
\cite{Catani:1999ss},
\beq
\bSgg{rs} \SME{m+2}{0}{p_r,p_s,\dots} =
\Big(\bSab{rs} + \bSnab{rs}\Big)\SME{m+2}{0}{p_r,p_s,\dots}\,.
\label{RRsoftggfact}
\eeq
The abelian term contains the product of two one-gluon soft functions,
$\cS_{ik}(r)$ given in \eqn{eq:Sikr},
\beq
\bSab{rs} \SME{m+2}{0}{p_r,p_s,\dots} =
(8\pi\as\mu^{2\eps})^2
\,\frac18 \sum_{i,j,k,l=1}^m \cS_{ik}(r)\,\cS_{jl}(s)
\,|{\cM}^{(0)}_{m,(i,k)(j,l)}(\dots)|^2
\:.
\label{RRsoftggfactab}
\eeq
The non-abelian contribution is written in terms of $\cS_{ik}(r, s)$,
the two-gluon soft function introduced in \Ref{Catani:1999ss},
\beq
\bSnab{rs} \SME{m+2}{0}{p_r,p_s,\dots} =
-(8\pi\as\mu^{2\eps})^2
\,\frac14 \CA \sum_{i,k=1}^m \cS_{ik}(r,
s)\,|{\cM}^{(0)}_{m,(i,k)}(\dots)|^2
\:.
\label{RRsoftggfactnab}
\eeq
The colour-correlated functions $|{\cM}^{(0)}_{m,(i,k)}|^2$ and
$|{\cM}^{(0)}_{m,(i,k)(j,l)}|^2$ are given in terms of $m$-parton
amplitudes obtained from the original $(m+2)$-parton amplitude by
omitting the gluons $r$ and $s$ (for the definitions see
\eqns{eq:colam2}{eq:colam22}, respectively).

It is useful to break the sum in \eqn{RRsoftggfactab} into sums
according to the number of hard partons in the eikonal factors,
\beeq
&&
\sum_{i,j,k,l=1}^m \cS_{ik}(p_r) \cS_{jl}(p_s)
|{\cM}^{(0)}_{m,(i,k)(j,l)}|^2 =
\nn \\ &&\qquad
=\sum_i \sum_{j \ne i} \sum_{k\ne i,j} \sum_{l\ne i,j,k}
\cS_{ik}(r) \cS_{jl}(s)
\bra{m}{(0)}{}
\,\{\bT_i\ldot\bT_k,\bT_j\ldot\bT_l\}\,
\ket{m}{(0)}{} +
\nn \\ &&\qquad\quad +
4 \sum_i \sum_{j \ne i} \sum_{k\ne i,j}
\cS_{ik}(r) \cS_{ji}(s)
\bra{m}{(0)}{}
\,\{\bT_i\ldot\bT_k,\bT_j\ldot\bT_i\}\,
\ket{m}{(0)}{} +
\nn \\ &&\qquad\quad +
2 \sum_i \sum_{k \ne i}
\cS_{ik}(r) \cS_{ik}(s)
\bra{m}{(0)}{}
\,\{\bT_i\ldot\bT_k,\bT_i\ldot\bT_k\}\,
\ket{m}{(0)}{}
\:.
\label{eq:softggab}
\eeeq
The sum in the first line requires at least four hard partons, the next
line involves at least three hard partons and the last sum contains
at least two hard partons. Therefore, for the case of two hard partons,
relevant to two-jet production in electron-positron annihilation, only
the last sum contributes.

For later convenience, we also record a new form of the two-gluon soft
function. The original expression of \Ref{Catani:1999ss} in our
notation reads 
\beq
\cS_{ik}(r, s) = \cS_{ik}^{(\rm s.o.)}(r, s) 
+ 4 \frac{s_{ir} s_{ks} + s_{is} s_{kr}}{s_{i(rs)} s_{k(rs)}}
\left[\frac{1-\eps}{s_{rs}^2} - \frac18 \cS_{ik}^{(\rm s.o.)}(r, s)\right]
- \frac{4}{s_{rs}} \cS_{ik}(rs)\ ,
\label{eq:softggnab}
\eeq
where
\beq
\cS_{ik}^{(\rm s.o.)}(r, s) = 
\cS_{ik}(s)\left(\cS_{is}(r) + \cS_{ks}(r) - \cS_{ik}(r)\right)
\label{eq:softggnabso}
\eeq
is the approximation of the soft function $\cS_{ik}(r,s)$ in the
strongly-ordered approximation and $\cS_{ik}(rs)$ is given by \eqn{eq:Sikr},
\beq
\cS_{ik}(rs) = \frac{2 s_{ik}}{s_{i(rs)} s_{k(rs)}}\,.
\label{eq:Sikrs}
\eeq
In order to exhibit the relation of this
expression to the triply-collinear splitting functions, we rewrite
\eqn{RRsoftggfactnab} using partial fractioning,
\beeq
&&\!\!\!\!\!\!\!\!\!\!\!\!
\frac14 \sum_{i,k=1}^m {\cS}_{ik}(p_r,p_s)
|{\cM}_{m(i,k)}^{(0)}(p_1,\dots p_m)|^2 =
\nn\\ && \qquad\!\!\!\!\!\!\!\!\!\!\!\!
=
\frac1{s_{rs}} \sum_i \frac1{s_{ir} + s_{is}} \sum_{k\ne i}
\bra{m}{(0)}{} \,\bT_i\ldot\bT_k\, \ket{m}{(0)}{} \times
\nn \\ &&\quad\quad\!\!\!\!\!\!
\times \Bigg[
\frac{1}{s_{rs}} (1 - \eps)
\left(\frac{s_{ir} s_{ks} + s_{kr} s_{is}}{s_{ir} + s_{is} + s_{kr} + s_{ks}}
- \frac{s_{ir} s_{is}}{s_{ir} + s_{is}}\right)-
\nn \\ &&\quad\quad\quad\!\!\!\!
- s_{ik} \left(\frac{4}{s_{ir} + s_{is} + s_{kr} + s_{ks}}
  - \frac{1}{s_{is} + s_{kr}}\right)-
\nn \\ &&\quad\quad\quad\!\!\!\!
- \frac{s_{rs} s_{ik}^2}{s_{ir} + s_{is} + s_{kr} + s_{ks}}
\left( \frac{2}{s_{ir} s_{kr}} + \frac{1}{s_{ir} s_{ks}} \right)+
\nn \\ &&\quad\quad\quad\!\!\!\!
+ \frac{s_{ik}(s_{ir} + s_{is})}{s_{ir} (s_{ir} + s_{ks})}
+ \frac{s_{ik}(s_{ir} + s_{is})}{s_{ir} (s_{ir} + s_{is} + s_{kr} + s_{ks})}+
\nn \\ &&\quad\quad\quad\!\!\!\!
+ \frac{s_{ik}(s_{ir} + s_{is})}
  {(s_{ir} + s_{ks})(s_{ir} + s_{is} + s_{kr} + s_{ks})}
\left(\frac{s_{is}}{s_{ir}}
- \frac{s_{is} (s_{kr} + s_{ks})}{s_{ks} (s_{ir} + s_{is})} \right)
+ (r \leftrightarrow s) \Bigg]\:.
\label{eq:softggnab1}
\eeeq

Finally we note that if the two soft partons are of different flavours,
the squared matrix element does not have a leading singularity,
\beq
\bS{rs} \SME{m+2}{0}{p_r,p_s,\dots} =
\delta_{f_rf_s}\bS{rs} \SME{m+2}{0}{p_r,p_s,\dots}\,.
\eeq

\subsection{Matching the doubly-unresolved limits}
\label{sec:matchingdoubleunresolved}

If we want to regularize the squared matrix elements in all
doubly-unresolved regions of the phase space, then we have to subtract
all possible collinear and soft limits. Analogously to \eqn{eq:A1}, we
define the subtraction term $\bA{2} |\cM_{m+2}^{(0)}|^2$, that
must contain the sum
\beq
\sum_r \sum_{s\ne r} \Bigg[
  \sum_{i\ne r,s}\Bigg(
  \sixth\,\bC{irs}
+ \half\,\bSCS{ir;s}
+ \sum_{j\ne i,r,s} \eighth\,\bC{ir;js}\Bigg)
+ \half\,\bS{rs} 
\Bigg]
|\cM_{m+2}^{(0)}{(p_i,p_r,p_j,p_s,\dots)}|^2\ .
\label{eq:A2part}
\eeq
The symmetry factors ensure that, after summation, all collinear
configurations are taken into account only once.  The various terms in
the sum (\ref{eq:A2part}) overlap in those regions of the phase space
where the limits overlap.  Therefore, \eqn{eq:A2part} contains double
and even triple subtractions. We observe that at most triple overlaps
may occur because the doubly- and triply-collinear regions do not
overlap. Let us first discuss the double subtractions.

The triply-collinear region overlaps with the soft-collinear one.
The soft-collinear limit ($z_s \to 0$, $z_r + z_i \to 1$) of the
triply-collinear factorization formula, \eqn{RR3collfact}, is 
\beeq
&&
\bSCS{ir;s}\bC{irs}|\cM_{m+2}^{(0)}(p_i,p_r,p_s,\dots)|^2 =
\nn\\ &&\quad
=(8\pi\as\mu^{2\eps})^2
\bT_{ir}^2\,\frac{2}{s_{ir}s_{(ir)s}} \frac{1}{z_s}
\bra{m}{(0)}{(p_{ir},\dots)}
\,\hP_{f_if_r}^{(0)}(z_i,z_r,\kT{r};\eps)
\,\ket{m}{(0)}{(p_{ir},\dots)}\,.
\qquad
\label{eq:SC-TC}
\eeeq
In the triply-collinear limit \eqn{RRsoftcollfact} behaves as
\beeq
&&
\bC{irs}\bSCS{ir;s}|\cM_{m+2}^{(0)}(p_i,p_r,p_s,\dots)|^2 =
\nn\\ &&\quad
=(8\pi\as\mu^{2\eps})^2
\bT_{ir}^2\,\frac{2}{s_{ir}s_{(ir)s}} \frac{z_{ir}}{z_s}
\bra{m}{(0)}{(p_{ir},\dots)}
\,\hP_{f_if_r}^{(0)}(z_i,z_r,\kT{r};\eps)
\,\ket{m}{(0)}{(p_{ir},\dots)}\,.
\qquad
\label{eq:TC-SC}
\eeeq
The right-hand sides of \eqns{eq:SC-TC}{eq:TC-SC} differ in subleading
terms in the soft-collinear limit ($z_{ir} = 1 - z_s = 1 + \O(\lambda)$).
We also record two equivalent forms of \eqns{eq:SC-TC}{eq:TC-SC} that
will be useful later. In the $i || r$ collinear limit $s_{is} = z_i
s_{(ir)s}$, therefore, \eqn{eq:SC-TC} can also be written as
\beeq
&&
\bSCS{ir;s}\bC{irs}|\cM_{m+2}^{(0)}(p_i,p_r,p_s,\dots)|^2 =
\nn\\ &&\quad
=(8\pi\as\mu^{2\eps})^2
\bT_{ir}^2\,\frac{2}{s_{ir}s_{is}} \frac{z_i}{z_s}
\bra{m}{(0)}{(p_{ir},\dots)}
\,\hP_{f_if_r}^{(0)}(z_i,z_r,\kT{r};\eps)
\,\ket{m}{(0)}{(p_{ir},\dots)}\,.
\qquad
\label{eq:SC-TC_2}
\eeeq
Taking the triply-collinear limit of the soft-collinear formula together with
\eqn{eq:Sj(ir)=Sji}, we get an identical expression to \eqn{eq:SC-TC_2},
\beeq
&&
\bC{irs}\bSCS{ir;s}|\cM_{m+2}^{(0)}(p_i,p_r,p_s,\dots)|^2 =
\nn\\ &&\quad
=(8\pi\as\mu^{2\eps})^2
\bT_{ir}^2\,\frac{2}{s_{ir}s_{is}} \frac{z_i}{z_s}
\bra{m}{(0)}{(p_{ir},\dots)}
\,\hP_{f_if_r}^{(0)}(z_i,z_r,\kT{r};\eps)
\,\ket{m}{(0)}{(p_{ir},\dots)}\,.
\qquad
\label{eq:TC-SC_2}
\eeeq
The fact that \eqns{eq:SC-TC_2}{eq:TC-SC_2} are identical while
\eqns{eq:SC-TC}{eq:TC-SC} differ explicitly by sub-leading terms
is due to the ambiguity in the definition of the soft-collinear
limit beyond the leading order, represented by \eqn{eq:Sj(ir)=Sji}.

The doubly-collinear region also overlaps with the soft-collinear one.
The soft-collinear limit of the doubly-collinear formula, 
\eqn{RR2collfact}, is
\beeq
&&
\bSCS{ir;s}\bC{ir;js}|\cM_{m+2}^{(0)}(p_i,p_r,p_j,p_s,\dots)|^2 =
\label{eq:SC-DC}
\\ &&\quad
=(8\pi\as\mu^{2\eps})^2
\bT_j^2\,\frac{2}{s_{ir}s_{js}} \frac{1}{z_s}
\bra{m}{(0)}{(p_{ir},p_j,\dots)}
\,\hP_{f_if_r}^{(0)}(z_i,z_r,\kT{r};\eps)
\,\ket{m}{(0)}{(p_{ir},p_j,\dots)}\,,
\nn
\eeeq
while in the doubly-collinear limit \eqn{RRsoftcollfact} behaves as
\beeq
&&
\bC{ir;js}\bSCS{ir;s}|\cM_{m+2}^{(0)}(p_i,p_r,p_j,p_s,\dots)|^2 =
\label{eq:DC-SC}
\\ &&\quad
=(8\pi\as\mu^{2\eps})^2
\bT_j^2\,\frac{2}{s_{ir}s_{js}} \frac{z_j}{z_s}
\bra{m}{(0)}{(p_{ir},p_j,\dots)}
\,\hP_{f_if_r}^{(0)}(z_i,z_r,\kT{r};\eps)
\,\ket{m}{(0)}{(p_{ir},p_j,\dots)}\,.
\nn
\eeeq
The difference between \eqns{eq:SC-DC}{eq:DC-SC} is again subleading 
in the soft-collinear limit ($z_j = 1 - z_s = 1 + \O(\lambda)$).
 
The region of splitting into a collinear $f_i\qb q$-triplet (parton $i$ can
be either a quark or a gluon) overlaps with the soft $q\qb$-emission.  In
this overlapping region,
\beeq
&&
\bC{irs}\bSqq{rs}|\cM_{m+2}^{(0)}(p_i,p_r,p_s,\dots)|^2 =
\label{eq:TC-DSqq}
\\ &&\qquad
=(8\pi\as\mu^{2\eps})^2 \,\bT_{i}^2\,\TR
\frac{2}{s_{i(rs)} s_{rs}}
\Bigg[\frac{z_i}{z_r + z_s} 
- \frac{(s_{ir} z_s - s_{is} z_r)^2}{s_{i(rs)} s_{rs} (z_r + z_s)^2}
\Bigg]
\,|\cM_m^{(0)}(p_i,\dots)|^2
.\quad~
\nn
\eeeq
In the soft-$\qb q$ limit, this formula differs by a subleading term
from the doubly-soft limit of the triply-collinear factorization
formula,
\beeq
&&
\bSqq{rs}\bC{irs}|\cM_{m+2}^{(0)}(p_i,p_r,p_s,\dots)|^2 =
\label{eq:DS-TCqq}
\\ &&\qquad
=(8\pi\as\mu^{2\eps})^2\,\bT_{i}^2\,\TR
\frac{2}{s_{i(rs)} s_{rs}}
\Bigg[\frac{1}{z_r + z_s} 
- \frac{(s_{ir} z_s - s_{is} z_r)^2}{s_{i(rs)} s_{rs} (z_r + z_s)^2}
\Bigg]
\,|\cM_m^{(0)}(p_i,\dots)|^2
.\quad~
\nn
\eeeq
The soft $q\qb$-emission does not overlap with any other
doubly-unresolved regions.

The phase space regions where soft gluon-pair emission occurs overlap
with all other doubly-unresolved parts of the phase space (except for
the soft-$\qb q$ emission).  In the overlapping region of splitting
into a collinear $f_i g g$-triplet (parton $i$ can be either a quark or
a gluon) and soft $gg$-emission we obtain
\beq
\bC{irs}\bSgg{rs}\M{m+2}{(0)} =
\bC{irs}\bSab{rs}\M{m+2}{(0)}
+ \bC{irs}\bSnab{rs}\M{m+2}{(0)}\,,
\eeq
where
\beeq
&&
\bC{irs}\bSab{rs}|\cM_{m+2}^{(0)}(p_i,p_r,p_s,\dots)|^2 =
(8\pi\as\mu^{2\eps})^2\,\bT_{i}^2
\bT_i^2\,\frac{4 z_i^2}{s_{ir} s_{is} z_r z_s}
|\cM^{(0)}_m(p_i,\dots)|^2
\,,\quad~
\label{eq:TC-DSabgg}
\\ &&
\bC{irs}\bSnab{rs}|\cM_{m+2}^{(0)}(p_i,p_r,p_s,\dots)|^2 =
(8\pi\as\mu^{2\eps})^2\,\bT_{i}^2\,\CA \times
\nn \\ &&\qquad\qquad\qquad
\times
\Bigg[
\frac{(1 - \eps)}{s_{i(rs)} s_{rs}}
\frac{(s_{ir} z_s - s_{is} z_r)^2}{s_{i(rs)} s_{rs} (z_r + z_s)^2}
- \frac{z_i}{s_{i(rs)} s_{rs}}
  \left(\frac{4}{z_r+z_s} - \frac1{z_r} \right) - 
\nn \\ && \qquad\qquad\qquad\qquad
- \frac{1}{s_{i(rs)}  s_{ir}} \frac{2z_i^2}{z_r (z_r + z_s)}
- \frac{z_i^2}{s_{i(rs)}  s_{is}} \frac{1}{z_r (z_r + z_s)} +
\nn \\ && \qquad\qquad\qquad\qquad
+ \frac{z_i}{s_{ir} s_{rs}}
  \left(\frac{1}{z_s} + \frac{1}{z_r + z_s}\right)
+ (r \leftrightarrow s) \Bigg]
|\cM^{(0)}_m(p_i,\dots)|^2
\,.~
\label{eq:TC-DSnabgg}
\eeeq
Computing the doubly-soft limits of the $\hP^{(\rm nab)}_{f_i g g}$
splitting kernels,%
\footnote{\eqn{eq:Pgqqab} and the azimuth-dependent terms in
\eqn{eq:Pgqqnab} and in the triple-gluon splitting function do not have
leading singular terms in the doubly-soft limit, as expected.}
we can derive that
\newpage
\beeq
&&\!\!\!\!\!
\bSgg{rs}\bC{irs}|\cM_{m+2}^{(0)}(p_i,p_r,p_s,\dots)|^2 =
\nn\\ &&\qquad\!\!\!\!\!
=(8\pi\as\mu^{2\eps})^2\,\bT_{i}^2
\Bigg\{
\bT_i^2\,\frac{4}{s_{ir} s_{is} z_r z_s}+
\nn \\ && \qquad\qquad\qquad\qquad\qquad\!\!\!\!\!
+ \CA \Bigg[
 \frac{(1-\eps)}{s_{i(rs)} s_{rs}}
\frac{(s_{ir} z_s - s_{is} z_r)^2}{s_{i(rs)} s_{rs} (z_r + z_s)^2}
- \frac{1}{s_{i(rs)} s_{rs}}
  \left(\frac{4}{z_r + z_s} - \frac1{z_r}\right) -
\nn \\ && \qquad\qquad\qquad\qquad\qquad\qquad\!\!\!\!\!
- \frac{2}{s_{i(rs)} s_{ir}} \frac{1}{z_r (z_r+z_s)}
- \frac{1}{s_{i(rs)} s_{is}} \frac{1}{z_r (z_r + z_s)} +
\nn \\ && \qquad\qquad\qquad\qquad\qquad\qquad\!\!\!\!\!
+ \frac1{s_{ir} s_{rs}} \left(\frac{1}{z_s} + \frac{1}{z_r + z_s} \right)
+ (r \leftrightarrow s) \Bigg] \Bigg\}
|\cM^{(0)}_m(p_i,\dots)|^2
\,.
\label{eq:DSgg-TC}
\eeeq
We see that in the doubly-soft limit \eqns{eq:TC-DSqq}{eq:DS-TCqq}
differ by subleading terms, as well as the sum of
\eqns{eq:TC-DSabgg}{eq:TC-DSnabgg} differ from \eqn{eq:DSgg-TC}
by subleading terms ($z_i = 1 - z_r - z_s = 1 + \O(\lambda)$).

In the doubly-collinear limit, the doubly-soft
factorization formula  simplifies to
\beeq
&&
\bC{ir;js}\bSgg{rs}|\cM_{m+2}^{(0)}(p_i,p_r,p_j,p_s,\dots)|^2 =
\nn\\&&\qquad
=(8\pi\as\mu^{2\eps})^2
\bT_i^2
\,\frac{2}{s_{ir}}\frac{z_i}{z_r}
\,\bT_j^2
\,\frac{2}{s_{js}}\frac{z_j}{z_s}
\,|{\cM}^{(0)}_m(p_i,p_j,\dots)|^2
\:.
\label{eq:DC-softgg}
\eeeq
In the region of the doubly-soft gluon emission, the limit of the
doubly-collinear formula,
\beeq
&&
\bSgg{rs}\bC{ir;js}|\cM_{m+2}^{(0)}(p_i,p_r,p_j,p_s,\dots)|^2 =
\nn\\&&\qquad
=(8\pi\as\mu^{2\eps})^2
\bT_i^2
\,\frac{2}{s_{ir} z_r}
\,\bT_j^2
\,\frac{2}{s_{js} z_s}
\,|{\cM}^{(0)}_m(p_i,p_j,\dots)|^2
\:,
\label{eq:softgg-DC}
\eeeq
differs from \eqn{eq:DC-softgg} by subleading terms  ($z_i = 1 - z_r = 1
+ \O(\lambda)$, $z_j = 1 - z_s = 1 + \O(\lambda)$).

The last doubly-overlapping region is the overlap of the soft-collinear
region with the emission of two soft gluons.  In the soft-collinear limit
\eqn{RRsoftggfact} becomes
\beeq
&&
\bSCS{ir;s}\bSgg{rs}|\cM_{m+2}^{(0)}(p_i,p_r,p_s,\dots)|^2 =
\nn\\&&\qquad
=- (8\pi\as\mu^{2\eps})^2
\frac2{s_{ir}}\,\frac{z_i}{z_r}\bT_i^2
\,\sum_j \sum_{l \ne j} \frac12 \cS_{jl}(s)
|\cM_{m;(j,l)}^{(0)}{(p_i,\dots)}|^2
\:.
\label{eq:SC-softgg}
\eeeq
Note that the non-abelian part of \eqn{RRsoftggfact} does not yield a 
leading singularity in this case.
In the same region, the doubly-soft limit of the soft-collinear formula is
\beeq
&&
\bSgg{rs}\bSCS{ir;s}|\cM_{m+2}^{(0)}(p_i,p_r,p_s,\dots)|^2 =
\nn\\&&\qquad
=- (8\pi\as\mu^{2\eps})^2
\frac2{s_{ir}}\,\frac{1}{z_r}\bT_i^2\,
\,\sum_j \sum_{l \ne j} \frac12 \cS_{jl}(s)
|\cM_{m;(j,l)}^{(0)}{(p_i,\dots)}|^2
\:.
\label{eq:softgg-SC}
\eeeq
Thus, in the doubly-soft region, \eqns{eq:SC-softgg}{eq:softgg-SC}
differ by subleading terms  ($z_i = 1 - z_r = 1 + \O(\lambda)$).

In addition to the pairwise overlapping doubly-unresolved regions,
there are two regions where three limits overlap. The first
one is the overlap of the triply-collinear, the soft-collinear
and the doubly-soft emissions. In this region,
\beeq
&&
\bC{irs} \bSCS{ir;s} \bS{rs}
|\cM_{m+2}^{(0)}(p_i,p_r,p_s,\dots)|^2 =
\nn\\&&\qquad
= (8\pi\as\mu^{2\eps})^2
\frac2{s_{ir}}\,\frac{z_i}{z_r}\,\bT_i^2
\,\frac2{s_{is}}\,\frac{z_i}{z_s}\,\bT_i^2\,
|\cM_m^{(0)}{(p_i,\dots)}|^2
\:.
\label{eq:TC-SC-softgg}
\eeeq
The other region of triple overlap is that of the doubly-collinear, the
soft-collinear and the doubly-soft emissions, where
\beeq
&&
\bC{ir;js} \bSCS{ir;s} \bS{rs}
|\cM_{m+2}^{(0)}(p_i,p_r,p_j,p_s,\dots)|^2 =
\nn\\&&\qquad
= (8\pi\as\mu^{2\eps})^2
\frac2{s_{ir}}\,\frac{z_i}{z_r}\,\bT_i^2
\,\frac2{s_{js}}\,\frac{z_j}{z_s}\,\bT_i^2\,
|\cM_m^{(0)}{(p_i,p_j,\dots)}|^2
\:.
\label{eq:DC-SC-softgg}
\eeeq
Using \eqnss{eq:SC-TC}{eq:DC-SC-softgg}, we can now easily check the
following relations,
\beeq
\label{eq:overlapstart}
\from{eq:SC-TC}{eq:TC-SC}
\bSCS{ir;s}\Big(\bC{irs} - \bC{irs}\bSCS{ir;s}\Big)
|\cM_{m+2}^{(0)}(p_i,p_s,p_r,\dots)|^2 &=& 0\,,
\\ 
\from{eq:SC-DC}{eq:DC-SC}
\bSCS{ir;s}\Big(\bC{ir;js} - \bC{ir;js}\bSCS{ir;s}\Big)
|\cM_{m+2}^{(0)}(p_i,p_r,p_j,p_s,\dots)|^2 &=& 0\,,
\\ 
\from{eq:TC-softgg}{eq:TC-SC-softgg}
\bSCS{ir;s}\Big(\bC{irs}\bS{rs}
- \bC{irs}\bSCS{ir;s}\bS{rs}\Big)
|\cM_{m+2}^{(0)}(p_i,p_r,p_s,\dots)|^2 &=& 0\,,
\\ 
\from{eq:DC-softgg}{eq:DC-SC-softgg}
\bSCS{ir;s}\Big(\bC{ir;js}\bS{rs}
- \bC{ir;js}\bSCS{ir;s}\bS{rs}\Big)
|\cM_{m+2}^{(0)}(p_i,p_r,p_j,p_s,\dots)|^2 &=& 0\,,
\\ 
\from{eq:TC-softgg}{eq:softgg-TC}
\bS{rs}\Big(\bC{irs} - \bC{irs}\bS{rs}\Big)
|\cM_{m+2}^{(0)}(p_i,p_r,p_s,\dots)|^2 &=& 0\,,
\\ 
\from{eq:DC-softgg}{eq:softgg-DC}
\bS{rs}\Big(\bC{ir;js} - \bC{ir;js}\bS{rs}\Big)
|\cM_{m+2}^{(0)}(p_i,p_r,p_j,p_s,\dots)|^2 &=& 0\,,
\\ 
\from{eq:SC-softgg}{eq:softgg-SC}
\bS{rs}\Big(\bSCS{ir;s} - \bSCS{ir;s}\bS{rs}\Big)
|\cM_{m+2}^{(0)}(p_i,p_r,p_s,\dots)|^2 &=& 0\,,
\\ 
\from{eq:TC-SC}{eq:TC-SC-softgg}
\bS{rs}\Big(\bC{irs}\bSCS{ir;s} - \bC{irs}\bSCS{ir;s}\bS{rs}\Big)
|\cM_{m+2}^{(0)}(p_i,p_r,p_s,\dots)|^2 &=& 0\,,
\\ 
\from{eq:DC-SC}{eq:DC-SC-softgg}
\bS{rs}\Big(\bC{ir;js}\bSCS{ir;s}
- \bC{ir;js}\bSCS{ir;s}\bS{rs}\Big)
|\cM_{m+2}^{(0)}(p_i,p_r,p_j,p_s,\dots)|^2 &=& 0\,.
\label{eq:overlapend}
\eeeq
With \eqnss{eq:overlapstart}{eq:overlapend} at hand it is not difficult
to verify that for the subtraction term,
\beeq
\bA{2} |\cM_{m+2}^{(0)}|^2 \aand=
\sum_r \sum_{s\ne r} \Bigg\{
  \sum_{i\ne r,s} \Bigg[
  \sixth\,\bC{irs}
+ \sum_{j\ne i,r,s} \eighth\,\bC{ir;js} +
\nn\\&&\qquad\qquad\qquad\quad
+ \half\,\Bigg(\bSCS{ir;s} - \bC{irs} \bSCS{ir;s}
- \sum_{j\ne i,r,s} \bC{ir;js} \bSCS{ir;s} \Bigg)\Bigg] +
\nn\\&&\qquad\qquad\quad
+ \half\,\bS{rs}
- \sum_{i\ne r,s} \Bigg[
    \bSCS{ir;s} \bS{rs}
  + \bC{irs} \left(\half\,\bS{rs} - \bSCS{ir;s} \bS{rs} \right) +
\nn\\&&\qquad\qquad\qquad\qquad\qquad\qquad
  + \sum_{j\ne i,r,s} \bC{ir;js} \left(\half\,\bS{rs}
  - \bSCS{ir;s} \bS{rs}\right)\Bigg]
\Bigg\}\times
\nn\\&&\qquad\qquad
\times
|\cM_{m+2}^{(0)}(p_i,p_r,p_j,p_s,\dots)|^2\,,
\label{eq:A2}
\eeeq
the following equations hold,
\beeq
\bC{irs}(1-\bA{2})\M{m+2}{(0)} \aand= 0\,,
\label{eq:CirsA2}
\\
\bC{ir;js}(1-\bA{2})\M{m+2}{(0)} \aand= 0\,,
\label{eq:Cir;jsA2}
\\
\bSCS{ir;s}(1-\bA{2})\M{m+2}{(0)} \aand= 0\,,
\label{eq:SCSir;sA2}
\\
\bS{rs}(1-\bA{2})\M{m+2}{(0)} \aand= 0\,.
\label{eq:SrsA2}
\eeeq
Therefore, \eqn{eq:A2} is free of double and triple subtractions.
Instead of \eqn{eq:A2}, the subtraction terms that can be derived from
antennae factorization, as proposed in
\Refs{Gehrmann-DeRidder:2004tv,Gehrmann-DeRidder:2005hi,Ridder:2005aw},
can also be used for regularizing the infra-red divergencies in the
doubly-unresolved regions. The advantage of antennae subtractions over
our proposal is that those avoid the problem of defining the momentum
fractions. Note however, that contrary to the original antennae
factorization terms of \Ref{Kosower:1997zr}, the spin-polarization of
the parent parton in the collinear regions is not yet taken into account
in \Refs{Gehrmann-DeRidder:2004tv,Gehrmann-DeRidder:2005hi,Ridder:2005aw}.
The spin-averaged antennae terms can only be used in a hybrid
slicing-subtraction method as applied in \Ref{Campbell:1998nn}.

We can simplify \eqn{eq:A2} by removing the terms that contain triple
limits.
Comparing \eqns{eq:DC-softgg}{eq:DC-SC-softgg} we see that in the last
term in \eqn{eq:A2}
\beq
\bC{ir;js} \left(\frac12 \bS{rs} - \bSCS{ir;s} \bS{rs}\right)
\M{m+2}{(0)}
= - \bC{ir;js} \frac12 \bS{rs} \M{m+2}{(0)}\,.
\eeq
Furthermore, comparing \eqns{eq:TC-DSabgg}{eq:TC-DSnabgg}
to \eqn{eq:TC-SC-softgg}, and using \eqn{eq:TC-DSqq}, and 
observing that the soft-collinear limit of \eqn{eq:TC-DSnabgg} is zero,
we see that
\beeq
&&
\bC{irs} \left(\frac12 \bS{rs} - \bSCS{ir;s} \bS{rs} \right)
\M{m+2}{(0)} =
\nn \\ && \qquad
=\frac12 \bC{irs}
\left[\Big(\bSnab{rs} - \bSab{rs} \Big)\,\delta_{f_rg}\,\delta_{f_sg}
+ \bS{rs}\,\delta_{f_rq}\,\delta_{f_s\qb} \right]
\M{m+2}{(0)}\,.
\eeeq
This equation shows that we can simplify the notation by introducing
\beq
\bom{\mathrm S}_{rs}^{(\rm A)} \equiv
\bSab{rs} \,\delta_{f_rg}\,\delta_{f_sg}
\label{eq:SA}
\eeq
and
\beq
\bom{\mathrm S}_{rs}^{(\rm N)} \equiv
  \bSnab{rs} \,\delta_{f_rg}\,\delta_{f_sg}
+ \bS{rs}\,\delta_{f_rq}\,\delta_{f_s\qb}\,.
\label{eq:SN}
\eeq
With this notation $\bS{rs} = \bSA{rs} + \bSN{rs}$.

\section{Matching the singly- and doubly-unresolved limits}
\label{sec:matchingsingledoubleunresolved}

Eqns.~(\ref{eq:CirsA2}--\ref{eq:SrsA2}) ensure that the difference
$|\cM_{m+2}^{(0)}|^2 - \bA{2} |\cM_{m+2}^{(0)}|^2$
does not contain leading singularities (therefore, it is integrable) over those
doubly-unresolved parts of the phase space where the factorized matrix
elements $\ket{m}{(0)}{}$ in the subtraction terms are finite.
Nevertheless, the regularized squared matrix element, $|\cM_{m+2}^{(0)}|^2 -
\bA{2}|\cM_{m+2}^{(0)}|^2$, contains leading singularities over the
singly-unresolved parts of the phase space.
We have a similar problem concerning the subtraction term
$\bA{1}\M{m+2}{(0)}$ that is used for regularizing the
singly-unresolved limits, namely, $\bA{1}\M{m+2}{(0)}$ becomes
singular over the doubly-unresolved parts of the phase space.
In a general subtraction scheme one wishes to regularize these unwanted
singularities, introduced by the subtraction terms, by defining further
subtractions that take into account the overlap of the singly- and
doubly-unresolved regions. The only consistent way to achieve that is to
introduce a third subtraction term $-\bA{12}\M{m+2}{(0)}$ that
simultaneously cancels all unwanted singularities in both the singly-
and doubly-unresolved parts of the phase space. As a result, the
combination
\beq
(\bA{1} + \bA{2} - \bA{12})\M{m+2}{(0)}
\label{eq:A1+A2-A12}
\eeq
may serve as a NNLO subtraction term that regularizes the squared matrix
element in all relevant unresolved regions of the phase space, i.e.~the
following equations must hold,
\beeq
\bC{ir}(\bA{1} + \bA{2} - \bA{12})\M{m+2}{(0)} \aand=
\bC{ir}\M{m+2}{(0)}\,,
\label{eq:Cir(A1+A2-A12)}
\\
\bS{r}(\bA{1} + \bA{2} - \bA{12})\M{m+2}{(0)} \aand=
\bS{r}\M{m+2}{(0)}\,,
\label{eq:Sr(A1+A2-A12)}
\\
\bC{irs}(\bA{1} + \bA{2} - \bA{12})\M{m+2}{(0)} \aand=
\bC{irs}\M{m+2}{(0)}\,,
\label{eq:Cirs(A1+A2-A12)}
\\
\bC{ir;js}(\bA{1} + \bA{2} - \bA{12})\M{m+2}{(0)} \aand=
\bC{ir;js}\M{m+2}{(0)}\,,
\label{eq:Cir;js(A1+A2-A12)}
\\
\bSCS{ir;s}(\bA{1} + \bA{2} - \bA{12})\M{m+2}{(0)} \aand=
\bSCS{ir;s}\M{m+2}{(0)}\,,
\label{eq:CSir;s(A1+A2-A12)}
\\
\bS{rs}(\bA{1} + \bA{2} - \bA{12})\M{m+2}{(0)} \aand=
\bS{rs}\M{m+2}{(0)}\,.
\label{eq:Srs(A1+A2-A12)}
\eeeq

In order to prove \eqnss{eq:Cir(A1+A2-A12)}{eq:Srs(A1+A2-A12)}, we
define $\bA{12}\M{m+2}{(0)}$ as $\bA{1}\bA{2}\M{m+2}{(0)}$. Explicitly,
\beeq
\bA{12}\M{m+2}{(0)}
&=& \sum_{t}\Bigg[\bS{t}\bA{2} + 
\sum_{k\ne t} \half\,\bC{kt}\bA{2} -
\sum_{k\ne t} \bC{kt}\bS{t}\bA{2}\Bigg]\M{m+2}{(0)}\, ,
\label{eq:A12}
\eeeq
where
\beeq
\bS{t}\bA{2}\M{m+2}{(0)} \aand=
\sum_{r\ne t}\Bigg\{
\sum_{i\ne r,t}\Bigg[
\half\,\Big(
\bS{t}\bC{irt} + \bS{t}\bSCS{ir;t} - \bS{t}\bC{irt}\bSCS{ir;t}
\Big) -
\nn\\ &&\qquad\qquad\qquad
-\bS{t}\bSCS{ir;t}\bS{rt} - \bS{t}\bC{irt}\bSnab{rt}
\Bigg] + \bS{t}\bS{rt}\Bigg\}\M{m+2}{(0)}\, ,
\label{eq:StA2}
\\
\bC{kt}\bA{2}\M{m+2}{(0)} \aand=
\Bigg\{\bC{kt}\bSN{kt}
+ \sum_{r\ne k,t}\Bigg[
\sum_{i\ne r,k,t}\Big(
\half\,\bC{kt}\bC{ir;kt} - \bC{kt}\bC{ir;kt}\bSCS{kt;r}\Big) +
\nn\\ && \qquad\qquad\qquad\qquad\qquad
+ \bC{kt}\bC{ktr}
+ \bC{kt}\bSCS{kt;r} -
\label{eq:CktA2} 
\\ && \qquad\qquad\qquad\qquad\qquad
 - \bC{kt}\bC{ktr}\bSCS{kt;r} -\bC{kt}\bC{rkt}\bSN{kt}
\Bigg]\Bigg\}
\M{m+2}{(0)}\,,
\nn
\eeeq
and finally
\beeq
\nn
\bC{kt}\bS{t}\bA{2}\M{m+2}{(0)} \aand=
\Bigg\{\bC{kt}\bS{t}\bSnab{kt} +
\\ &&\qquad
+ \sum_{r\ne k,t}\Bigg[
\sum_{i\ne r,k,t}\Big(
\half\,\bC{kt}\bS{t}\bSCS{ir;t} - \bC{kt}\bS{t}\bSCS{ir;t}\bS{rt}\Big) +
\nn\\ &&\qquad\qquad\qquad
+\bC{kt}\bS{t}\bC{krt}
+ \bC{kt}\bS{t}\bS{rt} - \bC{kt}\bS{t}\bSCS{kr;t}\bS{rt} - 
\nn\\ &&\qquad\qquad\qquad
-\bC{kt}\bS{t}\bC{rkt}\bSnab{kt}\Bigg]\Bigg\}\M{m+2}{(0)}\,.
\label{eq:CktStA2}
\eeeq
Note that the direct calculation of the terms on the right-hand side 
of \eqn{eq:A12} yields significantly larger expressions than those in 
\eqnss{eq:StA2}{eq:CktStA2}. In \sect{sec:singlyunresolvedA2} we shall 
derive some factorization formulae we used in obtaining 
\eqnss{eq:StA2}{eq:CktStA2}. In fact, all the terms of
\eqnss{eq:StA2}{eq:CktStA2} are singly-unresolved limits of
the factorization formulae for doubly-unresolved emission.
Most of them can be
computed easily using the explicit expressions of the previous
sections. The only non-trivial ones are those that are related to the
{\em  iterated singly-unresolved limits}, which we discuss first.

\section{Iterated singly-unresolved limits}
\label{sec:iteratedsingleunresolved}

In this section we derive the factorization formulae valid in the
singly-unresolved limits of \eqns{Rcollfactnew}{Rsoftfactnew} that we
call the iterated singly-unresolved limits. Thus we give the explicit
expressions of the various terms in \eqns{eq:CjsA1-Cjs}{eq:SsA1-Ss}.
{\em All the factorization formulae presented in this section are valid
for helicity- and colour-correlated squared matrix elements in
\eqns{Rcollfactnew}{Rsoftfactnew} written in terms of on-shell momenta.} 

\subsection{Collinear limit of the soft-factorization formula}
\label{sec:collsoftfact}

In the derivation of the collinear limit of the colour-correlated
squared matrix element $|\cM_{m+1;(j,l)}^{(0)}|^2$ of \eqn{Rsoftfact}, 
we have to distinguish two cases:
(i) the partons of the collinear pair $i,\, r$ are different from 
partons $j$ and $l$; (ii) one parton of the collinear pair is either 
$j$ or $l$, the other is different.
In principle there is a third case, when $j$ and $l$ become collinear.
The corresponding $1/s_{jl}$ pole is suppressed by the $s_{jl}$ factor in
the eikonal factor that multiplies $|\cM_{m+1;(j,l)}^{(0)}|^2$, therefore,
the contribution is subleading. Also, $|\cM_{m+1;(j,l)}^{(0)}|^2$ does
not have a leading singularity when $i$ and $r$ are both quarks or
antiquarks. 

In the first case, the colour-charge operators do not influence the
computation of the collinear limit, $|\cM_{m+1;(j,l)}^{(0)}|^2$
factorizes on the collinear poles as $|\cM_{m+1}^{(0)}|^2$ does. In the
second case, due to coherent soft-gluon emission from unresolved
partons, only the sum of
$|\cM_{m+1;(i,l)}^{(0)}|^2 + |\cM_{m+1;(r,l)}^{(0)}|^2$ (or 
$|\cM_{m+1;(j,i)}^{(0)}|^2 + |\cM_{m+1;(j,r)}^{(0)}|^2$) factorizes as
\beq
  \cS_{il}(s) |\cM_{m+1;(i,l)}^{(0)}|^2
+ \cS_{rl}(s) |\cM_{m+1;(r,l)}^{(0)}|^2
\simeq
8\pi\as\mu^{2\eps}\,\cS_{(ir)l}(s)\,\frac{1}{s_{ir}}\,
\bra{m}{(0)}{}\,\hP_{f_if_r}^{(0)}\,\bT_{ir}\cdot\bT_l\,\ket{m}{(0)}{}\,,
\eeq
where $\bT_{ir} = \bT_i + \bT_r$ and
${\cS}_{(ir)j}(s) = {\cS}_{j(ir)}(s)$ is defined in \eqn{Sj(ir)}.
Collecting all the terms, in the collinear limit $i||r$ we obtain
\beeq
&&
\bC{ir}\bS{s}|\cM_{m+2}^{(0)}(p_i,p_r,p_s,\dots)|^2 =
\label{eq:collsoftfact}
\\ && \qquad
=-(8\pi\as\mu^{2\eps})^2\times
\nn\\ &&\qquad\quad \times
\frac{1}{s_{ir}}\Bigg\{\sum_{j,l\ne (ir)}
\frac12 \cS_{jl}(s)\,
\bra{m}{(0)}{(p_{ir},\dots)}\,\hP_{f_if_r}^{(0)}(z_i,z_r)\,\bT_j\cdot\bT_l\,
\ket{m}{(0)}{(p_{ir},\dots)}+
\nn\\ &&\qquad\qquad\quad
+\sum_{l\ne (ir)}\cS_{(ir)l}(s)\,
\bra{m}{(0)}{(p_{ir},\dots)}\,\hP_{f_if_r}^{(0)}(z_i,z_r)\,\bT_{ir}\cdot\bT_l\,
\ket{m}{(0)}{(p_{ir},\dots)}\Bigg\}\,,
\nn
\eeeq
that is,
\beq
\bC{ir}\bS{s}|\cM_{m+2}^{(0)}|^2 = \bSCS{ir;s}|\cM_{m+2}^{(0)}|^2\,.
\label{eq:CirSs}
\eeq

\subsection{Soft limit of the soft-factorization formula}
\label{sec:softsoftfact}

We are interested in the limit of the right hand side of \eqn{Rsoftfact}
as a gluon of momentum $p_s =\lambda q_s$ becomes soft ($\lambda\to 0$ at
fixed $q_s$).  The soft limit of the colour-correlated squared matrix
elements, $|\cM_{m+1;(i,k)}^{(0)}|^2$, can be obtained in a similar way
as that of the squared matrix element.  We have to distinguish two cases:
(i) $s$ is different from $i$ and $k$, (ii) either $i$ or $k$ gets soft.

Let us start with case (i), when we can immediately use the
well known factorization of the matrix element
$\ket{m+1}{(0)}{(p_s,\dots)}$,
\beq
\la \nu \ket{m+1}{(0)}{(p_s,\dots)} \simeq
\gs \mu^\eps \bom{J}^\nu(p_s) \ket{m}{(0)}{(\dots)}\, ,
\label{eq:Msoftfact}
\eeq
where we neglected all contributions that are less singular than
$1/\lambda$. The factor $\bom{J}^\nu(p_s)$ is the eikonal current for the
emission of the soft gluon $s$,
\beq
\bom{J}^\nu(p_s) = \sum_{j} \bT_j \frac{p_j^\nu}{p_j\cdot p_s}\,.
\eeq
We substitute \eqn{eq:Msoftfact} into
$|\cM_{m+1;(i,k)}^{(0)}(p_s,\dots)|^2$ and obtain
\beq
\bS{s} |\cM_{m+1;(i,k)}^{(0)}(p_s,\dots)|^2 =
- 8\pi\as\mu^{2\eps}
\sum_{j,l} \frac12 \cS_{jl}(s)
\bra{m}{(0)}{(\ldots)}T^a_j T^b_i T^b_k T^a_l\ket{m}{(0)}{(\ldots)}\,.
\label{eq:Rsoftfacttmp1}
\eeq
We can use the colour algebra to derive the identity
\beeq
2T^a_j T^b_i T^b_k T^a_l \aand=
\{ \bT_j\cdot \bT_l,\bT_i\cdot \bT_k \} -
\nn\\ &&\quad
-\ri f_{abc}(T^a_j T^c_i T^b_k \delta_{li}+
T^a_j T^b_i T^c_k \delta_{lk}+T^a_l T^c_i T^b_k \delta_{jk}+
T^a_l T^b_i T^c_k \delta_{ji}) +
\nn\\ &&\quad
+\CA\bT_i\bT_k(\delta_{li}\delta_{jk}-\delta_{lk}\delta_{jk}-
\delta_{li}\delta_{ji}+\delta_{lk}\delta_{ji})\,,
\label{eq:color_alg}
\eeeq
which we substitute into \eqn{eq:Rsoftfacttmp1}.
Performing the summation over $j$ and $l$, we find that the terms
proportional to $f_{abc}$ in \eqn{eq:color_alg} are symmetric in $b$
and $c$, while $f_{abc}$ is antisymmetric, therefore, the sum of these
terms is zero. The remaining terms yield
\beeq
&&
\bS{s} |\cM_{m+1;(i,k)}^{(0)}(p_s,\dots)|^2 =
\nn\\ &&\qquad
=-8\pi\as\mu^{2\eps}
\frac14 \Bigg[\sum_{j,l}
\cS_{jl}(s) |\cM_{m;(i,k)(j,l)}^{(0)}(\dots)|^2
+2 \CA\,\cS_{ik}(s)
|\cM_{m;(i,k)}^{(0)}(\dots)|^2\Bigg]\,.
\label{eq:Miksoftfact}
\eeeq

In the second case, when for instance gluon $i$ gets soft, we first use 
colour conservation, $\bT_s = -\sum_{l\ne s}\bT_l$ to write
\beq
|\cM_{m+1;(s,k)}^{(0)}(p_s,\dots)|^2 =
- \sum_{l \ne s} |\cM_{m+1;(l,k)}^{(0)}(p_s,\dots)|^2\,.
\label{MsktoMlk}
\eeq
We obtain the factorization formulae for the $|\cM_{m+1;(l,k)}^{(0)}|^2$
colour-correlated matrix elements the same way as we did in case (i),
and substitute those into \eqn{MsktoMlk},
\beeq
&&
\bS{s} |\cM_{m+1;(s,k)}^{(0)}(p_s,\dots)|^2 =
\\ &&\qquad
=8\pi\as\mu^{2\eps}
\frac14 \sum_l\Bigg[\sum_{j,j'}
\cS_{jj'}(s) |\cM_{m;(l,k)(j,j')}^{(0)}(\dots)|^2
+2 \CA\,\cS_{lk}(s)
|\cM_{m;(l,k)}^{(0)}(\dots)|^2\Bigg]\,.
\nn
\eeeq
In the first term we can perform the summation over $l$ using colour
conservation, and find that the sum is zero. Therefore, 
\beq
\bS{s} |\cM_{m+1;(s,k)}^{(0)}(p_s,\dots)|^2 =
8\pi\as\mu^{2\eps}
\frac12 \CA \sum_l \cS_{lk}(s) |\cM_{m;(l,k)}^{(0)}(\dots)|^2\,.
\label{eq:Msksoftfact}
\eeq
We check explicitly the validity of \eqns{eq:Miksoftfact}{eq:Msksoftfact}
in Appendix A using the known colour-correlated squared matrix elements
for the $e^+ e^- \to q \qb + n g$ ($n= 1$, 2) processes.

We can now easily compute the soft-factorization formula for the right-hand 
side of \eqn{Rsoftfact}. We decompose it as
\beq
\sum_{i,k} \frac12 \cS_{ik}(r) |\cM_{m+1;(i,k)}^{(0)}|^2 =
\sum_{i,k \ne s} \frac12 \cS_{ik}(r) |\cM_{m+1;(i,k)}^{(0)}|^2
+ \sum_{k} \cS_{sk}(r) |\cM_{m+1;(s,k)}^{(0)}|^2\,.
\eeq
In the soft limit all the eikonal factors are finite, so we can simply
substitute \eqns{eq:Miksoftfact}{eq:Msksoftfact}, and obtain
\beeq
&&
\bS{s} \bS{r} |\cM_{m+2}^{(0)}(p_r,p_s,\dots)|^2 =
\nn\\ &&\qquad
=(8\pi\as\mu^{2\eps})^2
\frac14\Bigg[\sum_{i,k,j,l}
\frac12 \cS_{ik}(r) \cS_{jl}(s) |\cM_{m;(i,k)(j,l)}^{(0)}(\dots)|^2 -
\nn\\ &&\qquad\qquad\qquad\qquad\quad
-\CA\,\sum_{i,k=1}^m {\cS}_{ik}^{(\rm s.o.)}(r, s)
|\cM_{m;(i,k)}^{(0)}(\dots)|^2\Bigg]\,,
\label{eq:SsSrSME}
\eeeq
where
\beq
{\cS}_{ik}^{(\rm s.o.)}(r, s) =
\cS_{ik}(s) \Big(\cS_{is}(r) + \cS_{ks}(r) - \cS_{ik}(r)\Big)
\eeq
is the two-gluon soft function in the strongly-ordered approximation.
If either $r$ or $s$ is a quark, then the result is zero.

Finally, we consider the region where the soft momentum $p_s$
becomes also collinear to the momentum of another parton $j$. In this
region, we have
\beq
\frac14 \cS_{jk}(s) \Bigg(\cS_{js}(r) + \cS_{ks}(r) - \cS_{jk}(r)\Bigg)
\simeq \frac{s_{jk}}{s_{jr} s_{rs} s_{sk}} \simeq \O(1/\lambda^2)\,,
\eeq
thus the term proportional to $\CA$ is subleading.
The eikonal factor $\cS_{jl}(s)$ becomes independent of $l$,
\beq
\cS_{jl}(s) \simeq \frac{2}{s_{js}} \frac{z_j}{z_s}\,,
\eeq
therefore, using colour conservation, we can perform the summation over
$l$ in the first sum on the right-hand side of \eqn{eq:SsSrSME},
and obtain
\beq
\bC{js} \bS{s} \bS{r} |\cM_{m+2}^{(0)}(p_r,p_s,\dots)|^2 =
-(8\pi\as\mu^{2\eps})^2
\frac{2}{s_{js}} \frac{z_j}{z_s}
\bT_j^2 \sum_{i,k} \frac12 \cS_{ik}(r) |\cM_{m;(i,k)}^{(0)}(\dots)|^2\,.
\label{eq:softcollsoftfact}
\eeq
Comparing this result to \eqn{eq:collsoftfact}, using \eqn{eq:CirSs}
and
\beq
\bS{s}\bSCS{js;r}\SME{m+2}{0}{p_j,p_r,p_s,\dots} =
8\pi\as\mu^{2\eps}\,\frac{2}{s_{js}}\,\frac{1}{z_s}\,\bT_j^2\,
\bS{r}\SME{m+1}{0}{p_i,p_r,\dots}\, ,
\label{eq:SrCir;s}
\eeq
we see that
\beq
\bS{s} \Big(\bC{js} - \bC{js} \bS{s} \Big) \bS{r}
|\cM_{m+2}^{(0)}(p_r,p_s,\dots)|^2 = 0\,.
\eeq

\subsection{Collinear limit of the collinear-factorization formula}
\label{sec:collcollfact}

Next, we consider the limit of \eqn{Rcollfactnew} when parton $s$
becomes collinear with another parton $j$. The collinear-factorization
formula involves the helicity-dependent tensors,
\beq
\cT_{m+1}^{hh'} =
\bra{m+1}{(0)}{(p_{ir},p_j,p_s,\dots)} h\ra
\la h' \ket{m+1}{(0)}{(p_{ir},p_j,p_s,\dots)}\,,
\label{eq:Tmunu}
\eeq
which are the leading-order $(m+1)$-parton squared matrix elements not
summed over the spin polarizations of parton $(ir)$, i.e., if parton
$(ir)$ is a gluon, then
$-g_{\mu \nu} \cT_{m+1}^{\mu \nu} = |\cM_{m+1}^{(0)}|^2$,
and $\delta_{ss'} \cT_{m+1}^{ss'} = |\cM_{m+1}^{(0)}|^2 $ if $(ir)$ is
a quark.  

The computation of the collinear limit of the $\cT_{m+1}^{hh'}$ tensors
follows that for the collinear factorization of the squared matrix
element. When $j\ne (ir)$, we obtain
\beeq
&&
\bC{js}\cT_{m+1}^{hh'}(p_j,p_s,p_{ir},\dots) =
\label{Tjscollfact}
\\ &&\qquad
=8\pi\as\mu^{2\eps}
\frac{1}{s_{js}}\,
\bra{m}{(0)}{(p_{js},p_{ir},\dots)} h \ra
\,\hP_{f_jf_s}^{(0)}(z_j,z_s,\kT{s};\eps)
\,\la h' \ket{m}{(0)}{(p_{js},p_{ir},\dots)}\:.
\hspace*{2em}
\nn
\eeeq
If $j=(ir)$, then we find
\beeq
&&
\bC{(ir)s} \cT_{m+1}^{hh'}(p_{ir},p_s,\dots) =
\label{Tirscollfact}
\\ &&\qquad
=8\pi\as\mu^{2\eps}
\frac{1}{s_{(ir)s}}
\bra{m}{(0)}{(p_{(ir)s},\dots)} \,
\hP^{hh'}_{f_{ir}f_s}(z_s,\kT{s};p_{(ir)s},n_{(ir)s};\eps)\,
\ket{m}{(0)}{(p_{(ir)s},\dots)}\:.
\nn
\eeeq
The $\hP^{hh'}_{f_if_j}(z,\kT{};p,n;\eps)$ kernels (splitting tensors) are
related to the $d$-dimensional Altarelli-Parisi splitting
functions by summation over the spin indices $h$ and $h'$,
\beeq
\delta_{hh'}
\la s| \hP^{hh'}_{qg}(z,\kT{};p,n;\eps) |s'\ra \aand=
\la s| \hP^{(0)}_{qg}(1-z,z;\eps) |s'\ra\,,
\\
\delta_{hh'}
\la \mu| \hP^{hh'}_{q \qb}(z,\kT{};p,n;\eps) |\nu\ra \aand=
\la \mu| \hP^{(0)}_{q \qb}(1-z,z,\kT{};\eps) |\nu\ra\,,
\\
d_{\alpha\beta}(p,n)
\la s| \hP^{\alpha\beta}_{gq}(z,\kT{};p,n;\eps) |s'\ra \aand=
\la s| \hP^{(0)}_{qg}(z,1-z;\eps) |s'\ra\,,
\label{eq:ddotPgq}
\\
d_{\alpha\beta}(p,n)
\la \mu| \hP^{\alpha\beta}_{gg}(z,1-z,\kT{};p,n;\eps) |\nu\ra \aand=
\la \mu| \hP^{(0)}_{gg}(z,1-z,\kT{};\eps) |\nu\ra\,.
\label{eq:ddotPgg}
\eeeq
Here $d_{\alpha\beta}(p,n)$ is the gluon polarization tensor for
physical polarizations,
\beq
d_{\alpha\beta}(p,n)= 
-g_{\alpha\beta}+\frac{p_\alpha n_\beta + p_\beta n_\alpha}{p\cdot n}\,.
\label{eq:gpoltensor}
\eeq

In order to obtain the collinear limit of the collinear-factorization
formula, we insert \eqns{Tjscollfact}{Tirscollfact} into \eqn{Rcollfact}.
For $j \ne (ir)$ we obtain
\beeq
&&
\bC{js} \bC{ir} \SME{m+2}{0}{p_i,p_r,p_j,p_s,\dots} =
\nn\\ &&\qquad
=(8\pi\as\mu^{2\eps})^2
\,\frac{1}{s_{ir} s_{js}}\times
\label{Rcollcollfact}
\\ &&\qquad\quad
\times
\bra{m}{(0)}{(p_{ir},p_{js},\dots)}
\,\hP_{f_if_r}^{(0)}(z_i,z_r,\kT{r};\eps)
\,\hP_{f_jf_s}^{(0)}(z_j,z_s,\kT{j};\eps)
\,\ket{m}{(0)}{(p_{ir},p_{js},\dots)}\,.
\nn
\eeeq
The quark splitting function,
$\la h| \hP^{(0)}_{qg}(z_q,z_g,\kT{};\eps)|h'\ra$,
is diagonal in the spin indices (see \eqn{P0qg}), therefore,
\beq
\la h| \hP^{(0)}_{qg}(z_q,z_g,\kT{r};\eps)|h'\ra\,
\hP^{hh'}_{qf_s}(z_s,\kT{s};p_{(ir)s},n_{(ir)s};\eps) =
P^{(0)}_{qg}(z_q,z_g;\eps) \hP^{(0)}_{qf_s}(1-z_s,z_s;\eps)\,.
\eeq
As a result, the factorization formula in the case of parton $j = (ir)$
being a quark is
\beeq
&&
\bC{(ir)s} \bC{ir} \SME{m+2}{0}{p_i,p_r,p_s,\dots} =
\nn\\ &&\qquad
=(8\pi\as\mu^{2\eps})^2
\,\frac{1}{s_{ir} s_{(ir)s}}
\,P_{qg}^{(0)}(z_q,z_g,\kT{r};\eps)\times
\nn\\ &&\qquad\quad
\times
\bra{m}{(0)}{(p_{(ir)s},\dots)}
\,\hP_{qf_s}^{(0)}(1-z_s,z_s,\kT{ir};\eps)
\,\ket{m}{(0)}{(p_{(ir)s},\dots)}.\qquad~
\label{Rtriplecollfactq}
\eeeq
Thus we need explicit expressions only for the $\hP^{\alpha\beta}_{gf_s}$
splitting tensors,
\beeq
&&
\la s| \hP^{\alpha\beta}_{gq}(z,\kT{};p,n;\eps) |s'\ra =
\nn\\&&\qquad
=\CF\,\delta_{ss'}
\left[
-\frac{1-z}{2}g^{\alpha\beta}
- 2\frac{z}{1 - z}\frac{\kT{}^\alpha\kT{}^\beta}{\kT{}^2}
+ (1 - z) \frac{p^\alpha n^\beta + p^\beta n^\alpha}{2 p\cdot n}
\right]+\cdots\,,
\label{eq:Pgqalfabeta}
\\ &&
\la \mu| \hP^{\alpha\beta}_{gg}(z,\kT{};p,n;\eps) |\nu\ra =
\nn\\&&\qquad
=2\CA
\left[
  \frac{1-z}{z} g^{\alpha\mu}g^{\beta\nu}
+ \frac{z}{1 - z} g^{\mu\nu}\frac{\kT{}^\alpha\kT{}^\beta}{\kT{}^2}
- z(1 - z) d^{\alpha\beta}(p,n) \frac{\kT{}^\mu\kT{}^\nu}{\kT{}^2}
\right]+\cdots\,.~~~
\label{eq:Pggalfabeta}
\eeeq
In \eqns{eq:Pgqalfabeta}{eq:Pggalfabeta} the ellipses mean neglected
terms that do not contribute to the
contraction of the splitting functions and splitting tensors
\beq
\la \alpha| \hP^{(0)}_{gg}(z_i,z_r,\kT{r};\eps)|\beta\ra\,
\hP^{\alpha\beta}_{gf_s}(z_s,\kT{s};p_{(ir)s},n_{(ir)s};\eps)
\label{eq:SpfunctionSptensor}
\eeq
(summation
over the gluon polarization indices $\alpha$ and $\beta$ is understood),
which appear in the factorization formula for parton $j = (ir)$ being a
gluon,
\beeq
&&
\bC{(ir)s} \bC{ir} \SME{m+2}{0}{p_i,p_r,p_s,\dots} =
\nn\\ &&\qquad
=(8\pi\as\mu^{2\eps})^2
\frac{1}{s_{ir}} \frac{1}{s_{(ir)s}}
\la \alpha| \hP^{(0)}_{gg}(z_i,z_r,\kT{r};\eps)|\beta\ra \times
\nn\\ &&\qquad\quad\times
\bra{m}{(0)}{(p_{(ir)s},\dots)} \,
\hP^{\alpha\beta}_{gf_s}(z_s,\kT{s};p_{(ir)s},n_{(ir)s};\eps)\,
\ket{m}{(0)}{(p_{(ir)s},\dots)}\:.
\qquad~
\label{Rtriplecollfactg}
\eeeq
In Appendix B, we use the known $\cT^{\mu\nu}$ tensor for the process
$e^+e^- \to q \qb g$ to check the validity of
\eqns{eq:Pgqalfabeta}{Rtriplecollfactg} explicitly. In Appendix C,
we translate \eqn{Rtriplecollfactg} to a helicity basis and compare the
obtained expression to the corresponding one derived using the known
collinear factorization properties of helicity amplitudes.

We note that the collinear-factorization formula for the $\cT^{\mu\nu}$
tensor in \eqn{Tirscollfact} fulfills gauge invariance (only) in the
collinear limit,
\beq
p_{ir\,\alpha}
\bra{m}{(0)}{(p_{(ir)s},\dots)}
\hP^{\alpha\beta}_{gf_s}(z,\kT{};p,n;\eps)
\ket{m}{(0)}{(p_{(ir)s},\dots)} =
\O(\kT{})\,,
\label{gaueinv}
\eeq
where
$p_{ir}^\alpha = (1 - z) p_{(ir)s}^\alpha-\kT{}^\alpha+\O(\kT{}^2)$,
and in obtaining \eqn{gaueinv} for $f_s = g$, we used
$p_{(ir)s\,\mu} \cT_m^{\mu\nu} = 0$.
\mla
We kept the gauge terms in \eqns{eq:Pgqalfabeta}{eq:Pggalfabeta} only to
maintain this gauge invariance; those terms do not contribute to either
\eqns{eq:ddotPgq}{eq:ddotPgg} or \eqn{eq:SpfunctionSptensor}.

\subsection{Soft limit of the collinear-factorization formula}
\label{sec:softcollfact}

As in the case of the soft limit of the soft-factorization formula in
\sect{sec:softsoftfact}, in order to obtain the soft limit of the
collinear-factorization formula we use the factorization of the matrix
element, \eqn{eq:Msoftfact}, which we substitute into
\eqn{Rcollfact}. If the parent parton of the collinear splitting is
a gluon, the factorization formula contains non-trivial spin correlations
(see \eqns{P0gg}{P0qq}), therefore, we have to distinguish two
cases: (i) $(ir)$ is a hard parton; (ii) $(ir)$ is the soft gluon.
If the soft gluon is labelled with $s$, we find that
\begin{itemize}
\item if $(ir)$ is a hard parton, then
\beeq
&&
\bS{s}\bC{ir}\SME{m+2}{0}{p_i,p_r,p_s,\dots} =
\nn\\ &&\qquad
=-(8\pi\as\mu^{2\eps})^2\times
\nn\\ &&\qquad\quad \times
\frac{1}{s_{ir}}\,\Bigg\{\sum_{j,l\ne (ir)}
\frac12 \cS_{jl}(s)\,
\bra{m}{(0)}{(\dots)}\,\hP_{f_if_r}^{(0)}\,\bT_j\cdot\bT_l
\,\ket{m}{(0)}{(\dots)}+
\nn\\ &&\qquad\qquad\qquad
+\sum_{l\ne (ir)}\cS_{(ir)l}(s)\,
\bra{m}{(0)}{(\dots)}\,\hP_{f_if_r}^{(0)}\,\bT_{ir}\cdot\bT_l
\,\ket{m}{(0)}{(\dots)}\Bigg\}\,;
\label{eq:softcollfactq}
\eeeq
\item if $(ir) = s$ is the soft gluon, then
\beeq
\bS{(ir)}\bC{ir}\SME{m+2}{0}{p_i,p_r,\dots} &=&
(8\pi\as\mu^{2\eps})^2\times
\label{eq:softcollfactg}
\\ && \times
\frac{1}{s_{ir}}\,\sum_{j,l\ne s}
\frac12 \cS^{\mu\nu}_{jl}(s)\,
\la\mu|\hP_{f_if_r}^{(0)}|\nu\ra
|\cM_{m;(j,l)}^{(0)}(\dots)|^2
\,,
\nn
\eeeq
\end{itemize}
where 
\beq
\cS^{\mu\nu}_{jl}(s) =
4 \frac{p_j^\mu p_l^\nu}{s_{js} s_{sl}}\,,
\eeq
so $g_{\mu\nu} \cS^{\mu\nu}_{jl}(s) = \cS_{jl}(s)$. Notice that
the right-hand side of \eqn{eq:softcollfactq} coincides with that 
of \eqn{eq:collsoftfact},
\beq
\bS{s}\bC{ir}|\cM_{m+2}^{(0)}|^2 = \bSCS{ir;s}|\cM_{m+2}^{(0)}|^2\,.
\label{eq:SsCir}
\eeq

If the soft momentum $p_s$ becomes collinear to the momentum of another
parton $j$, then \eqn{eq:softcollfactq} becomes
\beeq
&&
\bC{js}\bS{s}\bC{ir}\SME{m+2}{0}{p_i,p_r,p_s,p_j,\dots} =
\nn\\ &&\qquad 
=(8\pi\as\mu^{2\eps})^2
\frac{1}{s_{ir}}\,\frac{2}{s_{js}}\,\bT_j^2\, \frac{z_j}{z_s}
\bra{m}{(0)}{(\dots)}\,\hP_{f_if_r}^{(0)}\,\ket{m}{(0)}{(\dots)}\,,
\label{eq:softcollcollfactq}
\eeeq
where $j$ can also be $(ir)$.

If $(ir) = s$ is the soft gluon, then in the $j||s$ limit
\eqn{eq:softcollfactg} becomes
\beeq
&&
\bC{j(ir)}\bS{(ir)}\bC{ir}\SME{m+2}{0}{p_i,p_r,\dots} =
\nn\\ &&\qquad
=(8\pi\as\mu^{2\eps})^2
\frac{1}{s_{ir}}\,\frac{2}{s_{j(ir)}}\,
\bT_j^2\,\frac{z_j}{z_{ir}}
\,2\CA\left(\frac{z_i}{z_r} + \frac{z_r}{z_i}\right)
|\cM_m^{(0)}(\dots)|^2
\label{eq:softcollcollfactgg}
\eeeq
if $i$ and $r$ are both gluons, and
\beeq
&&
\bC{j(ir)}\bS{(ir)_g}\bC{i_qr_\qb}\SME{m+2}{0}{p_i,p_r,\dots} =
\nn\\ &&\qquad
=(8\pi\as\mu^{2\eps})^2
\frac{1}{s_{ir}}\,\frac{2}{s_{j(ir)}}\,
\bT_j^2\,\frac{z_j}{z_{ir}}
\,\TR
|\cM_m^{(0)}(\dots)|^2
\label{eq:softcollcollfactgq}
\eeeq
if $i$ and $r$ are quark and antiquark, where we used
$\kT{}\cdot p_{(ir)} = 0$.  Comparing \eqn{Rcollcollfact}
to \eqn{eq:softcollcollfactq} and \eqn{Rtriplecollfactg} to
\eqns{eq:softcollcollfactgg}{eq:softcollcollfactgq}, we find that 
\beq
\bS{s}\Big(\bC{js} - \bC{js}\bS{s}\Big)\bC{ir}
\SME{m+2}{0}{p_i,p_r,p_s,p_j,\dots} = 0\,,
\eeq
no matter whether or not the soft parton $s$ is $(ir)$, or whether 
the soft parton becomes collinear with the hard parton $(ir)$.

We remark that the treatment of colour in the derivations of these
iterated singly-unresolved limits was different for the
spin-correlated matrix elements in \eqn{Rcollfact} from that for
the colour-correlated matrix elements in \eqn{Rsoftfact}.  
This means that soft factorization formulae do not exist for the
simultaneously spin- and colour-correlated squared matrix elements.

\subsection{Doubly-unresolved limits in the strongly-ordered approximation}
\label{sec:strongordered}

The factorization formulae of the iterated limits coincide with those
valid in the strongly-ordered doubly-unresolved parts of the phase space.
This statement can be used in proving
\eqnss{eq:Cir(A1+A2-A12)}{eq:Srs(A1+A2-A12)}, therefore, we discuss the
strongly-ordered limits briefly.

First we consider the strongly-ordered doubly-collinear limit, which can
occur in two ways. One is defined by the Sudakov parametrization of the
momenta of two separate splitting pairs $ir\to i+r$ and $js\to j+s$ as
in \sect{sec:doubleclim}, but with the strongly-ordered rescaling,
\beq
\kT{r} \to \lambda_r \kT{r}\:,\quad
\kT{s} \to \lambda_s \kT{s}\:,
\label{eq:so2clim}
\eeq
and studying the limit $\lambda_r,\,\lambda_s \to 0$, with
$\lambda_s \gg \lambda_r$. Keeping terms that are of
$\O[1/(\lambda_r^2\lambda_s^2)]$, the squared matrix element
$|{\cM}_{m+2}|^2$ behaves precisely as described by \eqn{RR2collfact}.

The other strongly-ordered doubly-collinear limit occurs when there are
two subsequent collinear splittings from the same hard parton,
$irs \to ir + s \to i + r + s$, which is the strongly-ordered limit of
the triply-collinear emission discussed in \sect{sec:tripleclim}.
In order to describe the $ir \to i+r$ collinear splitting, we
introduce the following Sudakov parametrization of the parton momenta,
\beeq
&&
p_i^\mu = \zeta_i p_{ir}^\mu - \kappa^\mu 
- \frac{\kappa^2}{\zeta_i}\frac{n_{ir}^\mu}{2p_{ir}n_{ir}}\:,
\qquad
p_r^\mu = \zeta_r p_{ir}^\mu + \kappa^\mu 
- \frac{\kappa^2}{\zeta_r}\frac{n_{ir}^\mu}{2p_{ir}n_{ir}}\:,
\nn\\&&
p_{ir}^\mu = (1 - z_s) p_{(ir)s}^\mu - \kT{s}^\mu
- \frac{\kT{s}^2}{1-z_s} \frac{n_{(ir)s}^\mu}{2p_{(ir)s}n_{(ir)s}}
\label{sosudakov}
\eeeq
where $p_{ir}^\mu$ is a light-like momentum that points towards the
collinear  direction of partons $i$ and $r$, $n_{ir}^\mu$ is an
auxiliary light-like vector that is orthogonal to the transverse
component of the third parton ($n_{ir}\kT{s} = 0$) and $\kappa$ are the
momentum components that are orthogonal to both $p_{ir}$ and $n_{ir}$.
Without loss of generality, we may choose $n_{ir} = n_{(ir)s}$.
The other vectors are defined in the same way as after
\eqn{triplesudakov}.  Comparing the two parametrizations of momenta
$p_i$ and $p_r$ in \eqns{triplesudakov}{sosudakov}, we deduce that
\beq
\zeta_{j} = \frac{z_j}{1-z_s}\,,\qquad j = i,r
\eeq
and
\beq
\kT{i} = - \zeta_i \kT{s} - \kappa\,,\qquad
\kT{r} = - \zeta_r \kT{s} + \kappa\,.
\eeq

The strongly-ordered collinear region is identified by performing the
rescaling 
\beq
\kappa \to \lambda \kappa
\,,\qquad
\kT{s} \to \lambda_s \kT{s}
\eeq
and studying the limit $\lambda,\,\lambda_s \to 0$, with
$\lambda_s\gg\lambda$. In this limit \eqnss{RR3collfact}{eq:Pgqqnab}, as
well as the triple-gluon splitting kernel take much simpler forms. In
fact, when neglecting terms that are less singular than
$1/(\lambda^2\lambda_s^2)$ the squared matrix element $|{\cM}_{m+2}|^2$
behaves as
\beeq
&&\!\!\!\!\!
\bC{(ir)s}\bC{ir} \SME{m+2}{0}{p_i,p_r,p_s,\dots} =
\nn\\ &&\qquad\!\!\!\!\!
=(8\pi\as\mu^{2\eps})^2
\frac{1}{s_{ir}} \frac{1}{s_{(ir)s}}\times
\label{RR3collfactso}
\\ && \qquad\quad\!\!\!\!\!
\times
\bra{m}{(0)}{(p_{irs},\dots)} \,
\hP^{{\rm s.o.}}_{f_if_rf_s}
(\zeta_i,\zeta_r,\kappa,z_s,\kT{s},s_{ir},s_{is},s_{rs};\eps)\,
\ket{m}{(0)}{(p_{irs},\dots)}\:,
\nn
\eeeq
where the kernels $\hP^{{\rm s.o.}}_{f_if_rf_s}$ are those in
\sect{sec:tripleclim} in the strongly-ordered approximation.
Explicitly,
\beeq
&&
P^{{\rm s.o.}}_{q_s\qb'_iq'_r}
(\zeta_i,\zeta_r,\kappa,z_s,s_{ir},s_{is},s_{rs};\eps) =
\nn \\ &&\quad
=\TR\Bigg[P^{(0)}_{q_s g}(z_s,1-z_s;\eps)
- 2 \CF\left(\zeta_i\,\zeta_r\,(1-z_s)
  + \frac{s_{\kappa s}^2}{s_{ir} s_{(ir)s}} \right)
\Bigg]\,,
\\[2mm] &&
P^{{\rm (ab)\,s.o.}}_{q_ig_rg_s} (\zeta_i,\zeta_r,z_s;\eps) =
P^{(0)}_{q g_s}(1-z_s,z_s;\eps)
P^{(0)}_{q_i g_r}(\zeta_i,\zeta_r;\eps)\,,
\\[2mm] &&
P^{{\rm (nab)\,s.o.}}_{q_sg_ig_r}
(\zeta_i,\zeta_r,\kappa,z_s,s_{ir},s_{is},s_{rs};\eps) =
\\ &&\quad
=2 \CA\Bigg[P^{(0)}_{q_s g}(z_s,1-z_s;\eps)
\left(\frac{\zeta_i}{\zeta_r} + \frac{\zeta_r}{\zeta_i}\right)
+ \CF\,(1 - \eps)\left(\zeta_i\,\zeta_r\,(1-z_s)
  + \frac{s_{\kappa s}^2}{s_{ir} s_{(ir)s}} \right)
\Bigg]\,,\quad~
\nn
\eeeq
where $s_{\kappa s} = 2 \kappa \cdot p_s$. Furthermore,
\beeq
&&
\la\mu|\hP^{{\rm (ab)\,s.o.}}_{g_iq_r\qb_s}
(\zeta_i,\zeta_r,z_s,\kT{s};\eps)|\nu\ra =
P^{(0)}_{q_r g_i}(\zeta_r,\zeta_i;\eps)
\la\mu|\hP^{(0)}_{\qb q}(z_s,1-z_s,\kT{s};\eps)|\nu\ra\,,
\\[2mm] &&
\la\mu|\hP^{{\rm (nab)\,s.o.}}_{g_sq_i\qb_r}
(\zeta_i,\zeta_r,\kappa,z_s,\kT{s},s_{ir},s_{is},s_{rs};\eps)|\nu\ra =
\nn\\ && \qquad
=2 \CA \TR\Bigg[
- g^{\mu\nu}\left(\frac{z_s}{1-z_s}+\frac{1-z_s}{z_s} \right)
+ g^{\mu\nu} \frac{s_{\kappa s}^2}{s_{ir} s_{(ir)s}}
+ 4 \zeta_i\,\zeta_r\frac{1 - z_s}{z_s}
\frac{\kappa^\mu \kappa^\nu}{\kappa^2}
\Bigg]-
\nn\\ && \qquad\quad
- 4\CA (1 - \eps)\,z_s\,(1-z_s)
P^{(0)}_{q_i \qb_r}(\zeta_i,\zeta_r,\kappa;\eps)
\frac{\kT{s}^\mu\kT{s}^\nu}{\kT{s}^2}
\,,
\\[2mm] &&
\la\mu|\hP^{{\rm s.o.}}_{g_ig_rg_s}
(\zeta_i,\zeta_r,\kappa,z_s,\kT{s},s_{ir},s_{is},s_{rs};\eps)|\nu\ra =
\nn\\ && \qquad
=4 \CA^2 \Bigg[
- g^{\mu\nu}\left(\frac{z_s}{1-z_s}+\frac{1-z_s}{z_s} \right)
  \left(\frac{\zeta_i}{\zeta_r} + \frac{\zeta_r}{\zeta_i}\right)
- g^{\mu\nu} \frac{1 - \eps}{2} \frac{s_{\kappa s}^2}{s_{ir} s_{(ir)s}}-
\nn\\ && \qquad\qquad\qquad
- 2 (1 - \eps)\,\zeta_i\,\zeta_r\,\frac{1 - z_s}{z_s}
\,\frac{\kappa^\mu \kappa^\nu}{\kappa^2}
\Bigg]-
\nn\\ &&\qquad\quad
- 4\CA\,(1 - \eps)\,z_s\,(1-z_s)\,P^{(0)}_{g g}\,(\zeta_i,\zeta_r)
\,\frac{\kT{s}^\mu\kT{s}^\nu}{\kT{s}^2}
\,.
\eeeq
In actual computations we can use
\beq
\kappa^\mu = \zeta_r p_i^\mu - \zeta_i p_r^\mu + \O(\kappa^2)\,,
\qquad
\kT{s}^\mu = z_s p_{ir}^\mu - z_{ir} p_s^\mu + \O(\kT{s}^2)\,.
\label{eq:kappakTs}
\eeq

The $\hP^{{\rm s.o.}}_{f_if_rf_s}$ kernels can also be written in a
unified form, 
\beeq
&&
\la\chi|
\hP^{(0)\,{\rm s.o.}}_{f_if_rf_s}
(\zeta_i,\zeta_r,\kappa,z_s,\kT{s},s_{ir},s_{is},s_{rs};\eps)
|\chi'\ra =
\nn \\ && \qquad
= \sum_{h,h'}
\la h| \hP^{(0)}_{f_if_r}(\zeta_i,\zeta_r,\kappa;\eps)|h'\ra\,
\la\chi|
\hP^{hh'}_{f_{ir}f_s}(z_s,\kT{s};p_{irs},n_{irs};\eps)
|\chi'\ra
\,,
\eeeq
where we used (see Appendix B)
\beq
4 \zeta_i\,\zeta_r\,\frac{z_s}{1-z_s}
\frac{(\kappa\cdot \kT{s})^2}{\kappa^2 \kT{s}^2} =
\frac{s_{\kappa s}^2}{s_{ir} s_{(ir)s}} + \O(\kT{s})\,.
\eeq
Therefore, \eqn{RR3collfactso} coincides with Eqs.~(\ref{Rcollcollfact}),
(\ref{Rtriplecollfactq}) and (\ref{Rtriplecollfactg}).

The strongly-ordered doubly-soft emission is defined by parametrizing the
momenta $p_r$ and $p_s$ as $p_r = \lambda_r q_r$ and $p_s = \lambda_s
q_s$, and taking the limit $\lambda_r,\,\lambda_s \to 0$, with
$\lambda_s \gg \lambda_r$. In this limit the squared matrix element
$|{\cM}_{m+2}|^2$ does not have a leading
$\O[1/(\lambda_r^2\lambda_s^2)]$ singularity when a soft $q\qb$-pair
is emitted. In the case of the emission of two soft gluons, the singular
behaviour of $|{\cM}_{m+2}|^2$ can be written as in \eqn{RRsoftggfact},
the sum of an abelian and a non-abelian contribution,
\beeq
&&\!\!\!\!
\bS{s}\bS{r} \SME{m+2}{0}{p_r,p_s,\dots} =
(8\pi\as\mu^{2\eps})^2
\,\frac14\,\Bigg(
\frac12 \sum_{i,j,k,l=1}^m {\cS}_{ik}(p_r) {\cS}_{jl}(p_s)
|{\cM}^{(0)}_{m,(i,k)(j,l)}|^2-
\nn\\&&\qquad\qquad\qquad\qquad\qquad\qquad\qquad\qquad\qquad
-\CA \sum_{i,k=1}^m {\cS}_{ik}^{(\rm s.o.)}(p_r, p_s)
|{\cM}^{(0)}_{m,(i,k)}|^2
\Bigg),\quad~
\label{RRsoftggfactso}
\eeeq
with ${\cS}_{ik}^{(\rm s.o.)}$ defined in \eqn{eq:softggnabso}.
Thus we see that the abelian term is the same as in \eqn{eq:softggab},
while the non-abelian term in the strong-ordered limit is
simplified a lot. We also see that the
strongly-ordered doubly-soft limit is the same as the iterated
doubly-soft limit recorded in \eqn{eq:SsSrSME}.

We remark that all the formulae in the strongly-ordered doubly-soft limit
are symmetric in $r$ and $s$, therefore, in the limit $\lambda_r \gg
\lambda_s$, exactly the same formulae are valid.

Finally, we discuss the strongly-ordered soft-collinear limit, which is
defined using a similar parametrization of the momenta as in
\sect{sec:s-clim}.  The soft limit is defined by the rescaling
$p_s = \lambda_s q_s$ with letting $\lambda_s \to 0$. The collinear
limit is defined by the usual Sudakov parametrization of
\eqn{eq:sudakov}, with rescaling of $\kT{r}$ as $\kT{r} = \lambda_r
\kT{r}$ and letting $\lambda_r \to 0$.  The limits
$\lambda_r,\,\lambda_s \to 0$ are taken such that either
$\lambda_s\gg\lambda_r$, or vice-versa.  In the limit when
$\lambda_r\gg\lambda_s$, the leading singular behaviour of the squared
matrix element can be written as in \eqn{eq:collsoftfact}, which
coincides with \eqn{RRsoftcollfact}.  In the limit when
$\lambda_s\gg\lambda_r$, the leading singular behaviour of the squared
matrix element can be written as in \eqn{eq:softcollfactq} which is
also the same as \eqn{RRsoftcollfact}.

There is yet another strongly-ordered soft-collinear limit that occurs
when a soft gluon of momentum $p_{(ir)} = \lambda_s q$ makes a collinear
splitting into a quark-antiquark or gluon pair of momenta $p_i$ and
$p_r$. The collinear splitting is parametrized as in \eqn{eq:sudakov} with
rescaling of $\kT{r}$ as $\kT{r} = \lambda_r \kT{r}$. We consider the
limit when $\lambda_s \to 0$ with $\lambda_s\gg\lambda_r$. When the
splitting of the gluon is into a $q\qb$ pair, we start with the
factorization formula for soft-$q\qb$ emission, \eqn{RRsoftqqfact},
suitably re-labelled,
\beeq
&&
\bSqq{ir} \SME{m+2}{0}{p_i,p_r,\dots} =
\label{RRsoftqqfactrelabelled}
\\ &&\qquad
=(8\pi\as\mu^{2\eps})^2
\frac{1}{s_{ir}^2}
\sum_{j \ne (ir)}  \sum_{l\ne j,(ir)}
\left(\frac{s_{ji} s_{lr} + s_{li} s_{jr} - s_{jl} s_{ir}}
           {s_{j(ir)} s_{l(ir)}}
- 2 \frac{s_{ji} s_{jr}}{s_{j(ir)}^2}\right)
\,\TR\,|\cM_{m;(j,l)}^{(0)}(\dots)|^2 \,.
\nn
\eeeq
The strongly-ordered limit can be obtained by substituting into
\eqn{RRsoftqqfactrelabelled} the Sudakov
parametrization of the momenta as in \eqn{eq:sudakov} , which leads to 
\beeq
&&
\bS{(ir)}\bC{ir} \SME{m+2}{0}{p_i,p_r,\dots} =
\nn \\ && \qquad
=(8\pi\as\mu^{2\eps})^2\,\delta_{f_i\qb}\,\delta_{f_rq}
\frac{1}{s_{ir}}\,\sum_{j,l\ne (ir)}
\frac12 \cS^{\mu\nu}_{jl}(ir)\,
\la\mu|\hP_{\qb q}^{(0)}|\nu\ra
|\cM_{m;(j,l)}^{(0)}(\dots)|^2
\,,
\label{eq:S(ir)gCir}
\eeeq
i.e. to the soft limit of the collinear factorization formula
(\eqn{eq:softcollfactg}).
When the
splitting of the gluon is into a gluon pair, we start with the
factorization formula for doubly-soft gluon emission, \eqn{RRsoftggfact},
again suitably relabelled. The abelian part does not have a leading
$\O[1/(\lambda_s^2 \lambda_r^2)]$ singularity. Substituting the Sudakov
parametrization of the momenta leads to an equation analogous to
\eqn{eq:S(ir)gCir} with the $\hP_{\qb q}^{(0)} \to \hP_{gg}^{(0)}$
substitution.

\section{Singly-unresolved limits of the factorization formulae for
doubly-\\unresolved emission}
\label{sec:singlyunresolvedA2}

We are now ready to list all the terms that are necessary for the explicit
construction of the subtraction term $\bA{12}\M{m+2}{(0)}$.

\subsection{Singly-collinear limits of the doubly-unresolved
factorization formulae}
\label{sec:singlycoll}

First we consider the limit of \eqn{RR3collfact} when partons $i$ and
$r$ become collinear.  In order to describe the $ir \to i+r$ collinear
splitting, we introduce the Sudakov parametrization of the
parton momenta as in \eqn{sosudakov}.
The collinear region is identified by performing the rescaling 
$\kappa \to \lambda \kappa$ and studying the limit $\lambda \to 0$.
This limit is the strongly-ordered doubly-collinear limit, therefore,
the factorization formula $\bC{irs}|{\cM}_{m+2}|^2$ factorizes as
\beq
\bC{ir}\bC{irs} \SME{m+2}{0}{p_i,p_r,p_s,\dots} =
\bC{(ir)s}\bC{ir} \SME{m+2}{0}{p_i,p_r,p_s,\dots}\,,
\label{eq:CirCirs}
\eeq
where the right-hand side is the iterated collinear limit of the collinear
factorization formula given in \eqnss{Rtriplecollfactq}{Rtriplecollfactg}.

We parametrize the collinear limit for the other doubly-unresolved
factorization formulae as in \eqn{eq:sudakov}.  The factorization
formula that is valid in the doubly-collinear limit (\eqn{RR2collfact})
is itself factorized in the two independent collinear splittings.
Therefore, the limit of \eqn{RR2collfact} when parton $i$ becomes
collinear to $r$ leaves that equation untouched, namely
\beq
\bC{ir}\bC{ir;js} \SME{m+2}{0}{p_i,p_r,p_j,p_s,\dots} =
\bC{ir}\bC{js} \SME{m+2}{0}{p_i,p_r,p_j,p_s,\dots}\,.
\label{eq:CirCirCjs}
\eeq
The same is true for \eqn{RRsoftcollfact},
\beq
\bC{ir}\bSCS{ir;s} \SME{m+2}{0}{p_i,p_r,p_s,\dots} =
\bSCS{ir;s} \SME{m+2}{0}{p_i,p_r,p_s,\dots}\,.
\label{eq:CirCir;s}
\eeq

The collinear limit of the factorization formula for the emission of a
soft $q\qb$-pair is clearly the strongly-ordered limit.  The same is
true for the non-abelian part of the two-gluon soft emission, thus,
\beq
\bC{ir}\bSN{ir} \SME{m+2}{0}{p_i,p_r,\dots} =
\bS{(ir)}\bC{ir}\SME{m+2}{0}{p_i,p_r,\dots}\,,
\label{eq:CirSgg}
\eeq
where the right-hand side is the soft limit of the collinear-factorization 
formula given in \eqn{eq:softcollfactg}. The abelian part
of the two-gluon soft emission cancels in $\bC{kt}\bA{2}\M{m+2}{(0)}$,
therefore, we do not record it.

The computation of the remaining terms in \eqn{eq:CktA2} is
straightforward. The action of $\bC{ir}$ on \eqn{eq:TC-SC} reproduces
the same,
\beq
\bC{ir}\bC{irs}\bSCS{ir;s}\SME{m+2}{0}{p_i,p_r,p_s,\dots} =
\bC{irs}\bSCS{ir;s}\SME{m+2}{0}{p_i,p_r,p_s,\dots}\,,
\eeq
and so does it when applied on \eqn{eq:DC-SC},
\beq
\bC{ir}\bC{ir;js}\bSCS{ir;s}\SME{m+2}{0}{p_i,p_r,p_j,p_s,\dots} =
\bC{ir;js}\bSCS{ir;s}\SME{m+2}{0}{p_i,p_r,p_j,p_s,\dots}\,.
\eeq
{}From \eqn{eq:TC-DSqq} we get
\beeq
\lefteqn{\bC{rs}\bC{irs}\bSqq{rs}
\SME{m+2}{0}{p_i,p_r,p_s,\dots}= }
\nn\\ && 
= (8\pi\as\mu^{2\eps})^2
\bT_i^2\,\TR\,\frac{2}{s_{rs}s_{i(rs)}}
\left(
\frac{z_i}{z_r + z_s} - \frac{s_{i\kT{s}}^2}{s_{rs}s_{i(rs)}}
\right)\SME{m}{0}{p_i,\dots}\,,
\eeeq
and from \eqn{eq:TC-DSnabgg}, with $z_r = 1 - z_s$, we find
after some algebra,
\beeq
\lefteqn{\bC{r s}\bC{i r s}\bSnab{r s}
\SME{m+2}{0}{p_i,p_r,p_s,\dots} = } 
\nn\\ &&\quad
=(8\pi\as\mu^{2\eps})^2 \times\ 
\nn\\ &&\quad\quad \times\bT_i^2\,\CA\,\frac{2}{s_{rs}s_{i(rs)}}
\left[
\frac{2 z_i}{z_r + z_s}\left(\frac{z_r}{z_s} + \frac{z_s}{z_r}\right)
+ (1 - \eps) \frac{s_{i\kT{s}}^2}{s_{rs}s_{i(rs)}}
\right]\SME{m}{0}{p_i,\dots}\,.
\eeeq

\subsection{Singly-soft limits of the doubly-unresolved factorization
formulae}
\label{sec:singlysoft}

We consider the limit when the soft momentum is parametrized
as $p_r=\lambda q_r$ and let $\lambda\to 0$ at fixed $q_r$.
In this limit the triple-parton splitting kernels factorize into a soft
function times the Altarelli-Parisi kernel obtained from the triple-parton
kernel by deleting the soft parton. Therefore, we obtain
\beeq
&&
\bS{r}\bC{irs}\SME{m+2}{0}{p_i,p_r,p_s,\dots} =
\nn\\&&\qquad
= 8\pi\as\mu^{2\eps} 
\,P^{\rm (S)}_{f_i f_r f_s}(z_i,z_r,z_s,s_{ir},s_{is},s_{rs};\eps)
\,\bC{is}\SME{m+1}{0}{p_i,p_s,\dots}\, ,
\label{eq:SrCirs}
\eeeq
with
\beeq
P^{\rm (S)}_{q_i \qb'_r q'_s}(z_i,z_r,z_s,s_{ir},s_{is},s_{rs};\eps) \aand=
0\,,
\label{eq:SrPqqq}
\\
P^{\rm (S)}_{q_i g_r g_s}(z_i,z_r,z_s,s_{ir},s_{is},s_{rs};\eps) \aand=
\CF\,\frac{2}{s_{ir}} \,\frac{z_i}{z_r}
+ \CA\left(
  \frac{s_{is}}{s_{ir} s_{rs}}
+ \frac{1}{s_{rs}}\,\frac{z_s}{z_r}
- \frac{1}{s_{ir}}\,\frac{z_i}{z_r}\right)\,,
\label{eq:SrPqgg}
\\
P^{\rm (S)}_{g_r q_s \qb_i}(z_i,z_r,z_s,s_{ir},s_{is},s_{rs};\eps) \aand=
\CF\,\frac{2 s_{is}}{s_{ir} s_{rs}}
+ \CA\left(
  \frac{1}{s_{rs}}\,\frac{z_s}{z_r}
+ \frac{1}{s_{ir}}\,\frac{z_i}{z_r}
- \frac{s_{is}}{s_{ir} s_{rs}}\right)\,,
\label{eq:SrPgqq}
\\
P^{\rm (S)}_{g_i g_r g_s}(z_i,z_r,z_s,s_{ir},s_{is},s_{rs};\eps) \aand=
  \CA \left(\frac{s_{is}}{s_{ir}s_{rs}}
+ \frac{1}{s_{ir}}\,\frac{z_i}{z_r}
+ \frac{1}{s_{rs}}\,\frac{z_s}{z_r}\right)
\:.
\label{eq:SrPggg}
\eeeq

The soft limit of the soft-collinear factorization formula in
\eqn{RRsoftcollfact} leaves it unchanged if $p_s \to 0$,
\beq
\bS{s}\bSCS{ir;s}\SME{m+2}{0}{p_i,p_r,p_s,\dots} =
\bSCS{ir;s}\SME{m+2}{0}{p_i,p_r,p_s,\dots}\,.
\label{eq:SsCir;s}
\eeq
However, it is crucial that in \eqn{eq:SsCir;s} we use \eqn{eq:Sj(ir)=Sji}.
When only either parton $r$ or $s$ becomes soft,
the asymptotic behaviour of the factorization formulae for the emission
of a soft $\qb q$ pair (in \eqn{RRsoftqqfact}) or of a gluon pair
(in \eqn{RRsoftggfact}) is
\beeq
&&
\bS{r}\bSqq{rs}\SME{m+2}{0}{p_r,p_s,\dots} = 0\,,
\label{eq:SrSrqsq}
\\ &&
\bS{r}\bSgg{rs}\SME{m+2}{0}{p_r,p_s,\dots} =
\bS{r}\bS{s}\SME{m+2}{0}{p_r,p_s,\dots}\,,
\label{eq:SrSrgsg}
\eeeq
i.e., the iterated soft limit of \eqn{eq:SsSrSME}.

The computation of the remaining terms in \eqn{eq:StA2} is
straightforward. From \eqn{eq:TC-DSnabgg} we get
\beeq
&&
\bS{s}\bC{irs}\bSnab{rs}\SME{m+2}{0}{p_i,p_r,p_s,\dots} =
\nn\\ &&\qquad
=(8\pi\as\mu^{2\eps})^2\times
\nn \\ &&\qquad\quad
\times\,\bT_i^2\,\CA\,\Bigg(
\frac{1}{s_{ir}s_{rs}}\,\frac{2z_i}{z_s} +
\frac{1}{s_{is}s_{rs}}\,\frac{2z_i}{z_r}
- \frac{1}{s_{ir}s_{is}}\,\frac{2z_i^2}{z_r z_s}\Bigg)
\SME{m}{0}{p_i,\dots}\,.
\label{eq:SsCirsSnabrs}
\eeeq
The action of $\bS{s}$ on \eqn{eq:TC-SC_2} leads to the same,
\beq
\bS{s}\bC{irs}\bSCS{ir;s}\SME{m+2}{0}{p_i,p_r,p_s,\dots} =
\bC{irs}\bSCS{ir;s}\SME{m+2}{0}{p_i,p_r,p_s,\dots}\,,
\eeq
and so does it when it is applied on \eqn{eq:SC-softgg},
\beq
\bS{s}\bSCS{ir;s}\bS{rs}\SME{m+2}{0}{p_i,p_r,p_s,\dots} =
\bSCS{ir;s}\bS{rs}\SME{m+2}{0}{p_i,p_r,p_s,\dots}\,.
\eeq

Finally, we consider the limit when the soft momentum $r$ is also
collinear to the momentum of parton $i$. Clearly, in \eqn{eq:SrCirs}
the operator $\bC{ir}$ acts only on the $P^{\rm (S)}_{i r s}$ functions,
\beeq
&&
\bC{ir}\bS{r}\bC{irs}\SME{m+2}{0}{p_i,p_r,p_s,\dots} =
\nn\\&&\qquad
= 8\pi\as\mu^{2\eps} 
\,\Big[\bC{ir}P^{\rm (S)}_{i r s}(z_i,z_r,z_s,s_{ir},s_{is},s_{rs};\eps)\Big]
\,\bC{is}\SME{m+1}{0}{p_i,p_s,\dots}\,.
\label{eq:CirSrCirs}
\eeeq
In this limit, for the non-vanishing $P^{\rm (S)}_{i r s}$ functions we
obtain
\beq
\bC{ir} P^{\rm (S)}_{i r s}(z_i,z_r,z_s,s_{ir},s_{is},s_{rs};\eps) =
\bT_i^2\,\frac{2}{s_{ir}} \,\frac{z_i}{z_r}
\:.
\label{eq:CirSrPirs}
\eeq
The action of $\bC{js}$ on \eqn{eq:SsCir;s} is 
\beeq
&&
\bC{js}\bS{s}\bSCS{ir;s}\SME{m+2}{0}{p_i,p_j,p_r,p_s,\dots} =
\nn\\ && \qquad
=(8\pi\as\mu^{2\eps})^2
\frac{1}{s_{ir}s_{js}}\,\frac{2z_j}{z_s}\,\bT_j^2
\,\bra{m}{(0)}{(p_{ir},\ldots)}\,\hP_{f_i f_r}^{(0)}\,\ket{m}{(0)}{(p_{ir},\ldots)}\,,
\label{eq:CjsSsCir;s}
\eeeq
while on \eqn{eq:SrSrgsg} it acts like
\beq
\bC{js}\bS{r}\bSgg{rs}\SME{m+2}{0}{p_j,p_r,p_s,\dots} =
\bS{r}\bC{js}\bS{s}\SME{m+2}{0}{p_j,p_r,p_s,\dots}\,,
\label{eq:CirSrSrs}
\eeq
which is given explicitly in \eqn{eq:softcollsoftfact}.
Furthermore,%
\footnote{In the $\bC{rs}\bS{s}$ limit the momentum $p_r$ is the correct
variable in the factorized matrix element (see \eqn{Rcsoftfact2}).}
\beeq
&&
\bC{rs}\bS{s}\bSnab{rs}\SME{m+2}{0}{p_r,p_s,\dots} =
\nn\\ && \qquad
=-(8\pi\as\mu^{2\eps})^2
\CA \frac{1}{s_{rs}}\,\frac{2z_r}{z_{s}}\sum_{j}\sum_{l\ne j}
\,\frac12\,\cS_{jl}(r)\SME{m,(j,l)}{0}{\ldots}\,.
\label{eq:CrsSrSrs}
\eeeq
Applying the operator $\bC{js}\bS{s}$ on \eqn{eq:SC-softgg} we get
\beeq
&&
\bC{js}\bS{s}\bSCS{ir;s}\bS{rs} \SME{m+2}{0}{p_i,p_j,p_r,p_s,\dots} =
\nn\\ && \qquad
=(8\pi\as\mu^{2\eps})^2
\,\bT_i^2\,\frac{1}{s_{ir}}\,\frac{2z_i}{z_{r}}
\,\bT_j^2\,\frac{1}{s_{js}}\,\frac{2z_j}{z_{s}}
\,\SME{m}{0}{p_i,p_j,\ldots}\,,
\label{eq:CjsSsCir;sSigrg}
\eeeq
which holds also for $j = i$. Finally, the action of $\bC{rs}$ on
\eqn{eq:SsCirsSnabrs} is 
\beq
\bC{rs}\bS{s}\bC{irs}\bSnab{rs} \SME{m+2}{0}{p_i,p_r,p_s,\dots} =
(8\pi\as\mu^{2\eps})^2
\bT_i^2\,\CA\,
\frac{1}{s_{ir}s_{rs}}\,\frac{4z_i}{z_s}
\SME{m}{0}{p_i,\dots}\,.
\eeq

As we noted at the end of \sect{sec:matchingsingledoubleunresolved},
the factorization formulae of this
section, with similar ones that are not listed, can be used to write
the terms on the right-hand side of \eqn{eq:A12} in the form given in
\eqnss{eq:StA2}{eq:CktStA2}.

\section{The singly- and doubly-unresolved limits of the NNLO subtraction 
term}
\label{sec:matching}

We now turn to the proof of \eqnss{eq:Cir(A1+A2-A12)}{eq:Srs(A1+A2-A12)}.
The explicit check of these equations involves straightforward, but
lengthy algebra. Therefore, we only outline the proof, which uses
the fact that $\bA{12}\M{m+2}{(0)} = \bA{1}\bA{2}\M{m+2}{(0)}$, and let 
the interested readers go through the details.

Let us first consider the singly-unresolved limits. Grouping the terms
as in \eqns{eq:CjsA1}{eq:SsA1}, we obtain
\beq
\bom{X}_s\bA{1}\bA{2}\M{m+2}{(0)} =
\Big(\bom{X}_s \bA{2} + (\bom{X}_s\bA{1} - \bom{X}_s) \bA{2} \Big)
\M{m+2}{(0)}\,,
\eeq
where $\bom{X}_s$ stands for either the collinear limit $\bC{js}$, or
the soft one $\bS{s}$. As a result,
\beeq
\bom{X}_s(\bA{1} + \bA{2} -\bA{1}\bA{2})\M{m+2}{(0)} \aand=
\bom{X}_s\M{m+2}{(0)}
\nn \\ && \quad
+\;(\bom{X}_s\bA{1} - \bom{X}_s) (1 - \bA{2})\M{m+2}{(0)}\,.
\label{eq:Xs(A1+A2-A12)}
\eeeq
The factor in the first parenthesis on the right-hand side contains
operators that take the iterated singly-unresolved limits, which give
vanishing contributions on $(1 - \bA{2})\M{m+2}{(0)}$,
\beq
(\bom{X}_s\bA{1} - \bom{X}_s) (1 - \bA{2})\M{m+2}{(0)} = 0\,.
\label{eq:(XsA1-Xs)(1-A2)}
\eeq
We checked \eqn{eq:(XsA1-Xs)(1-A2)} explicitly for all the terms in
\eqns{eq:CjsA1}{eq:SsA1}. It is a consequence of
\eqnss{eq:CirsA2}{eq:SrsA2}, which hold also in the strongly-ordered
doubly-unresolved (or equivalently in the iterated singly-unresolved)
regions.  Thus only the first term on the right-hand side of 
\eqn{eq:Xs(A1+A2-A12)} remains,
and that is the relevant singly-unresolved limit of the squared matrix
element. Therefore, \eqns{eq:Cir(A1+A2-A12)}{eq:Sr(A1+A2-A12)} hold. 

Turning to the doubly-unresolved limits, we first observe that
according to \eqnss{eq:CirsA2}{eq:SrsA2} we have
\beq
\bom{Y}_{rs}\bA{2}\M{m+2}{(0)} = \bom{Y}_{rs}\M{m+2}{(0)}\,,
\eeq
where $\bom{Y}_{rs}$ stands for any of the operators $\bC{irs}$,
$\bC{ir;js}$, $\bSCS{ir;s}$ and $\bS{rs}$. Thus, all we have to prove is
the following relation:
\beq
\bom{Y}_{rs}(\bA{1} - \bA{12})\M{m+2}{(0)} = 0\,.
\label{eq:Yrs(A1-A12)}
\eeq
We checked the validity of \eqn{eq:Yrs(A1-A12)} explicitly for the four
different limits. In general it follows from
\beq
\bom{Y}_{rs}(\bA{1} - \bA{12})\M{m+2}{(0)}
= \bom{Y}_{rs}\bA{1}(1 - \bA{2})\M{m+2}{(0)}\,,
\label{eq:Yrs(A1-A12)new}
\eeq
and observing that in any term $\bom{X}_{s} \M{m+2}{(0)}$ in the NLO
subtraction $\bA{1}\M{m+2}{0)}$ the dependence on $s$ is already
factorized, therefore, the terms $\bom{Y}_{rs} \bom{X}_{s} \M{m+2}{(0)}$
are factorization formulae valid in the iterated singly-unresolved
limit, or according to the results of \sect{sec:strongordered}, 
in the strongly-ordered doubly-unresolved regions. In these regions 
\eqnss{eq:CirsA2}{eq:SrsA2} hold, therefore, 
\beq
\bom{Y}_{rs} \bA{1}(1 - \bA{2}) \M{m+2}{(0)} = 0\,,
\label{eq:YrsA1(1-A2)}
\eeq
which leads to \eqn{eq:Yrs(A1-A12)}.

It is worth noting that in \eqns{eq:(XsA1-Xs)(1-A2)}{eq:Yrs(A1-A12)} we
have to take unresolved limits of factorization formulae that are valid
in different unresolved regions of the phase space. For instance,
\eqn{eq:(XsA1-Xs)(1-A2)} contains terms such as
$\bS{s}\bC{sr}\bS{r}\bA{2}\M{m+2}{(0)}$. According to \eqn{eq:CktStA2}
such terms contain the factorization formula
$\bC{sr}\bS{r}\bC{irs}\M{m+2}{(0)}$, given explicitly by
\beeq
&&
\bC{sr}\bS{r}\bC{irs}\SME{m+2}{0}{p_i,p_r,p_s,\dots} =
\nn\\ &&\qquad
=(8\pi\as\mu^{2\eps} )^2\times
\nn\\&&\qquad\quad
\times
\frac{2}{s_{rs}}\frac{z_s}{z_r}\,\bT_s^2
\,\frac{1}{s_{is}}\,
\bra{m}{(0)}{(p_{is},\dots)} \hP_{f_i f_s}^{(0)}(z_i,\kT{s};\eps)
\ket{m}{(0)}{(p_{is},\dots)}\,.
\label{eq:CsrSrCirs}
\eeeq
In the region where $p_s\to 0$, this expression contains the ambiguous
limit $\lim_{z_s \to 0} z_s/z_s$. This limit can be computed using the
constraints the momentum fractions have to obey. The $z_s$ in the
numerator is relevant in the collinear limit of partons $s$ and $r$,
where it fulfills the constraint $z_s + z_r = 1$. The $z_s$ in the
denominator is relevant
in the region where the partons $i$ and $s$ are collinear when $z_s +
z_i = 1$.  Thus the momentum fractions, needed to define the
various collinear limits, have to be extended to other parts of the
phase space such that they fulfill different constraints in different
regions. For instance, using the energy fractions, mentioned at the end
of \sect{sec:tripleclim}, we find that the soft limit of
\eqn{eq:CsrSrCirs} is 
\beq
\bS{s}\bC{sr}\bS{r}\bC{irs}\SME{m+2}{0}{p_i,p_r,p_s,\dots} =
(8\pi\as\mu^{2\eps} )^2
\,\frac{2}{s_{rs}s_{is}}\,\frac{z_i}{z_r}\,\CA\,\bT_i^2
\SME{m}{0}{p_{i},\dots}\,.
\eeq
The cancellations discussed in \eqns{eq:(XsA1-Xs)(1-A2)}{eq:Yrs(A1-A12)}
all take place with momentum fractions defined as energy fractions.

\section{A possible general subtraction method for computing NNLO
corrections to jet cross sections}
\label{sec:method}

Building on our studies in the previous sections concerning the
matching of the various singly- and doubly-unresolved limits of the
squared matrix element, in this section we outline a possible general
subtraction method for computing NNLO corrections to jet cross sections.

The full cross section to NNLO accuracy can schematically be
written as 
\beq
\sigma_{{\rm NNLO}} = \tsig{LO} + \tsig{NLO} + \tsig{NNLO}\:.
\eeq
Assuming an $m$-jet quantity, the leading-order contribution is the integral of
the fully differential Born cross section $\dsig{B}_m$ of $m$
final-state partons over the available $m$-parton phase space defined by
the jet function $J_m$,
\beq
\tsig{LO} = \int_m\!\dsig{B}_m J_m\:.
\eeq
The NLO contribution is a sum of two terms, the real and virtual
corrections,
\beq
\tsig{NLO} =
\int_{m+1}\!\dsig{R}_{m+1} J_{m+1} + \int_m\!\dsig{V}_m J_m\:.
\label{eq:sigmaNLO}
\eeq
Here the notation for the integrals indicate that the real correction
involves $m+1$ final-state partons, one of those being unresolved, while
the virtual correction has $m$-parton kinematics. The NNLO correction is
a sum of three contributions, the double-real, the one-loop
single-unresolved real-virtual and the two-loop double-virtual terms,
\beq
\tsig{NNLO} =
\int_{m+2}\!\dsig{RR}_{m+2} J_{m+2}
+ \int_{m+1}\!\dsig{RV}_{m+1} J_{m+1}
+ \int_m\!\dsig{VV}_m J_m\:.
\label{eq:sigmaNNLO}
\eeq
Here the notation for the integrals indicate that the double-real
corrections involve $m+2$ final-state partons, the real-virtual
contribution involves $m+1$ final-state partons and the double virtual
term is an integral over the phase space of $m$ partons, and the phase
spaces are restricted by the corresponding jet functions $J_n$ that
define the physical quantity.

The various contributions to the NLO and NNLO corrections are divergent.
We regularize these singularities using dimensional regularization, that
is defining the integrals in $d = 4 - 2\eps$ dimiensions.
After ultraviolet (UV) renormalization of the one- and two-loop matrix
elements involved in $\dsig{V}$, $\dsig{RV}$ and $\dsig{VV}$ all the
contributions to $\sigma_{{\rm NNLO}}$ are UV finite. In general, the
jet function of $m+n$ partons vanishes if $n+1$ or more partons become
unresolved. Explicitly, $J_m$ vanishes when one parton becomes soft or
collinear to another one,
\beq
J_m(p_1,\ldots,p_m ) \to 0\:, \quad {\rm if} \quad
p_i \ldot p_j \to 0 \:,
\label{JLO1}
\eeq
and $J_{m+1}$ vanishes when two partons become simultaneously soft
and/or collinear, 
\beq
J_{m+1}(p_1,\ldots,p_{m+1} ) \to 0\:, \quad {\rm if} \quad
p_i \ldot p_j\; {\rm and}\; p_k \ldot p_l \to 0 \quad (i \ne k) \:.
\label{JNLO}
\eeq
Therefore, the Born contribution $\dsig{B}_m$ is integrable over the
one- or two-parton infrared (IR) regions of the phase space. Also the
real and virtual contributions $\dsig{R}_{m+1} J_{m+1}$ and $\dsig{V}_m
J_m$ vanish when we integrate over the two-parton IR region of the
phase space, but are separately divergent in $d = 4$ dimensions when
only one (real or virtual) parton becomes soft or collinear to another
parton.  Their sum is finite for {\em infrared-safe} observables.
Formally, infrared safety is expressed by the following requirements for
the jet function,
\beq
\label{Jm1soft}
J_{n+1}(p_1,\ldots,p_j=\lambda q,\ldots,p_{n+1} ) \to
J_n(p_1,\ldots,p_{n+1} )  \quad {\rm if} \quad\lambda \to 0 \:,
\eeq
\beq
\label{Jm1coll}
J_{n+1}(p_1,\ldots,p_i,\ldots,p_j,\ldots,p_{n+1} ) \to
J_n(p_1,\ldots,p,\ldots,p_{n+1} ) \quad {\rm if} \quad p_i \to zp
\:,\; p_j \to (1-z)p
\eeq
for all $n \geq m$.  Equations (\ref{Jm1soft}) and (\ref{Jm1coll})
respectively guarantee that the jet observable is infrared and
collinear safe for {\em any\/} number $n$ of final-state partons, i.e.\
to {\em any\/} order in QCD perturbation theory if there is only
one infrared parton.  The $n$-parton jet function $J_n$ on the
right-hand side of \eqn{Jm1soft} is obtained from the original
$J_{n+1}$ by removing the soft parton $p_j$, and that on the right-hand
side of \eqn{Jm1coll} by replacing the collinear partons
$\{p_i,p_j\}$ by $p = p_i + p_j$.  For infrared-safe observables,
process- and observable-independent techniques exist to cancel
the IR divergences by devising an approximate fully differential
cross section over the $m$-parton phase space, $\dsig{R,A}_{m+1}$ such
that (i) $\dsig{R,A}_{m+1}$ matches the pointwise singular behaviour of
$\dsig{R}_{m+1}$ in the one-parton IR regions of the phase space in any
dimensions (ii) and it can be integrated analytically over the
one-parton phase space, so we can combine it with $\dsig{V}_m$ before
integration. We then write
\beq
\tsig{NLO} =
\int_{m+1}\!\left[\dsig{R}_{m+1} J_{m+1} - \dsig{R,A}_{m+1} J_m\right]
+ \int_m\!\left[\dsig{V}_m + \int_1\!  \dsig{R,A}_{m+1}\right] J_m\:,
\label{eq:sigmaNLO2}
\eeq
where both integrals on the right-hand side are finite in $d = 4$
dimensions (in order to prove the finiteness of the first integral, we
have to use \eqns{Jm1soft}{Jm1coll}).  The final result is that we were
able to rewrite the two NLO contributions in \eqn{eq:sigmaNLO} as a sum
of two finite integrals,
\beq
\tsig{NLO} =
\int_{m+1}\!\dsig{NLO}_{m+1} + \int_m\!\dsig{NLO}_m\:,
\label{eq:sigmaNLOfin}
\eeq
that are integrable in four dimensions using standard numerical
techniques. 

The three contributions to the NNLO corrections in \eqn{eq:sigmaNNLO}
are separately divergent in both the one- and the two-parton IR regions,
but their sum is finite for infrared-safe observables.  To NNLO
accuracy, the formal requirement of infrared safety is expressed by the
following relations,
\beeq
&&
J_{n+2}(p_1,\ldots,p_j=\lambda q,\ldots,p_k=\lambda r,\ldots,p_{n+2} ) \to
J_n(p_1,\ldots,p_{n+2} )  \quad {\rm if} \quad\lambda \to 0 \:,
\label{Jm2soft}
\\ &&
J_{n+2}(p_1,\ldots,p_j=\lambda q,\ldots,p_k,\ldots,p_l,\ldots,p_{n+2} ) \to
J_n(p_1,\ldots,p_{kl},\ldots,p_{n+2})
\nn \\ && \qquad\quad
 {\rm if} \quad\lambda \to 0\quad
 {\rm and} \quad p_k \to z_k p_{kl} \:,\; p_l \to (1-z_k) p_{kl}\:,
\label{Jm2softcoll}
\\ &&
J_{n+2}(p_1,\ldots,p_i,\ldots,p_j,\ldots,p_k,\ldots,p_l,\ldots,p_{n+2} ) \to
J_n(p_1,\ldots,p_{ij},\ldots,p_{kl},\ldots,p_{n+2})
\nn \\ && \qquad\quad
 {\rm if} \quad p_i \to z_i p_{ij} \:,\; p_j \to (1-z_i) p_{ij}\quad
 {\rm and} \quad p_k \to z_k p_{kl} \:,\; p_l \to (1-z_k) p_{kl}\:,
\hspace*{2em}
\label{Jm2dcoll}
\\ &&
J_{n+2}(p_1,\ldots,p_i,\ldots,p_j,\ldots,p_k,\ldots,p_{n+2} ) \to
J_n(p_1,\ldots,p_{ijk},\ldots,p_{n+2})
\nn \\ && \qquad\quad
 {\rm if} \quad p_i \to z_i p_{ijk} \:,\quad p_j \to z_j p_{ijk}\quad
 {\rm and} \quad p_k \to (1-z_i-z_j) p_{ijk}
\label{Jm2tcoll}
\eeeq
for all $n \geq m$ (as implicitly mentioned before, $J_{m+2}$
vanishes if three or more partons become unresolved). 
\eqnss{Jm2soft}{Jm2tcoll} guarantee that the jet observable is infrared
safe for {\em any\/} number $n$ of final-state partons, i.e.\ to {\em
any\/} order in QCD perturbation theory if there are exactly two
unresolved partons.  The $n$-parton jet function $J_n$ on the
right-hand side of \eqn{Jm2soft} is obtained from the original
$J_{n+2}$ by removing the soft partons $p_j$ and $p_k$, that on the
right-hand side of \eqn{Jm2softcoll} by removing the soft parton $p_j$
and replacing the collinear partons $\{p_k,p_l\}$ by $p_{kl} = p_k +
p_l$, that on the right-hand side of \eqn{Jm2dcoll} by replacing the
collinear partons $\{p_i,p_j\}$ by $p_{ij} = p_i + p_j$,  and
$\{p_k,p_l\}$ by $p_{kl} = p_k + p_l$ and that on the right-hand side
of \eqn{Jm2tcoll} by replacing the collinear partons $\{p_i,p_j,p_k\}$
by $p_{ijk} = p_i + p_j + p_k$.  

In order to cancel the one- and two-parton singularities between the
double-real, real-virtual and double-virtual terms in
\eqn{eq:sigmaNNLO}, we proceed like in \eqn{eq:sigmaNLO2},
namely, we introduce approximate cross sections $\dsiga{RR}{2}_{m+2}$ and
$\dsiga{RV}{1}_{m+1}$ and rewrite \eqn{eq:sigmaNNLO} as
\beeq
&&
\tsig{NNLO} =
\int_{m+2}\!\left[\dsig{RR} J_{m+2} - \dsiga{RR}{2}_{m+2} J_m\right]
+ \int_{m+1}\!\left[\dsig{RV} J_{m+1} - \dsiga{RV}{1}_{m+1} J_m\right]+
\nn \\ && \qquad
+ \int_m\!\dsig{VV} J_m
+ \int_{m+2}\!\dsiga{RR}2_{m+2} J_m + \int_{m+1}\!\dsiga{RV}1_{m+1} J_m
\:.
\label{eq:sigmaNNLO2}
\eeeq
The fully differential approximate cross section to the double-real
emission is constructed such that it has the same pointwise singular
behaviour in $d$ dimensions as $\dsig{RR}_{m+2}$ in the two-parton
infrared regions (hence the subscript 2 on A in the superscript).
Then \eqnss{Jm2soft}{Jm2tcoll} guarantee that the $(m+2)$-parton
integral in \eqn{eq:sigmaNNLO2} is finite over the doubly-unresolved
regions of the phase space. However, this integral is still divergent
in the singly-unresolved regions, because both the fully differential
cross sections $\dsig{RR}$ and $\dsiga{RR}{2}_{m+2}$ as well as the jet
functions $J_{m+2}$ and $J_m$ have different limits in the
singly-unresolved regions.

The real-virtual contribution has two types of singularities: (i) explicit
$\eps$ poles in the loop amplitude and (ii) singular behaviour in the
singly-unresolved regions of the phase space. To regularize the latter,
the fully differential approximate cross section, $\dsiga{RV}1_{m+1}$,
to the real-virtual contribution must have the same pointwise singular
behaviour in $d$ dimensions as $\dsig{RV}_{m+1}$ itself in the
one-parton infrared regions (hence the subscript 1 on A in the
superscript). Using such a regulator, the difference 
$\left[\dsig{RV}_{m+1} J_{m+1} - \dsiga{RV}1_{m+1} J_m\right]$ 
is still not integrable over the full
$(m+1)$-parton phase space in $d = 4$ dimensions, simply because the
integrand still contains $\eps$ poles in phase space regions away
from the one-parton unresolved infrared regions. The pole structures of
$\dsig{RV}_{m+1}$ and $\dsiga{RV}1_{m+1}$ are different.

In order to cancel the remaining singularities in the $(m+2)$-parton and
$(m+1)$-parton integrals in the first line of \eqn{eq:sigmaNNLO2}, we
perform a subtraction procedure between these two integrals, which is very
similar to that used in a NLO computation,
\eqn{eq:sigmaNLO2}. This amounts to devising fully differential
approximate cross sections to both $\dsig{RR}_{m+2}$ and
$\dsiga{RR}{2}_{m+2}$ that match the singular behaviour of these cross
sections in the singly-unresolved infrared regions in $d$ dimensions,
which are denoted by $\dsiga{RR}{1}_{m+2}$ and $\dsiga{RR}{12}_{m+2}$,
respectively.  Thus, the full NNLO cross section is written as
\beeq
&&
\tsig{NNLO} =
\int_{m+2}\!\left[\dsig{RR} J_{m+2} - \dsiga{RR}{2}_{m+2} J_m
- \dsiga{RR}1_{m+2} J_{m+1} + \dsiga{RR}{12}_{m+2} J_m \right]+
\nn \\ && \qquad\quad
+ \int_{m+1}\!\left[\dsig{RV} J_{m+1} - \dsiga{RV}{1}_{m+1} J_m\right]
+ \int_{m+2}\!\left[\dsiga{RR}1_{m+2} J_{m+1} - \dsiga{RR}{12}_{m+2} J_m \right]+
\nn \\ && \qquad\quad
+ \int_m\!\dsig{VV} J_m 
  + \int_{m+2}\!\dsiga{RR}2_{m+2} J_m + \int_{m+1}\!\dsiga{RV}1_{m+1} J_m
\:.
\label{eq:sigmaNNLO3}
\eeeq
The first integral on the right-hand side of this equation is finite in
$d = 4$ dimensions by construction.  Hence, this first integral can be
performed numerically in four dimensions using standard Monte Carlo
techniques.  

The singularities are associated to the integrals in the second and
third lines of \linebreak
\eqn{eq:sigmaNNLO3}. We shall show in a separate
publication that the integration of $[\dsiga{RR}1_{m+2} J_{m+1}-$\linebreak
$\dsiga{RR}{12}_{m+2} J_m]$ can be performed analytically over the
one-parton phase space of the unresolved parton of the $(m+2)$-parton
phase space, leading to $\eps$ poles, which cancel exactly the
remaining $\eps$ poles in the integrand $[\dsig{RV}_{m+1} J_{m+1} -
\dsiga{RV}{1}_{m+1} J_m]$. As a result the integrals in the
second line of \eqn{eq:sigmaNNLO3} can be combined into a single
$(m+1)$-parton integral, where the integrand is free of $\eps$ poles
and can be integrated numerically in four dimensions. We emphasize that
this statement does not follow from unitarity, but has to be proven.

Once we have shown the finiteness of the combination of integrals in the
second line of \eqn{eq:sigmaNNLO3}, it follows from the 
Kinoshita-Lee-Nauenberg theorem that
the combination of the integrals in the third line must be finite, though
separately all three are divergent. If one is able to carry out
analytically the integration of $\dsiga{RR}2_{m+2}$ over the two-parton
subspace and that of $\dsiga{RV}1_{m+1}$ over the one-parton subspace of
the doubly- and singly-unresolved regions of the $(m+2)$-parton and
$(m+1)$-parton phase spaces, respectively, leading to $\eps$ poles, we
can combine these poles with those in the double-virtual contribution. As
a result all the divergences cancel and the remaining $m$-parton integral can
be performed numerically in four dimensions.

The final result of the above manipulations is that we rewrite
\eqn{eq:sigmaNNLO} as 
\beq
\tsig{NNLO}=
\int_{m+2}\!\dsig{NNLO}_{m+2}
+ \int_{m+1}\!\dsig{NNLO}_{m+1}
+ \int_m\!\dsig{NNLO}_m\:,
\label{eq:sigmaNNLOfin}
\eeq
that is, a sum of three integrals,
\beq
\dsig{NNLO}_{m+2} =
\left[\dsig{RR}_{m+2} J_{m+2} - \dsiga{RR}2_{m+2} J_m 
- \dsiga{RR}1_{m+2} J_{m+1} + \dsiga{RR}{12}_{m+2} J_m \right]_{\eps =0}\:,
\label{eq:sigmaNNLOm+2}
\eeq
\beq
\dsig{NNLO}_{m+1} =
\left[\dsig{RV}_{m+1} J_{m+1} - \dsiga{RV}{1}_{m+1} J_m
+ \int_1\!\left(\dsiga{RR}1_{m+2} J_{m+1}
  - \dsiga{RR}{12}_{m+2} J_m \right) \right]_{\eps =0}\:,
\label{eq:sigmaNNLOm+1}
\eeq
and
\beq
\dsig{NNLO}_m =
\left[\dsig{VV} + \int_2\!\dsiga{RR}2_{m+2}
 + \int_1\!\dsiga{RV}{1}_{m+1}\right]_{\eps = 0} J_m
\:,
\label{eq:sigmaNNLOm}
\eeq
each integrable in four dimensions using standard numerical techniques.

The factorization formulae presented in the previous sections can be
used in constructing the approximate cross sections
$\dsiga{RR}1_{m+2}$, $\dsiga{RR}2_{m+2}$ and $\dsiga{RR}{12}_{m+2}$. 
However, we emphasize that the momenta in the
matrix elements on the right hand sides of the factorization formulae
are unambigously defined only in the strict unresolved limits.
Therefore the factorization formulae cannot be directly used as true
subtraction formulae, unless we can explicitly specify which partons
become unresolved --- either by partitioning the squared matrix element
as in \Ref{Ellis:1980wv}, or using the measurement function as in
\Ref{Frixione:1995ms}, or by partitioning the phase space as in
\Ref{Nagy:1996bz}. These procedures lead to the so-called residue
methods which are well understood in NLO computations. If one wishes to
avoid specifying the unresolved partons, then one has to implement
exact factorization of the phase space, such that the integrations over
the unresolved momenta can be carried out and momentum conservation is
maintained as done in the case of dipole subtraction method of
\Ref{Catani:1996vz}, which is worked out completely for NLO computations.

\section{Conclusions}
\label{sec:concl}

In this paper we studied the infrared structure of the known
factorization formulae for tree-level QCD squared matrix elements in
all the possible soft and collinear limits. We presented new
factorization formulae for the colour-correlated and the
spin-correlated squared matrix elements that we termed iterated
singly-unresolved limits. We pointed out that  soft factorization
formulae do not exist for the simultaneously spin- and
colour-correlated squared matrix elements which indicates that
in the general subtraction scheme envisaged by us, the azimuthally
correlated singly-collinear subtraction terms must not contain
colour correlations. This can be achieved naturally for those processes
when the colour charges in the colour-correlated squared matrix elements
can be factorized completely (see Appendix A).
We derived the factorization formulae of
the squared matrix element that are valid in the strongly-ordered
doubly-unresolved regions of the phase space and showed that these are
equal to the iterated singly-unresolved factorization formulae. We
dealt only with final state singularities. Using crossing symmetry, it
is possible to extend our results to cases when initial state partons
become unresolved, which we shall present in a separate publication.

We introduced a new formal notation for denoting the various limits in
the overlapping unresolved parts of the phase space. Using this
notation, we explicitly constructed subtraction terms that regularize
the kinematical singularities of the squared matrix element in all
singly- and doubly-unresolved parts of the phase space and demonstrated
that the subtraction terms avoid all possible double and triple
subtractions.  The relevant subtraction terms were presented in 
Sects.~\ref{sec:doubleunresolved} and \ref{sec:matchingsingledoubleunresolved}.
As a result the regularized squared matrix element is integrable over all
phase space regions where at most two partons become unresolved. We
emphasize however, that the subtraction terms derived in this paper can
be used as true subtraction terms only if the unresolved momenta are
specified, like in \Refs{Frixione:1995ms,Nagy:1996bz} in the case of
NLO computations. In an approach, where any parton can become
unresolved, the factorization formulae have to be extended over the
whole phase space.  This extension requires a phase-space factorization
that maintains momentum conservation exactly, as done in
\Ref{Catani:1996vz} in the case of NLO computations, but such that in
addition it respects the structure of the delicate cancellations among
the various subtraction terms. 
So far none of these methods has been demonstrated to be applicable in
practical computations of NNLO corrections.  The kinematics of two-jet
production in electron-positron annihilation is sufficiently simple so
that we were able to implement the subtraction scheme presented in
this paper for the case of $m = 2$. We computed the cross sections in
\eqns{eq:sigmaNNLOm+2}{eq:sigmaNNLOm+1} and found numerically stable
results for any observable. The details of the computations will be
given elsewhere.

\section*{Acknowledgments}
ZT thanks the INFN, Sez. di Torino, VDD thanks the Nucl.~Res.~Inst.~of
the HAS for their kind hospitality during the long course of this work.
We are grateful to S. Catani for his comments on the manuscript.
This research was supported in part by
the Hungarian Scientific Research Fund grant OTKA T-038240.

\newpage
\appendix
\section{Explicit computation of the soft limits of known
colour-correlated squared matrix elements} 
\label{sec:checkMikfact}

In this appendix we check the validity of the soft factorization formulae
\eqns{eq:Miksoftfact}{eq:Msksoftfact} using the colour-correlated squared
matrix elements for the $e^+e^- \to q \qb + n g$ processes ($n = 1$, 2).

\subsection{Soft limit of $|\cM_{3;(i,k)}^{(0)}|^2$}

Using colour conservation, we can compute the product of the colour
charges acting on the three-parton colour space in terms of Casimir
invariants \cite{Catani:1996vz}, 
\beq
\bT_q \cdot \bT_\qb = \frac{\CA}{2} - \CF\,,\qquad 
\bT_g \cdot \bT_q = \bT_g \cdot \bT_\qb = -\frac{\CA}{2}\,.
\label{3partoncolalg}
\eeq
Thus the colour charges in the colour-correlated
squared matrix elements for the $e^+e^- \to q \qb g$ process can be
factorized completely \cite{Ellis:1980wv}. If the quark, antiquark and gluon is
labelled (in this order) 1, 2 and 3, then we have
\beq
|\cM_{3;(1,2)}^{(0)}|^2 =
\left(\frac{\CA}{2} - \CF\right) |\cM_3^{(0)}|^2 \,.
\eeq
and
\beq
|\cM_{3;(1,3)}^{(0)}|^2 = |\cM_{3;(2,3)}^{(0)}|^2 =
-\frac{\CA}{2} |\cM_3^{(0)}|^2 \,,
\eeq
Consequently, the limit of the colour-correlated squared matrix
elements when $p_3$ gets soft is simply colour factors times the soft
limit of the squared matrix element~\cite{Catani:2000pi},
\beq
|\cM_{3;(1,2)}^{(0)}(p_1,p_2,p_3)|^2 \simeq
-8\pi \as \mu^{2\eps}\,
\left(\frac{\CA}{2} - \CF\right) \cS_{12}(3) |\cM_{2;(1,2)}^{(0)}(p_1,p_2)|^2
\,,
\eeq
which --- using
$\bT_1 \cdot \bT_2 \ket{2}{(0)}{(p_1,p_2)} = -\CF \ket{2}{(0)}{(p_1,p_2)}$
--- can also be obtained from \eqn{eq:Miksoftfact},
and 
\beq
|\cM_{3;(1,3)}^{(0)}(p_1,p_2,p_3)|^2 \simeq 
8\pi \as \mu^{2\eps}\,
\frac{\CA}{2} \cS_{12}(3)|\cM_{2;(1,2)}^{(0)}(p_1,p_2)|^2 \,,
\eeq
which can also be obtained from \eqn{eq:Msksoftfact}.

\subsection{Soft limit of $|\cM_{4;(i,k)}^{(0)}|^2$}

The colour-correlated squared matrix elements for the 
$e^+e^- \to q_1 \qb_2 g_3 g_4$ process were computed in Appendix B of
\Ref{Nagy:1998bb}.%
\footnote{Note a typographic error in Eqns.~(B11--B13) of \Ref{Nagy:1998bb}:
the labelling convention given in that paper is correct for the process
$e^+e^- \to q_1 \qb_2 g_3 g_4$ we consider here. In order to get the
$M_0^{ik}$, $M_x^{ik}$ and $M_{xx}^{ik}$ matrices relevant to the
labelling convention used in \Ref{Nagy:1998bb} the 2, 3 and 4 indices of the
$M_0^{ik}$ matrices should be cyclicly permuted, (2,3,4) $\to$ (4,2,3).}
We rewrite the expressions presented there using our notation:
\beq
\left|{\cal M}_{4;(i,k)}^{(0)}\right|^2 =
-\frac12 \Nc \CF \left(
\CF^2 M_0^{ik} + \CA \CF M_x^{ik} + \CA^2 M_{xx}^{ik}\right)\,,
\label{M04ik}
\eeq
where the non-vanishing elements of the matrices $M_0^{ik}$, $M_x^{ik}$,
$M_{xx}^{ik}$ are given by
\beeq
&&
M_0^{12} = 2 |m_3|^2\:,
\label{M0ik}
\\ &&
M_x^{12} = -3 |m_3|^2 + |m_1|^2 + |m_2|^2\:,\qquad
M_x^{13} = M_x^{14} = M_x^{23} = M_x^{24} = |m_3|^2\:,
\\ &&
M_{xx}^{34} =  \frac12 (|m_1|^2 + |m_2|^2)\:,\qquad\qquad\quad
M_{xx}^{12} = |m_3|^2 - M_{xx}^{34}\:,
\\ &&
M_{xx}^{13} = M_{xx}^{24} = -\frac12 (|m_3|^2 - |m_1|^2)\:,\quad
M_{xx}^{14} = M_{xx}^{23} = -\frac12 (|m_3|^2 - |m_2|^2)\:,
\label{Mxxik}
\eeeq
and the $|m_i|^2$ functions are the helicity-summed squared helicity
amplitudes with all coupling factors included. In terms of the amplitudes 
defined in Appendix A of \Ref{Nagy:1998bb}
\beq
|m_1|^2 = \sum_{\rm hel} |m(1,3,4,2)|^2\,,\qquad
|m_2|^2 = \sum_{\rm hel} |m(1,4,3,2)|^2\,,
\eeq
and $m_3 = m_1 + m_2$.  Using the soft factorization formula for the
helicity amplitudes, it is not difficult to obtain the following
factorization formulae for the $|m_i|^2$ functions when $p_4 \to 0$:
\beeq
&&
|m_1(p_1, p_2, p_3, p_4)|^2 \simeq 8\pi\as \mu^{2\eps} \cS_{23}(4)
\frac{1}{\Nc \CF} |\cM_3^{(0)}(p_1,p_2,p_3)|^2\:,
\\ &&
|m_2(p_1, p_2, p_3, p_4)|^2 \simeq 8\pi\as \mu^{2\eps} \cS_{13}(4)
\frac{1}{\Nc \CF} |\cM_3^{(0)}(p_1,p_2,p_3)|^2\:,
\\ &&
|m_3(p_1, p_2, p_3, p_4)|^2 \simeq 8\pi\as \mu^{2\eps} \cS_{12}(4)
\frac{1}{\Nc \CF} |\cM_3^{(0)}(p_1,p_2,p_3)|^2\:.
\label{S3softfact}
\eeeq
Substituting \eqnss{M0ik}{S3softfact} into \eqn{M04ik},
we obtain the soft factorization formulae for the colour-correlated
squared matrix elements, when $p_4 \to 0$:
\beeq
&&
\left|{\cal M}_{4;(i,k)}^{(0)}(p_1,p_2,p_3,p_4)\right|^2 \simeq
-8\pi\as \mu^{2\eps} \left|{\cal M}_3^{(0)}(p_1,p_2,p_3)\right|^2
\cF_{ik}(p_1,p_2,p_3,p_4)\,,
\label{M04iksoftfact}
\eeeq
where
\beeq
&&
\cF_{12}(p_1,p_2,p_3,p_4) =
\left(\CF - \frac{\CA}{2}\right) \left[ (\CF - \CA)\cS_{12}(4)
+ \frac{\CA}{2}\left(\cS_{13}(4) + \cS_{23}(4)\right) \right]\,,~~~~
\label{F12}
\\ &&
\cF_{13}(p_1,p_2,p_3,p_4) =
\frac{\CA}{2} \left[ \left(\CF - \frac{\CA}{2}\right)\cS_{12}(4) +
\frac{\CA}{2}\cS_{23}(4)\right]\,,
\label{F13}
\\ &&
\cF_{14}(p_1,p_2,p_3,p_4) =
\frac{\CA}{2} \left[ \left(\CF - \frac{\CA}{2}\right)\cS_{12}(4) +
\frac{\CA}{2}\cS_{13}(4)\right]\,,
\label{F14}
\\ &&
\cF_{34}(p_1,p_2,p_3,p_4) =
\left(\frac{\CA}{2}\right)^2
\left[\cS_{13}(4) + \cS_{23}(4)\right]\,.
\label{F34}
\eeeq
Using \eqn{3partoncolalg} it is straightforward to check that the
results in \eqn{M04iksoftfact}, \eqns{F12}{F13} can also be obtained
from \eqn{eq:Miksoftfact}, and those in Eqs.~(\ref{M04iksoftfact}),
(\ref{F14}) and (\ref{F34}) can also be obtained from \eqn{eq:Msksoftfact}.

\section{Collinear limit of the spin-polarisation tensor for the process
\hbox{$e^+e^- \to q \qb g$}}

Consider the {\em tree-level} matrix element for the production of a
quark-antiquark pair and a gluon in electron-positron annihilation,
\beq
\cM_{q_1,\qb_2,g_3}^{c_1,c_2,c_3;s_1,s_2,\mu_3}(p_1,p_2,p) =
\la c_1 c_2 c_3|\otimes \la s_1 s_2 \mu_3| \ket{3}{(0)}{(p_1,p_2,p)}\,,
\label{Mqqbg}
\eeq
where $\{c_1,c_2,c_3\}$, $\{s_1,s_2,\mu_3\}$ and
$\{q_1,\bar{q}_2,g_3\}$ are the colour, spin and flavour indices of the
partons.  We define the spin-polarization tensor as in \eqn{eq:Tmunu}:
\beq
\cT_{q\bar{q}g}^{\mu\nu}(p_1,p_2,p) =
\bra{3}{(0)}{(p_1,p_2,p)} \mu\ra \la \nu \ket{3}{(0)}{(p_1,p_2,p)}
\,,
\label{Tmn}
\eeq
so $-g_{\mu\nu}\cT_{q\qb g}^{\mu\nu}(p_1,p_2,p)=\SME{3}{0}{p_1,p_2,p}$.

In this Appendix, we examine the collinear limit of
$\la\mu|\hP_{gg}^{(0)}(z,\kT{};\eps)|\nu\ra\,\cT_{q\qb g}^{\mu\nu}(p_1,p_2,p)$
(we suppress the second argument of the \AP kernel, which equals $1-z$).
{}From Appendix D of \Ref{Catani:1996vz} we have (in $d=4$ dimensions)
\beq
\cT_{q\qb g}^{\mu\nu}(p_1,p_2,p) =
-\frac{1}{y_{1p}^2+y_{2p}^2+2y_{12}}\,\SME{3}{0}{p_1,p_2,p}T^{\mu\nu},
\label{eq:Tqqbgmunu}
\eeq
where $y_{ij} = s_{ij}/Q^2 \equiv s_{ij}/(p_1+p_2+p)^2$ and
\beeq
T^{\mu\nu} &=&
+ 2\frac{p_1^{\mu} p_2^{\nu}}{Q^2} + 2\frac{p_2^{\mu} p_1^{\nu}}{Q^2} 
- 2\frac{y_{2p}}{y_{1p}}\frac{p_1^{\mu} p_1^{\nu}}{Q^2}
- 2\frac{y_{1p}}{y_{2p}}\frac{p_2^{\mu} p_2^{\nu}}{Q^2}+
\nn\\ &&
+ \frac{y_{12}-(y_{12}+y_{2p})^2}{y_{1p}}
  \left[\frac{p_1^{\mu} p^{\nu}}{Q^2}+\frac{p^{\mu} p_1^{\nu}}{Q^2}\right]
+ \frac{y_{12}-(y_{12}+y_{1p})^2}{y_{2p}}
  \left[\frac{p_2^{\mu} p^{\nu}}{Q^2}+\frac{p^{\mu} p_2^{\nu}}{Q^2}\right]+
\nn\\ &&
+ \frac{1}{2}\left(y_{1p}^2+y_{2p}^2\right)g^{\mu\nu}.
\eeeq
We also have (in $d=4$ dimensions)
\beq
\SME{3}{0}{p_1,p_2,p}=\CF\,\frac{8\pi\alps}{Q^2}
\,\frac{y_{1p}^2+y_{2p}^2+2y_{12}}{y_{1p}y_{2p}}\,\M{2}{(0)}.
\label{eq:SME3}
\eeq
Here $\M{2}{(0)}$ is the squared matrix element for the production
of a quark-antiquark pair in electron-positron annihilation, averaged
over event orientation (so it has no dependnece on parton momenta).
Substituting \eqn{eq:SME3} into \eqn{eq:Tqqbgmunu}, we find:
\beq
\cT_{q\qb g}^{\mu\nu}(p_1,p_2,p) =
-8\pi\alps\,\CF\,\frac{1}{Q^2}\,\M{2}{(0)}\,\frac{1}{y_{1p}y_{2p}}\,T^{\mu\nu}.
\eeq

The tree-level Altarelli-Parisi splitting kernel for $g\to gg$
splitting, $\la\mu|\hP_{gg}^{(0)}(\zeta,\kappa;\eps)|\nu\ra $ is given in
\eqn{P0gg}.  The case of interest to us is when $p$ in the
spin-polarisation tensor is the collinear direction of the $g\to gg$
splitting.  In particular this means that $p\cdot \kappa=0$.  

With the explicit expressions for $\cT_{q\qb g}^{\mu\nu}(p_1,p_2,p)$ and
$\la\mu|\hP_{gg}^{(0)}(\zeta,\kappa;\eps)|\nu\ra $,
we can carry out the contraction
\beeq
&&
\la\mu|\hP_{gg}^{(0)}(\zeta,\kappa;\eps)|\nu\ra 
\,{\cal T}_{q\qb g}^{\mu\nu}(p_1,p_2,p) = 
\nn\\ &&\qquad
=2\CA
\Bigg\{
\SME{3}{0}{p_1,p_2,p}
\,\left(\frac{\zeta}{1-\zeta}+\frac{1-\zeta}{\zeta}\right)+
\label{PT}
\\ &&\qquad\qquad\qquad
+ 8\pi\as\CF\,\M{2}{(0)}
\,\frac{1-\eps}{s_{34}}
\left[\left(\frac{s_{1\kappa}}{s_{1p}}-\frac{s_{2\kappa}}{s_{2p}}\right)^2
+ \frac{s_{34}}{Q^2}\,\zeta(1-\zeta)
\,\frac{s_{1p}^2+s_{2p}^2}{s_{1p}s_{2p}}\right]
\Bigg\}\,,
\nn
\eeeq
where we have introduced the following notation:
\beq
s_{i\kappa}=2p_i\kappa,\quad s_{ip}=2p_ip,\quad i=1,2\,, \qquad
s_{34}=-\frac{\kappa^2}{\zeta(1-\zeta)}\,.
\eeq

We now derive the $p_1||p$ collinear limit of \eqn{PT}, which is
precisely defined by the following Sudakov parametrization of the
momenta:
\beeq
&&
p_1^{\mu}=z_1 P^{\mu}+\kT{}^{\mu}
-\frac{\kT{}^2}{z_1}\frac{n^{\mu}}{2Pn}\,,\quad
p^{\mu}=(1-z_1) P^{\mu}-\kT{}^{\mu}
-\frac{\kT{}^2}{1-z_1}\frac{n^{\mu}}{2Pn}\,,
\nn\\ &&
s_{1p} = -\frac{\kT{}^2}{z_1(1-z_1)} = \O(\kT{}^2)\,,
\qquad \kT{}\to 0\,,
\label{s1p}
\label{colllimit}
\eeeq
where $P\kT{}=n\kT{}=P^2=n^2=0$.  When calculating $s_{1\kappa}$, we
use $p\kappa=0$ to eliminate the $P\kappa$ term from $s_{1\kappa}$. Then we get
\beq
s_{1\kappa} = \frac{2}{1-z_1}\kT{}\cdot\kappa
+ 2\left(\frac{z_1}{(1-z_1)^2}-\frac{1}{z_1}\right)
\frac{n\kappa}{2nP}\kT{}^2 = \O(\kT{}).
\label{s1kappa}
\eeq
Then the collinear limit of \eqn{PT} reads
\beeq
&&
\bC{1p}
\la\mu|\hP_{gg}^{(0)}(\zeta,\kappa;\eps)|\nu\ra 
\,{\cal T}_{q\qb g}^{\mu\nu}(p_1,p_2,p) =
\nn\\ &&\qquad
=8\pi\alps\,\SME{2}{0}{P,p_2}\,2\CA \,\frac{1}{s_{1p}}\times
\label{direct_c1pPT}
\\ &&\qquad\quad
\times
\Bigg\{
\left(\frac{\zeta}{1-\zeta}+\frac{1-\zeta}{\zeta}\right) P^{(0)}_{qg}(z_1;\eps)
+ \CF\,\frac{1-\eps}{s_{34}}
\left[\frac{s_{1\kappa}^2}{s_{1p}}+s_{34}\,\zeta(1-\zeta)\,(1-z_1)\right]
\Bigg\}
\,.\quad~
\nn
\eeeq

We can compute the same limit using the expression derived for the
collinear limit of the spin-polarisation tensor in
\sect{sec:collcollfact}.  Defining the $p_1||p$ limit as in
\eqn{colllimit}, we have (see \eqn{Tirscollfact}):
\beq
\cT^{\mu\nu}_{q\qb g}(p_1,p_2,p) \simeq \epas
\frac{1}{s_{1p}}\,\bra{2}{(0)}{(P,p_2)}
\,\hP^{\mu\nu}_{gq}(z_1,\kT{};P,n,\eps)
\,\ket{2}{(0)}{(P,p_2)},
\label{Tlim}
\eeq
where $\la r|\hP^{\mu\nu}_{gq}(\zeta,\kappa;p,n,\eps)|s\ra$ is given in
\eqn{eq:Pgqalfabeta}.
Using \eqn{Tlim} we obtain for the collinear limit of \eqn{PT}
\beeq
&&
\bC{1p}
\,\la\mu|\hP_{gg}^{(0)}(\zeta,\kappa;\eps)|\nu\ra 
\,{\cal T}_{q\qb g}^{\mu\nu}(p_1,p_2,p) =
\nn\\ &&\qquad
=\epas\,\SME{2}{0}{P,p_2} \,2\CA \,\frac{1}{s_{1p}}\times
\label{new_c1pPT}
\\&&\qquad\quad
\times
\Bigg\{
\left(\frac{\zeta}{1-\zeta}+\frac{1-\zeta}{\zeta}\right) P^{(0)}_{qg}(z_1;\eps)+
\nn\\&&\qquad\qquad\quad
+\,2\CF\,\frac{1-\eps}{s_{34}} \left[
 - \frac{1-z_1}{2}\kappa^2
 - 2\frac{z_1}{1-z_1} \frac{(\kT{}\cdot\kappa)^2}{\kT{}^2}
 + (1-z_1)\frac{(P\kappa)(n\kappa)}{Pn}\right]
\Bigg\}
\,.\quad~
\nn
\eeeq
To see that \eqns{direct_c1pPT}{new_c1pPT} are in fact equal up to
subleading terms, note the following. From $p\kappa = 0$, we get
\beq
P\kappa = \frac{1}{1-z_1}\kT{}\cdot\kappa
+ \frac{z_1 \kT{}^2}{(1-z_1)^2}\frac{n\kappa}{2Pn}
=\O(\kT{})\,,
\label{Pkappa}
\eeq
so the last term in the square brackets in \eqn{new_c1pPT} is subleading.
Furthermore, we see from \eqns{s1p}{s1kappa} that
\beq
\frac{(\kT{}\cdot\kappa)^2}{\kT{}^2} = 
-\frac{1-z_1}{4z_1}\frac{s_{1\kappa}^2}{s_{1p}}+\O(\kT{})\,.
\label{Kkappa}
\eeq
Using \eqns{Pkappa}{Kkappa} in \eqn{new_c1pPT} we obtain
\eqn{direct_c1pPT}.  This gives a direct check of result
\eqn{eq:Pgqalfabeta} for the splitting tensor.

\section{The collinear limit of the spin-polarization
tensors using a helicity basis} 

In this appendix we record the collinear limit of the $\cT^{hh'}$ tensors
using a helicity basis. There are two ways to find the factorization
formula. One is to work directly with helicity amplitudes and use the
known factorization formulae with the splitting functions factorized. The
other is to project the factorization formulae of \sect{sec:collcollfact}
onto a helicity basis. We explore both ways and show that these lead to the
same expressions. The splitting functions are defined in $d = 4$
dimensions, therefore, in this appendix we set $\eps = 0$ everywhere.

Let us start with the former. In our notation the helicity amplitudes are
obtained as
\beq
\cAtree{m}(1^{\lambda_1},\dots,m^{\lambda_m}) =
\la \lambda_1,\dots ,\lambda_m\ket{m}{(0)}{(p_1,\dots,p_m)}\,.
\eeq
The helicity amplitudes can be written as a sum of products of colour
factors $C_\sigma$ and partial amplitudes $\Atree{m+1}$,
\beq
\cAtree{m}(1^{\lambda_1},\dots,m^{\lambda_m}) =
\gs^{m-2}\sum_\sigma C_{\sigma{(\{1,\dots,m\})}}
\Atree{m}\Big(\sigma(\{1^{\lambda_1},\dots,m^{\lambda_m}\})\Big)\,,
\eeq
where $\sigma{(\{1,\dots,m\})}$ denotes a certain permutation of the
labels. The allowed permutations are not important in our present
calculation.

The collinear-factorization formula for the partial amplitudes when
partons $i$ and $r$ become collinear reads (see
e.g.~\Ref{Mangano:1990by})
\beq
\Atree{m+1}(i^{\lambda_i},r^{\lambda_r},\dots) =
\Sp{-\lambda}{i^{\lambda_i}}{r^{\lambda_r}}
\Atree{m}(p_{ir}^{\lambda},\dots) + \cdots\,,
\label{eq:Atreecollfact}
\eeq
where the ellipses mean subleading terms, which are neglected in the
collinear limit.  Substituting \eqn{eq:Atreecollfact} into the
definition of the helicity-dependent tensor
\beq
\cT_{m+1}^{\lambda_i\lambda_i'}(p_i,p_r,\dots) =
\bra{m+1}{(0)}{(p_i,p_r,\dots)} \lambda_i\ra
\la \lambda_i' \ket{m+1}{(0)}{(p_i,p_r,\dots)}\,,
\label{eq:Tlalap}
\eeq
we obtain the factorization formula
\beq
\bC{ir}
\cT_{m+1}^{\lambda_i\lambda_i'}(p_i,p_r,\dots)
= 8\pi\as\,\bT_{ir}^2\sum_{\lambda\lambda'}
\la\lambda|\hP_{g_if_r}^{\lambda_i\lambda_i'}(z_r,\spa{ir})|\lambda'\ra
\cAtree{m}(p_{ir}^{\lambda},\dots)
\Big[\cAtree{m}(p_{ir}^{\lambda'},\dots)\Big]^*
\label{eq:Tlalapcollfact}
\eeq
where
\beq
\la\lambda|\hP_{f_if_r}^{\lambda_i\lambda_i'}(z_r,\spa{ir})|\lambda'\ra
= \frac12\sum_{\lambda_r\lambda_r'}
\Sp{-\lambda}{i^{\lambda_i}}{r^{\lambda_r}}
\Big[\Sp{-\lambda'}{i^{\lambda_i'}}{r^{\lambda_r'}}\Big]^*
\eeq
is the representation of the $\hP^{\alpha\beta}_{f_if_r}$ splitting
tensors on a helicity basis and $\spa{ir}$ is the spinor product as defined,
e.g., in \Ref{Mangano:1990by}. Using the known splitting functions for the
processes $q \to q_r + g_i$ and $g \to g_r + g_r$, we can easily compute
the tensors $\la \lambda | \hP_{g_iq_r}^{\lambda_i\lambda_i'} |\lambda'\ra$ and
$\la \lambda | \hP_{g_ig_r}^{\lambda_i\lambda_i'} |\lambda'\ra$.
For the case of quark splitting we have 
\beq
\la \lambda | \hP_{g_iq_r}^{\lambda_i\lambda_i'} |\lambda'\ra
= \delta_{\lambda\lambda'} P_{g_iq_r}^{\lambda_i\lambda_i'}\,,
\eeq
therefore, we have to compute only the following cases,
\beeq
P_{g_iq_r}^{++} \aand = P_{g_iq_r}^{--} =
\frac{1}{s_{ir}}\,\frac{1+z^2}{2(1-z)} =
\frac{1}{s_{ir}} \left(\frac{z}{1-z} + \frac{1-z}{2}\right) \,,
\nn \\
P_{g_iq_r}^{+-} \aand = \Big(P_{g_iq_r}^{-+}\Big)^* =
\frac{1}{\spa{ir}^2}\,\frac{z}{1-z}
\,.
\label{eq:Pgqtensor}
\eeeq
For the gluon splitting we find
\beeq
\la+|\hP_{g_ig_r}^{++}(z_r,\spa{ir})|+\ra \aand=
\la-|\hP_{g_ig_r}^{--}(z_r,\spa{ir})|-\ra =
\frac{1}{s_{ir}}\,\frac{z^3}{1-z}\,,
\nn \\
\la+|\hP_{g_ig_r}^{+-}(z_r,\spa{ir})|-\ra \aand=
\la-|\hP_{g_ig_r}^{-+}(z_r,\spa{ir})|+\ra = 0\,,
\nn \\
\la+|\hP_{g_ig_r}^{--}(z_r,\spa{ir})|+\ra \aand=
\la-|\hP_{g_ig_r}^{++}(z_r,\spa{ir})|-\ra =
\frac{1}{s_{ir}} \left(\frac{(1-z)^3}{z} + \frac{1}{z(1-z)}\right)\,,
\nn \\
\la+|\hP_{g_ig_r}^{-+}(z_r,\spa{ir})|-\ra \aand=
\la-|\hP_{g_ig_r}^{+-}(z_r,\spa{ir})|+\ra =
\frac{2}{s_{ir}}\,\frac{1-z}{z}\,,
\nn \\
\la+|\hP_{g_ig_r}^{+-}(z_r,\spa{ir})|+\ra \aand=
\la-|\hP_{g_ig_r}^{+-}(z_r,\spa{ir})|-\ra =
\la+|\hP_{g_ig_r}^{-+}(z_r,\spa{ir})|+\ra^* =
\nn \\
\aand= \la-|\hP_{g_ig_r}^{-+}(z_r,\spa{ir})|-\ra^* =
\frac{1}{\spa{ir}^2}\,\frac{z}{1-z}\,,
\nn \\
\la+|\hP_{g_ig_r}^{++}(z_r,\spa{ir})|-\ra \aand=
\la+|\hP_{g_ig_r}^{--}(z_r,\spa{ir})|-\ra =
\la-|\hP_{g_ig_r}^{++}(z_r,\spa{ir})|+\ra^* =
\nn \\
\aand= \la-|\hP_{g_ig_r}^{--}(z_r,\spa{ir})|+\ra^* =
\frac{1}{\spa{ir}^2}\,z(1-z)\,.
\label{eq:Pggtensor}
\eeeq

We can derive the factorization formula in \eqn{eq:Tlalapcollfact} from 
\eqn{Tirscollfact} by inserting the resolution of the identity in terms
of projection operators,
\beq
\sum_\lambda |\lambda\ra\la \lambda| = 1\,,
\eeq
and projecting the Lorentz indices onto helicity states with polarization
vectors $\varepsilon_\mu^\lambda \equiv \la \lambda|\mu \ra$,
\beeq
&&
\bC{ir} \cT_{m+1}^{\lambda_i\lambda_i'}(p_i,p_r,\dots) =
\nn\\ &&\qquad
=8\pi\as
\frac{1}{s_{ir}} \sum_{\lambda\lambda'}
\bra{m}{(0)}{(p_{ir},\dots)} \lambda \ra\,\la \lambda | \mu \ra \times
\nn\\ &&\qquad\quad\times
\la \lambda_i | \alpha \ra\,\la \mu |
\hP^{\alpha\beta}_{g_if_r}(z_r,\kT{r};p_{ir},n_{ir})
\ra | \nu \ra\,\la \beta| \lambda_i' \ra\,\la \nu | \lambda' \ra
\,\la \lambda' \ket{m}{(0)}{(p_{ir},\dots)}\:.
\eeeq
Comparing it to \eqn{eq:Tlalapcollfact}, we see that 
\beq
\bT_{ir}^2\,\la\lambda|\hP_{g_if_r}^{\lambda_i\lambda_i'}(z_r,\spa{ir})|\lambda'\ra
= \frac{1}{s_{ir}}
\la \lambda | \mu \ra \la \lambda_i | \alpha \ra\,\la \mu |
\hP^{\alpha\beta}_{g_if_r}(z_r,\kT{r};p_{ir},n_{ir})
\ra | \nu \ra\,\la \beta| \lambda_i' \ra\,\la \nu | \lambda' \ra\,.
\label{eq:Pgftensor}
\eeq
In order to compute the right-hand side of the equation above from
\eqns{eq:Pgqalfabeta}{eq:Pggalfabeta}, we note first that the gauge terms
do not contribute. Secondly, in \eqn{eq:Pggalfabeta} we left out terms
that do not contribute to either \eqns{eq:ddotPgq}{eq:ddotPgg} or
\eqn{eq:SpfunctionSptensor}. Including these terms and neglecting the
gauge ones, we have
\beeq
&&
\la s| \hP^{\alpha\beta}_{gq}(z,\kT{};p,n;\eps) |s'\ra =
\CF\,\delta_{ss'}
\left[
-\frac{1-z}{2}g^{\alpha\beta}
- 2\frac{z}{1 - z}\frac{\kT{}^\alpha\kT{}^\beta}{\kT{}^2}
\right]+\cdots\,,
\label{eq:Pgqalfabeta2}
\\ &&
\la \mu| \hP^{\alpha\beta}_{gg}(z,\kT{};p,n;\eps) |\nu\ra =
\nn\\&&\qquad
=2\CA
\Bigg[
  \frac{1-z}{z} g^{\alpha\mu}g^{\beta\nu}
+ \frac{z}{1 - z} g^{\mu\nu}\frac{\kT{}^\alpha\kT{}^\beta}{\kT{}^2}
+ z(1 - z) g^{\alpha\beta} \frac{\kT{}^\mu\kT{}^\nu}{\kT{}^2}+
\nn\\&&\qquad\qquad
+ \frac{z}{\kT{}^2}\left(
  g^{\mu\alpha}\kT{}^\beta\kT{}^\nu
+ g^{\beta\nu}\kT{}^\mu\kT{}^\alpha
- g^{\alpha\nu}\kT{}^\mu\kT{}^\beta
- g^{\beta\mu}\kT{}^\nu\kT{}^\alpha
\right)
\Bigg]+\cdots\,.
\label{eq:Pggalfabeta2}
\eeeq
Using the conventions for spinor products and polarization vectors
of \Ref{Mangano:1990by} and the representation of the transverse momenta
in \eqn{eq:kappakTs}, we can derive the following relations, valid in
the collinear limit,
\beq
\la \lambda | \mu \ra\,g^{\mu\nu}\,\la \nu | \lambda' \ra =
- \delta_{\lambda\lambda'}\,,
\qquad
\la \lambda | \mu \ra\,\la \lambda' | \alpha \ra\,g^{\mu\alpha} =
- \delta_{\lambda(-\lambda')}\,,
\eeq
\beq
-2\,\la \lambda | \mu \ra
\,\frac{\kT{}^\mu \kT{}^\nu}{\kT{}^2}
\,\la \nu | \lambda' \ra =
-2\,\la \lambda | \alpha \ra
\,\frac{\kT{}^\alpha \kT{}^\beta}{\kT{}^2}
\,\la \beta | \lambda' \ra =
  \delta_{\lambda\lambda'}
+ \delta_{\lambda(-\lambda')} \frac{s_{js}}{\spa{js}^2}
\eeq
and
\beq
-2\,\la \lambda | \mu \ra \,\la -\lambda | \alpha \ra
\,\frac{\kT{}^\mu \kT{}^\alpha}{\kT{}^2} = 1\,.
\eeq
Using these relations when inserting
\eqns{eq:Pgqalfabeta2}{eq:Pggalfabeta2} into \eqn{eq:Pgftensor}, we
obtain the same results as in \eqns{eq:Pgqtensor}{eq:Pggtensor}.


\begin{thebibliography}{99}

\bibitem{Giele:1991vf}
W.~T.~Giele and E.~W.~N.~Glover,
Phys.\ Rev.\ D {\bf 46} (1992) 1980.

\bibitem{Giele:1993dj}
W.~T.~Giele, E.~W.~N.~Glover and D.~A.~Kosower,
Nucl.\ Phys.\ B {\bf 403} (1993) 633
[arXiv:hep-ph/9302225].

\bibitem{Frixione:1995ms}
S.~Frixione, Z.~Kunszt and A.~Signer,
Nucl.\ Phys.\ B {\bf 467} (1996) 399
[arXiv:hep-ph/9512328].

\bibitem{Nagy:1996bz}
Z.~Nagy and Z.~Tr\'ocs\'anyi,
Nucl.\ Phys.\ B {\bf 486} (1997) 189
[arXiv:hep-ph/9610498].

\bibitem{Frixione:1997np}
S.~Frixione,
Nucl.\ Phys.\ B {\bf 507} (1997) 295
[arXiv:hep-ph/9706545].

\bibitem{Catani:1996vz}
S.~Catani and M.~H.~Seymour,
Nucl.\ Phys.\ B {\bf 485} (1997) 291
[Erratum-ibid.\ B {\bf 510} (1997) 291]
[hep-ph/9605323].

\bibitem{Moch:2004pa}
S.~Moch, J.~A.~M.~Vermaseren and A.~Vogt,
Nucl.\ Phys.\ B {\bf 688}, 101 (2004)
[arXiv:hep-ph/0403192], 
A.~Vogt, S.~Moch and J.~A.~M.~Vermaseren,
Nucl.\ Phys.\ B {\bf 691}, 129 (2004)
[arXiv:hep-ph/0404111].

\bibitem{Hamberg:1990np}
R.~Hamberg, W.~L.~van Neerven and T.~Matsuura,
Nucl.\ Phys.\ B {\bf 359} (1991) 343
[Erratum-ibid.\ B {\bf 644} (2002) 403].

\bibitem{Harlander:2002wh}
R.~V.~Harlander and W.~B.~Kilgore,
Phys.\ Rev.\ Lett.\  {\bf 88} (2002) 201801
[arXiv:hep-ph/0201206].

\bibitem{Anastasiou:2003yy}
C.~Anastasiou, L.~J.~Dixon, K.~Melnikov and F.~Petriello,
Phys.\ Rev.\ Lett.\  {\bf 91} (2003) 182002
[arXiv:hep-ph/0306192].

\bibitem{Anastasiou:2003ds}
C.~Anastasiou, L.~Dixon, K.~Melnikov and F.~Petriello,
Phys.\ Rev.\ D {\bf 69} (2004) 094008
[arXiv:hep-ph/0312266].

\bibitem{Anastasiou:2002yz}
C.~Anastasiou and K.~Melnikov,
Nucl.\ Phys.\ B {\bf 646} (2002) 220
[arXiv:hep-ph/0207004].

\bibitem{Anastasiou:2004xq}
C.~Anastasiou, K.~Melnikov and F.~Petriello,
arXiv:hep-ph/0409088.

\bibitem{Anastasiou:2005qj}
C.~Anastasiou, K.~Melnikov and F.~Petriello,
arXiv:hep-ph/0501130.

\bibitem{Anastasiou:2004qd}
C.~Anastasiou, K.~Melnikov and F.~Petriello,
Phys.\ Rev.\ Lett.\  {\bf 93} (2004) 032002
[arXiv:hep-ph/0402280].

\bibitem{Gehrmann-DeRidder:2004tv}
A.~Gehrmann-De Ridder, T.~Gehrmann and E.~W.~N.~Glover,
Nucl.\ Phys.\ B {\bf 691} (2004) 195
[arXiv:hep-ph/0403057].

\bibitem{Gehrmann-DeRidder:2004xe}
A.~Gehrmann-De Ridder, T.~Gehrmann and E.~W.~N.~Glover,
Nucl.\ Phys.\ Proc.\ Suppl.\  {\bf 135}, 97 (2004)
[arXiv:hep-ph/0407023].

\bibitem{Heinrich:2002rc}
G.~Heinrich,
Nucl.\ Phys.\ Proc.\ Suppl.\  {\bf 116} (2003) 368
[arXiv:hep-ph/0211144].

\bibitem{Anastasiou:2003gr}
C.~Anastasiou, K.~Melnikov and F.~Petriello,
Phys.\ Rev.\ D {\bf 69} (2004) 076010
[arXiv:hep-ph/0311311].

\bibitem{Binoth:2004jv}
T.~Binoth and G.~Heinrich,
Nucl.\ Phys.\ B {\bf 693} (2004) 134
[arXiv:hep-ph/0402265].

\bibitem{Heinrich:2004jv}
G.~Heinrich,
Nucl.\ Phys.\ Proc.\ Suppl.\  {\bf 135} (2004) 290
[arXiv:hep-ph/0406332].

\bibitem{Kosower:1997zr}
D.~A.~Kosower,
Phys.\ Rev.\ D {\bf 57} (1998) 5410
[arXiv:hep-ph/9710213].

\bibitem{Campbell:1998nn}
J.~M.~Campbell, M.~A.~Cullen and E.~W.~N.~Glover,
Eur.\ Phys.\ J.\ C {\bf 9} (1999) 245
[arXiv:hep-ph/9809429].

\bibitem{Weinzierl:2003fx}
S.~Weinzierl,
JHEP {\bf 0303} (2003) 062
[arXiv:hep-ph/0302180].

\bibitem{Weinzierl:2003ra}
S.~Weinzierl,
JHEP {\bf 0307} (2003) 052
[arXiv:hep-ph/0306248].

\bibitem{Kosower:2003bh}
D.~A.~Kosower,
Phys.\ Rev.\ D {\bf 71}, 045016 (2005)
[arXiv:hep-ph/0311272].

\bibitem{Gehrmann-DeRidder:2003bm}
A.~Gehrmann-De Ridder, T.~Gehrmann and G.~Heinrich,
Nucl.\ Phys.\ B {\bf 682} (2004) 265
[arXiv:hep-ph/0311276].

\bibitem{Frixione:2004is}
S.~Frixione and M.~Grazzini,
arXiv:hep-ph/0411399.

\bibitem{Gehrmann-DeRidder:2005hi}
A.~Gehrmann-De Ridder, T.~Gehrmann and E.~W.~N.~Glover,
Phys.\ Lett.\ B {\bf 612}, 36 (2005)
[arXiv:hep-ph/0501291].

\bibitem{Ridder:2005aw}
A.~Gehrmann-De Ridder, T.~Gehrmann and E.~W.~N.~Glover,
Phys.\ Lett.\ B {\bf 612}, 49 (2005)
[arXiv:hep-ph/0502110].

\bibitem{Campbell:1997hg}
J.~M.~Campbell and E.~W.~N.~Glover,
Nucl.\ Phys.\ B {\bf 527} (1998) 264
[arXiv:hep-ph/9710255].

\bibitem{Catani:1998nv}
S.~Catani and M.~Grazzini,
Phys.\ Lett.\ B {\bf 446} (1999) 143
[arXiv:hep-ph/9810389].

\bibitem{DelDuca:1999ha}
V.~Del Duca, A.~Frizzo and F.~Maltoni,
Nucl.\ Phys.\ B {\bf 568} (2000) 211
[arXiv:hep-ph/9909464].
 
\bibitem{Kosower:2002su}
D.~A.~Kosower,
Phys.\ Rev.\ D {\bf 67} (2003) 116003
[arXiv:hep-ph/0212097].



\bibitem{BCM}
A.~Bassetto, M.~Ciafaloni and G.~Marchesini,
Phys.\ Rept.\  {\bf 100}, 201 (1983).

\bibitem{AP}
G.~Altarelli and G.~Parisi,
Nucl.\ Phys.\ B {\bf 126}, 298 (1977).

\bibitem{Catani:1999ss}
S.~Catani and M.~Grazzini,
Nucl.\ Phys.\ B {\bf 570}, 287 (2000)
[arXiv:hep-ph/9908523].

\bibitem{Ellis:1980wv}
R.~K.~Ellis, D.~A.~Ross and A.~E.~Terrano,
Nucl.\ Phys.\ B {\bf 178} (1981) 421.

\bibitem{Catani:2000pi}
  S.~Catani and M.~Grazzini,
  Nucl.\ Phys.\ B {\bf 591} (2000) 435
  [arXiv:hep-ph/0007142].

\bibitem{Nagy:1998bb}
Z.~Nagy and Z.~Tr\'ocs\'anyi,
Phys.\ Rev.\ D {\bf 59} (1999) 014020
[Erratum-ibid.\ D {\bf 62} (1999) 099902]
[hep-ph/9806317].

\bibitem{Mangano:1990by}
M.~L.~Mangano and S.~J.~Parke,
Phys.\ Rept.\  {\bf 200}, 301 (1991).
\end{thebibliography}
\end{document}